\documentclass[%
reprint,amsmath,amssymb,aps,nofootinbib]{revtex4-1}

\usepackage{comment} 
\usepackage{subcaption,graphicx}
\usepackage{dcolumn}
\usepackage{bm}
\usepackage{amsmath}
\usepackage{slashed}
\usepackage{bm}%
\usepackage[colorlinks=true,linkcolor=blue]{hyperref}%
\expandafter\ifx\csname package@font\endcsname\relax\else
 \expandafter\expandafter
 \expandafter\usepackage
 \expandafter\expandafter
 \expandafter{\csname package@font\endcsname}%
\fi
\usepackage{comment}
\usepackage[normalem]{ulem}
\usepackage{color}

\newcommand{\beq}{\begin{equation}}
\newcommand{\eeq}{\end{equation}}
\def\barr{\begin{array}}
\def\earr{\end{array}}
\def\dis{\displaystyle}
\def\lapp{\mathrel{\rlap{\raise.5ex\hbox{$<$}}
                    {\lower.5ex\hbox{$\sim$}}}}
\def\gapp{\mathrel{\rlap{\raise.5ex\hbox{$>$}}
                    {\lower.5ex\hbox{$\sim$}}}}

\newcommand{\bsp}{\begin{split}}
\newcommand{\esp}{\end{split}}
\newcommand{\bit}{\begin{itemize}}
\newcommand{\eit}{\end{itemize}}
\definecolor{darkcyan}{cmyk}{1,0,0,0.4}

\graphicspath{ {./} }

\begin{document}

\title{Axion Icebergs: Clockwork ALPs at hadron colliders}
\author{Srimoy Bhattacharya}
\email{bhattacharyasrimoy@gmail.com}
\affiliation{School of Physics and Institute for Collider Particle Physics, University of the Witwatersrand, Johannesburg, Wits 2050, South Africa}
\author{Debajyoti Choudhury}
\email{debchou.physics@gmail.com}
\affiliation{Department of Physics and Astrophysics, University of Delhi, Delhi 110007, India}
\author{Suvam Maharana}
\email{suvam.maharana\textunderscore528@tifr.res.in (Corresponding author)}
\affiliation{Department of Physics and Astrophysics, University of Delhi, Delhi 110007, India}
\affiliation{Department of Theoretical Physics, Tata Institute of Fundamental Research, Homi Bhabha Road, Mumbai 400005, India}
\author{Tripurari Srivastava}
\email{tripurarisri022@gmail.com}
\affiliation{Department of Physics and Astrophysics, University of Delhi, Delhi 110007, India}
\affiliation{Institute of Particle Physics and Key Laboratory of Quark and Lepton Physics (MOE),
Central China Normal University, Wuhan, Hubei 430079, China}

\preprint{APS/123-QED}


\begin{abstract}
Scenarios with multiple pseudoscalars are interesting as they usually
tend to provide a framework to naturally realize a light axion with a
large decay constant which has rich applications in cosmology,
especially in the context of inflation and light dark matter
physics. On the other hand, from a particle physics perspective, this
facilitates a solution to the strong CP problem with a low
Peccei-Quinn symmetry breaking scale. One such realization is
afforded within the framework of the \emph{clockwork} mechanism where
the axion can have suppressed couplings with the gluons or photons
while its companion axion-like particles (ALPs) have relatively
unsuppressed couplings, thereby facilitating detectability. We study a
minimal clockwork model for the QCD axion invoking a KSVZ-like setup
and examine the visibility of its unique multi-ALP $(a_n)$ signature
at the LHC, the most sensitive channel being $p p \to a_n \, (+ \,
{\rm additional \, jets})$ followed by $a_n \to \gamma \gamma$. In
congruence with the astrophysical and cosmological bounds for the
axion, a striking feature emerges for the case of light ALPs $(m \sim
\mathcal{O}(10 \, {\rm GeV}))$ wherein the mass-splittings among the
former are so small that the signal profile mimics that of a single
broad resonance, or an \emph{axion iceberg}. The scenario is found to
be imminently testable by the end of LHC's Run 3 phase for an
integrated luminosity of $\sim 300 {\rm \, fb^{-1}}$. A larger average
ALP mass, on the other hand, results in multiple closely-spaced peaks
with a characteristic signal profile, and would be expected to be seen
at the forthcoming HL-LHC. Possible additional signals are also
listed.
  \end{abstract}

\maketitle
\flushbottom
\section{Introduction}
\label{sec:intro}
The strong CP problem is, perhaps, one of the strongest motivations to
look for new dynamics beyond the Standard Model (SM) of particle
physics. Of particular interest has been establishing the existence of
the light pseudoscalar particle --- popularly called an axion or, more
specifically, the QCD axion --- predicted by the Peccei-Quinn (PQ)
mechanism \cite{PQPhysRevLett, WeinbergPhysRevLett,
  WilczekPhysRevLett} which is, arguably, the simplest and the most
elegant solution\footnote{There exist interesting alternatives to the
PQ scenario, most notable among them being solutions based on a
massless up quark
\cite{Georgi:1981be,Kaplan:1986ru,Choi:1988sy,Banks:1994yg,
  Dine:2014dga} and those invoking spontaneous CP violation ({\em a
  la} the Nelson-Barr mechanism)~\cite{Nelson:1983zb,
  Barr:1984qx}. However, the $m_u=0$ solution within the standard
theory currently stands in tension with the lattice QCD result of the
topological mass contribution \cite{Alexandrou:2020bkd}. Similarly,
models based on the Nelson-Barr mechanism are also subject to several
theoretical challenges \cite{Dine:2015jga}.} to the problem proposed
to date.

Some theoretical and cosmological issues such as the quality problem
and domain wall formation do question the mechanism's general
viability, although there have been several noteworthy developments
towards mitigating these issues
\cite{Lazarides:1982tw,Chatterjee:2019rch,Ibe:2019yew,Zhang:2023gfu}.
Nevertheless, the axion solution offers an interesting and important
domain to explore phenomenologically, more so because the axion can
potentially also play the role of an ultralight dark matter candidate
\cite{Duffy:2009ig,Chadha-Day:2021szb}.

The axion is typically expected to have a very large decay constant ($f \gtrsim \mathcal{O}(10^{10} \,
  \mbox{GeV})$) ---with the lower limit depending on its mass---so as
to be consistent with the current limits (drawn from astrophysical and
cosmological observations as well as from direct search experiments)
on its effective couplings with photons (as also electrons and
nucleons) \cite{ParticleDataGroup:2024cfk,AxionLimits}. In other
words, this essentially mandates that the PQ symmetry breaking scale
($f_{PQ}$) must be hierarchically larger than the electroweak (EW)
scale.  Furthermore, the QCD axion mass typically scales inversely
with the decay constant and, hence, a large $f$ also implies a very
light mass for the axion. The extremely large value of $f_{PQ}$ (and,
the attendant smallness of the axion mass), thus, introduces a new
hierarchy problem.

In addition, the tiny mass of the minimal QCD axion and its extremely
feeble couplings render it nearly invisible to most of the current and
future experimental probes, especially in the context of high energy
experiments. Consequently, recent collider searches have tended to
concentrate, instead, either on axion-like particles (ALPs), that have
no role in the strong CP issue
\cite{Bauer:2018uxu,dEnterria:2021ljz,Florez:2021zoo}, or on heavy QCD
axions obtained by way of introducing new or extended confining
sectors
\cite{Rubakov:1997vp,Berezhiani:2000gh,Fukuda:2015ana,Hook:2019qoh,Agrawal:2017ksf,Gherghetta:2020keg}
The latter case, although interesting in its own right, typically does
not capture the cosmological attributes of a light axion, \emph{i.e.} a role as a possible dark matter candidate and/or as
the inflaton field in the so-called \emph{natural inflation} scenario
\cite{Freese:1990rb, Adams:1992bn}. In view of this, a question
naturally arises as to whether the discovery of a heavy axion would
necessarily negate the existence of the light QCD axion. This aspect
assumes particular significance as there are hints of several local
excesses in searches for such scalar/pseudoscalar resonances
\cite{ATLAS:2022abz,CMS-PAS-HIG-20-002,ATLAS:2021hbr,CMS:2018nsh,ATLAS:2024itc,Schmieden:2021pvm}
without these being robust enough to merit a discovery claim.

Scenarios with multiple aligned axions \cite{Kim:2004rp,Choi:2014rja}
could, in principle, address these issues without assuming a large
hierarchy in the individual decay constants and yet allowing for a
quasi-flat direction in the field space, thereby leading to an
effective decay constant $f_a \gg f_{PQ}$ for the QCD
axion. Ref.\cite{Higaki:2015jag} sought to exploit
this with $f_{PQ}$ close to the EW scale and ALPs with masses $\gtrsim
m_{EW}$.

In this work, we seek to present a theoretically better-motivated
model that also leads to a very \emph{distinctive} multi-ALP
spectrum. With ALP masses $\lesssim m_{EW}$, this exposition would not
only be complementary to the scenarios of
refs.\cite{Higaki:2015jag,Farina:2016tgd} but would also confront
features in existant data at the
LHC~\cite{ATLAS:2022abz,CMS-PAS-HIG-20-002,ATLAS:2021hbr,CMS:2018nsh,ATLAS:2024itc,Schmieden:2021pvm}. Furthermore, the light QCD axion inherent to the
scenario, apart from being a viable DM candidate, would likely be
testable at the current and/or next generation of dedicated
experiments. To this end, we focus on the clockwork paradigm,
originally proposed in refs.\cite{Choi:2014rja,Choi:2015fiu,Kaplan:2015fuy,Giudice:2016yja}, as it
automatically generates hierarchical couplings or mass scales through
the localization of the lightest particle in the theory space lattice
defined by $N$ fields which interact with each other through
nearest-neighbour couplings with a strength characterized by an ${\cal
  O}(1)$ parameter $q$. The massive modes, on the other hand, are
delocalized over the entire lattice.
  
To realize a QCD axion within such a
construction, we consider a model with a clockwork sector comprising
$N+1$ complex scalars $\Phi_{j}(j=0,...,N)$, each charged under a
unique $U(1)$ symmetry. Inspired by the KSVZ model
\cite{Kim:1979if,Shifman:1979if}, we introduce a pair of new,
$SU(2)_L$ singlet, coloured Weyl fermions ($\Psi_{L,R}$), chiral under
the $N$th group $U(1)_N$, which couple to the $N$-th complex
scalar. With only a single species of new coloured fermions, the
putative domain wall problem, on account of a spontaneous breaking of
the discrete symmetry present in the QCD-induced potential for the
axion, is well under control, and such a KSVZ-like model is free of
the problems that often beset, say, a DFSZ-like
\cite{Zhitnitsky:1980tq, Dine:1981rt} scenario.

A spontaneous breaking of the full symmetry ($U(1)^{N+1}$) at a scale
$f$ (an analogue of $f_{PQ}$) in the clockwork sector, alongwith an
explicit breaking of the symmetry to a single factor $U(1)_{CW}$ due
to the nearest-neighbour interactions characterized by an effective
mass scale $m \ll f$, leads to one massless Nambu-Goldstone Boson (NGB) and $N$ pseudo-NGBs in the
spectrum. The chiral anomaly, then, induces a coupling of the light
NGB to the gluonic topological operator ($\Tilde{G}G$) with an
exponentially large effective decay constant $\sim q^N f$, thanks to
the localization mechanism. Below the confinement scale, this leads to
a small mass for the light NGB through nonperturbative
effects. Therefore, for $q>1$ and a sizable number of CW fields, the
light NGB can be identified with the QCD axion even for the fundamental SSB scale $f$ being as low as a few
TeVs. The pNGBs, on the other hand, act as ALPs with masses of the
order $\sim m q$ (thereby creating a mass gap) and couple to gluons
with an effective decay constant comparable to the scale $f$.  The
nonzero hypercharge of $\Psi_{L,R}$ not only allows them to decay, but
also engenders loop-induced couplings of the ALPs to a pair of photons
and $Z$'s. At a hadron collider (current or future), ALPs are
dominantly produced through gluon-gluon fusion, and their most
prominent, yet unique, signature would be the occurrence of multiple
closely spaced resonance peaks in the $\gamma \gamma$, $ZZ$ and $Z
\gamma$ channels. The closeness in the ALP masses emerges naturally in
the model as a consequence of the clockwork construction wherein the
mass-splitting varies as $\sim 2m/N$, (with $N \gtrsim 10$ for
adequate enhancement of the axion decay constant).  Understandably,
for lighter ALPs ($m \lesssim 100$ GeV), the mass-splittings would be
small, and in certain cases so much so that the currently operating
detectors at the LHC might not be able to completely resolve the
individual peaks, leading to the appearance of the entire ALP spectrum
as a single broad resonance. We argue that both signals (multiple
closely spaced narrow peaks or a single anomalously broad peak) are
unique to the clockwork paradigm and can be generated neither by a
single, phenomenologically viable ALP candidate, nor by a generic
model of multiple ALPs without invoking an \emph{ad hoc} fine-tuning
in the parameter space. Additionally, such signatures are
complementary to the possibility of observing \emph{periodic} (in invariant mass) signals in the case of relatively heavier
ALPs, \emph{e.g.} in the spirit of the analyses carried out in
  the context of the gravity model in refs.\cite{Giudice:2017fmj,
    Beauchesne:2019tpx,ATLAS:2023hbp}. Driven by this notion, we
assume benchmark points with ALP masses at or below the EW scale and
investigate their detection prospects in the diphoton channel at the
LHC. The most interesting outcome is found in the case of the mass
parameter $m$ being small, where although the individual resonances
are barely visible over the background, the significance of the
conjoint signal profile, composed of multiple overlapping peaks, is
projected to reach the discovery threshold even as early as by the end
of the LHC's Run 3 phase. An analysis of the diphoton
final state, which is arguably the cleanest channel to probe the ALPs,
is also motivated by the number of dedicated collider searches for
light scalar resonances, near and below the EW scale, performed by the
ATLAS \cite{ATLAS:2022abz, ATLAS:2023pja} and CMS
\cite{CMS-PAS-HIG-20-002} collaborations in the recent years. Thus,
our study also appeals to the case for further experimental probes of
light scalar or pseudoscalar particles in the diphoton channel,
scanning not only for one but multiple closely-spaced candidates or
even in the form of a broad resonance.

Note that although the introduction of a clockwork sector does not
alter or improve upon the basic features of the PQ mechanism, it does
provide ways to mitigate certain issues present in the minimal QCD
axion models, which in turn makes this avenue even more desirable to
explore. For one, the possibility of a low scale $f_{PQ}$ helps
improve the \emph{quality} of the axion solution against putative
quantum gravity effects\footnote{Since the degree of the global
symmetry breaking depends on the exact manifestation of quantum
gravity effects near the Planck scale, a robust estimate of the
quality problem in our model, given the complexity of the global
symmetry involved, is difficult to obtain. Nevertheless, under the
assumption that the leading QG effect arises as a dim-5 operator of
the form $(g/M_{Pl})\left(|\Phi_j|^2 \right)^2 \Phi_{j'}$ in the IR, the required
fine-tuning for an acceptable PQ solution can be characterized in
terms of an upper limit on the coupling $|g|$, which, in a model with
$f_{PQ} \sim \mathcal{O}(1 \, \mbox{TeV})$, can be estimated to be
$|g|\lesssim 10^{-12}/N^{4}$. E.g., $N \sim 30$ implies $|g| \lesssim
10^{-18}$. Although this seems a very stringent upper limit in the
absolute sense, it is nevertheless many orders
of magnitude larger than what is required in the minimal models with
$f_{PQ}\sim 10^{12}$ GeV, namely $|g|\lesssim 10^{-55}$
\cite{Kamionkowski:1992mf}. Thus, even with the aforementioned
ambiguities, it can be qualitatively argued that the clockwork
scenario offers some respite from the severe quality issues of the
typical axion models, if not a solution. The severity may be relaxed
further with the imposition of additional discrete symmetries in the
UV in order to restrict lower dimensional operators. However, we
  desist from doing so.}
\cite{Kamionkowski:1992mf,Higaki:2015jag}. Secondly, the geometric
progression of the light axion's distribution over the lattice can be
exploited to obtain an enhanced coupling with photons, independent of
the coupling to the gluons, thereby opening a window in the parameter
space that falls outside of the usual QCD band on the photon-coupling
vs axion-mass plane. This aspect was studied in detail in
ref.\cite{Farina:2016tgd}.

The structure of the paper is as follows. In section \ref{sec:model},
we define a minimal clockwork model for a KSVZ-like QCD axion and demonstrate how it addresses the
strong CP problem. We identify, in section \ref{sec:colliders}, four
benchmark points of the model suitable for probing the ALPs through
the gluon-fusion channel at the LHC. In section \ref{sec:sig_bkg}, we
discuss the corresponding signal and background profiles for
$\sqrt{s}=13$ TeV and an integrated luminosity $\mathcal{L}=138 \,
{\rm fb}^{-1}$. Section \ref{sec:VLQ}, on the other hand, briefly
describes the implications of the vector-like quark and the heavy
radial scalars present in the model. Finally, we summarize the results
and conclude in section \ref{sec:conclusion}.

\section{The model} \label{sec:model}
We begin by constructing a clockwork realization of the minimal QCD
axion, adopting a KSVZ-like~\cite{Kim:1979if,Shifman:1979if}
scenario. In the ensuing, we first detail the structure and spectrum
of the clockwork sector and then proceed towards describing how the
KSVZ setup can be implemented in the context of the clockwork
mechanism.
\subsection{Clockwork scalars}
We define the clockwork (CW) sector in terms of $N+1$ complex scalars
$\Phi_j$, each charged under a global Abelian group $U(1)_j$ described
by the Lagrangian
\beq
\barr{rcl}
{\cal L}_{\rm CW} & = & \dis {\cal L}_1 + {\cal L}_2 \, ,
     \\[1.5ex]
{\cal L}_1 & \equiv & \dis \sum_{j=0}^{N}
\left[(\partial_{\mu} \Phi_j^\dagger) (\partial^{\mu} \Phi_j) - \lambda \left(
  \Phi_j^{\dagger} \Phi_j - \frac{f^2}{2}\right)^2\right] \, ,
\\[2.5ex]
{\cal L}_2 & \equiv & \lambda'
\Lambda^{3-q}\sum_{j=0}^{N-1} \Phi_j^{\dagger} \Phi_{j+1}^q +
\mbox{h.c.} \ ,
\earr
\eeq
where $\lambda, \lambda'$ are dimensionless real constants and
$\Lambda \, (\ll f)$ is a characteristic scale associated with ${\cal
  L}_2$ (the exponent $q$ being, as yet, unspecified). To
  understand better the pieces ${\cal L}_{1,2}$, let us examine them
  in isolation.  In the absence of ${\cal L}_2$ ({\em i.e.,} in the
limit $\lambda' \to 0$), the Lagrangian has a $U(1)^{N+1}$ global
symmetry, the spontaneous breaking of which, at the scale $f$ (through
$\langle \Phi_j^{\dagger} \Phi_j \rangle = f^2/2$ for all $j$) would
lead to $(N+ 1)$ Goldstone bosons $\pi_j$, with the complex scalars
being representable as
  \beq
  \Phi_j =\frac{1}{\sqrt{2}}(\phi_j + f)e^{i \pi_j /f} \ ,
  \label{eq:PNGB}
\eeq
where $\phi_j$ are the corresponding massive scalars.  Note that this
breaking mechanism still retains the additional global discrete
symmetry $\Phi_j \leftrightarrow \Phi_k$ inherent in ${\cal L}_1$.

The large global symmetry, discrete or continuous (and the attendant
plethora of Goldstones in the broken phase) is neither well-motivated
nor desirable. Clearly, it can be broken by arbitrary terms expressing
interactions between the scalars. On the other hand, at least a single
$U(1)$ needs to be present\footnote{As previously mentioned, an exact
global symmetry is unrealistic because even in the absence of new
symmetry breaking effects one would expect the global $U(1)$ to be
explicitly broken by quantum gravity effects. The Planck suppressed
symmetry violating terms thus generated can, in general, upset the PQ
solution through corrections to the QCD axion potential to be
discussed later \cite{Kamionkowski:1992mf}. While the small $f_{PQ}$ in our scenario already alleviates this
so-called \emph{quality problem}, it may be further assuaged by
ensuring that the Planck suppressed operators in the theory appear
only at a sufficiently high mass-dimension, {\em e.g.}, by considering
the CW scalars to be composites of entities belonging to new confining
sector(s) \cite{Randall:1992ut,Coy:2017yex}.} so that its spontaneous
breaking could lead to the axion. The interaction Lagrangian ${\cal
  L}_2$ serves exactly this purpose\footnote{The requirement $\Lambda
\ll f$ at best introduces a \emph{little} hierarchy and for $q=2$, as
we would adopt, it represents only a soft breaking of the
symmetry.}. Representing perhaps the simplest set of operators which
preserve a single $U(1)$ while exhibiting the clockwork mechanism, it
can be understood in terms of nearest-neighbour interaction
terms\footnote{In general, one could also have non-nearest neighbour
interactions while keeping the CW symmetry intact, an example being a
term of the form $\Phi^{\dagger}_j
\Phi^{q/p}_{j+1}...\Phi^{q^{p}/p}_{j+p}$. With reference to
eq.(\ref{eq: eigval}) such a term adds, to the average mass scale of
the pNGBs, a contribution $m_a^{(p)}\sim m (q^{p}/p)$
\cite{Ben-Dayan:2017rvr}. For the case $q=2$ (which would be our
primary choice as a benchmark for the collider analysis to be
discussed) renormalizability necessitates $p=1,2$ for which the
overall pNGB mass scale changes only slightly with $m_a \sim
\sqrt{2}m_a^{(1)}$. The case for $q=3$ is a little different with $m_a
\sim 3 m_a^{(1)}$, although the qualitative aspects of our results and
conclusions in this work would be applicable just as well.} that
exhibit locality in a theory space and, hence, the lattice defined by
the fields $\Phi_j$ (each specifying a \textit{site} in the theory
space) can be regarded as the fifth dimension\footnote{A continuum
theory that reproduces this theory on discretization, though, requires
the introduction of certain additional features and we desist from
discussing it.}  As for the exponent $q$ in ${\cal L}_2$, a
non-integer value would imply compositeness, thereby adding a further
layer of complications.  The clockwork mechanism, which will be
employed in the following discussions, necessitates $q>1$. On the
other hand, for $q > 3$ one would obtain nonrenormalizable operators
in $\mathcal{L}_{CW}$ implying that the explicit symmetry breaking
takes place due to new dynamics operational at the heavy scale
$\Lambda \gg f$ and beyond. Furthermore, $q>3$ would also tend to
destabilize the vacuum. To avoid such issues and for simplicity we
consider a renormalizable theory with $1 < q \leq 3$.

With the introduction of ${\cal L}_2$, the erstwhile discrete symmetry
is completely broken, whereas the $U(1)^{N+1}$ symmetry breaks
explicitly to one combination $U(1)_{CW}$ corresponding to the
generator,
\beq \label{eq:cwgen} \mathcal{Q}_{CW}=\sum_j
\frac{\mathcal{Q}_j}{q^j}, \eeq
with $\mathcal{Q}_j$ being the generators (charges) corresponding to
the individual $U(1)_j$'s. Consequently, only one combination of the
$\pi_j$'s would now be a true Goldstone boson, while all the rest
would gain masses much smaller than $f$. In other words, the latter
are now only
  pNGBs. To be specific, the full spin-$0$
Lagrangian, in the broken phase, can be written as,
\begin{widetext}
\beq \label{eq:lagscal}
\barr{rcl}
\mathcal{L}^{Full}_{\Phi}
 = \sum_{j=0}^{N} \Big[\frac{1}{2}\partial_{\mu} \phi_j \partial^{\mu} \phi_j - \frac{1}{4}\lambda \left( \phi_j + f \right)^4\Big] +
2^{(1-q)/2}\lambda' \Lambda^{3-q} \dis \sum_{j=0}^{N-1}(\phi_j + f)(\phi_{j+1} + f)^q \cos{\frac{\pi_j - q\pi_{j+1}}{f}} \,.
\earr
\eeq
\end{widetext}
The mass eigenvalues of the pseudoscalar system are given by
\beq \label{eq: eigval} m_n^2 = \begin{cases} 0 \, & n=0 \\ m^2 \,
  \left[1+q^2 - 2q\cos{\frac{n \pi}{N+1}}\right] \, & n \neq 0
\end{cases},
\eeq
where $m^2 \equiv 2^{(1-q)/2}\lambda' \Lambda^{3-q} f^{q-1}$ (thus,
$m_n^2 \ll f^2$ as was expected).  The transformation relation between
the unphysical basis $(\pi_j)$ and the physical basis $(a_n)$ is
specified by a matrix $C$ ({\em viz.} $a_n \equiv \sum_{j=0}^{N} C_{n
  j} \pi_j $) with elements
\beq \label{eq:eigvec}
\begin{split}
C_{0 j} &= \mathcal{N}_0 q^{-j} \, , \\
C_{n j} &= \mathcal{N}_n
\left[ q \sin{\frac{j n \pi}{N+1}} - \sin \frac{(j+1) n
    \pi}{N+1}\right].
\end{split}
\eeq
Here $\mathcal{N}_0$ and $\mathcal{N}_n$ are the normalization
factors
\beq \mathcal{N}_0 =
\sqrt{\frac{q^2-1}{q^2-q^{-2N}}} \ , \qquad \quad \mathcal{N}_{n>0} =
\frac{m}{m_n}\sqrt{\frac{2 }{N+1}} \, .
\eeq
It is evident from eq.(\ref{eq:eigvec}) that while the heavy
pseudoscalars are nearly delocalized over the entire lattice,
the massless mode --- which will subsequently assume the role of the
QCD axion $(a_0)$ --- is localized towards the $j=0$ site, {\em i.e.}
it has an exponentially suppressed overlap with $\pi_N$ by virtue of
the clockwork mechanism. Interestingly, this feature will eventually
ensure that the light axion has a hierarchically large decay
constant\footnote{Note that the usage of the terminology ``decay
constant'' here is different from that encountered in the context of
processes such as $\pi^{\pm} \to l^{\pm}\nu$, where $\pi^{\pm}$ are
the charged QCD pions.} as compared to the heavy axions in the
spectrum.

\subsection{A KSVZ axion}
Now that we have a massless pseudoscalar in the clockwork spectrum, we
would like to promote it to the status of a realistic QCD axion. This
is most easily carried out by embedding the clockwork sector in any
one of the usual QCD axion models. Owing to the singlet nature of the
CW scalars, the KSVZ construction is the simplest one. To this end, we
introduce one generation of Weyl fermions $\Psi_L, \Psi_R$ with the
following nontrivial charges under the SM gauge group and the $N$-th
(global) Abelian group\footnote{The chiral configuration only at the
$N$-site ensures that the color anomaly is localized at a single site
which enables us to exploit the CW mechanism to generate a large
effective decay constant for the axion, as we will see shortly. On the
other hand, while it is possible to have the $\Psi_{L,R}$ charged
nonchirally under all the additional $U(1)$s, this only adds a layer
of complexity without any qualitative changes in the phenomenology.}
of the CW sector ($G_{SM} \times U(1)_N$),
\begin{center}
\begin{tabular}{||c c c c c||} 
 \hline
 Field & $SU(3)_c$ & $SU(2)_L$ & $U(1)_Y$ & $U(1)_N$ \\ [0.5ex] 
 \hline\hline
 $\Phi_N$ & $1$ & $1$ & $0$ & $\xi$ \\ 
 \hline
 $\Psi_L$ & $3$ & $1$ & $Y_{\Psi}$ & $\xi_L$ \\ 
 \hline
 $\Psi_R$ & $3$ & $1$ & $Y_{\Psi}$ & $\xi_R$ \\ 
 \hline
\end{tabular} \, ,
\end{center}
with the restriction that $\xi=\xi_L - \xi_R$. This is reminiscent of
the chiral PQ charges typically assigned to the new ({\em i.e.},
non-SM) fermions in KSVZ models. The SM fields are assumed to be
uncharged under the new $U(1)$'s. For $\xi_L \neq \xi_R$, the new
singlet quark cannot have a bare mass term. On the other hand, for the
aforementioned choice of $\xi=\xi_L - \xi_R$, the new fermion may now
interact with the CW sector through a Yukawa term localized at the
$N$-th site of the lattice. However, with just this interaction, the
new quark would be absolutely stable and, hence, phenomenologically
untenable. The situation can be remedied only if it couples with the
SM fermions which, for our charge assignments, can be most simply
ensured when $\xi_R=0$ (and, thus, $\xi_L=\xi$) alongwith an
appropriate choice of $Y_\Psi$. To be specific, we choose $Y_\Psi =
2/3$, thereby allowing $\Psi$ to mix with the up-type quarks.  The
complete set of Yukawa interactions is, thus, given by
\beq \label{eq:yukawa}
\begin{split}
& \mathcal{L}_{\Psi} \supset  -\lambda_{\Psi} \Phi_N
\Bar{\Psi}_L \Psi_R \\
&-
\sum_{\alpha=1}^{3}y^{(\alpha)}_{\Psi}\Bar{Q}^{(\alpha)}_L \Tilde{H}\Psi_R -\sum_{\alpha=1}^{3}y^{'(\alpha)}_{\Psi}\Phi_N \Bar{\Psi}_L u^{(\alpha)}_R +
\mbox{h.c.} \, ,
\end{split}
\eeq
\\
where $Q^{(\alpha)}_L$ are the SM quark doublets, $u_R^{(\alpha)}$ the
up-type singlets and $H$ the SM Higgs doublet. As for the
dimensionless couplings $\lambda_{\Psi}$, $y^{(\alpha)}_{\Psi}$ and
$y^{'(\alpha)}_{\Psi}$, the first determines the mass of the $\Psi$
while the other two determine its decay rate. For the sake of
simplicity, we assume that $\Psi_{L, R}$ couple to only the
third-generation quarks, thereby precluding sizable flavour-changing
neutral currents involving the first two generations.

In addition, the CW sector can also couple with the SM Higgs
through\footnote{For simplicity, we assume uniform value of the
  coupling $\lambda_{\Phi-H}$ for all $j$. Relaxing this only adds
  layers of complications without any qualitative change.}
\beq \mathcal{L}_{\Phi-H} = -\lambda_{\Phi H} \sum_{j=0}^{N}
\Phi_j^{\dagger} \Phi_j H^{\dagger}H \, .
\eeq
We will come back to reviewing the dynamics of the heavy radial
scalars and the quarks, alongwith the flavour constraints on the
Yukawa couplings, in a later section. Alluding to the core objective
of the study, we first describe the low-energy physics of the
pseudoscalars in the discussions below.

After SSB in the clockwork sector, we have
\begin{widetext}
\beq \label{eq:lagpsi2}
\mathcal{L}_{\Psi} \supset -\frac{1}{\sqrt{2}}\lambda_{\Psi} (\phi_N + f)e^{i\xi \pi_N /f} \Bar{\Psi}_L \Psi_R - y_{\Psi}\Bar{Q}^{(3)}_L \Tilde{H}\Psi_R - \frac{1}{\sqrt{2}}y'_{\Psi}(\phi_N + f)e^{i\xi \pi_N /f}\Bar{\Psi}_L u^{(3)}_R + \mbox{h.c.} \, .
\eeq
\end{widetext}
A rephasing of the new fermions with respect to the Goldstone field
$\pi_N$ leads to the following interaction terms with the SM gauge
bosons via the chiral anomaly \cite{Alonso-Alvarez:2018irt} (see appendix \ref{sec:appanomaly} for a
discussion)
\beq \label{eq:lagpivv}
\mathcal{L}_{\pi VV} = - g_{\pi GG} \, \pi_N G^{A \mu \nu} \Tilde{G}^{A}_{\mu \nu}-g_{\pi BB} \, \pi_N B^{\mu \nu} \Tilde{B}_{\mu \nu},
\eeq
with the coefficients given by
\beq
|g_{\pi GG}|= \frac{g^2_s }{32 \pi^2 f_{\rm eff}} \, , \quad |g_{\pi BB}| = \frac{2 N_c g'^2 Y^2_{\Psi}}{32 \pi^2 f_{\rm eff}} \, ,
\eeq
and an effective scale
  defined through $f_{\rm eff} \, \equiv f/|\xi|$. Note that while
$f$ continues to determine the masses of the $\Psi$ and the $\Phi_N$,
it is $f_{\rm eff}$ that encapsulates the pNGB decay constants.

\subsubsection{The Peccei-Quinn Mechanism} \label{subsubsec:pqm}
\begin{figure*}[tbp] 
\centering
      \includegraphics[scale=0.3,keepaspectratio=true]{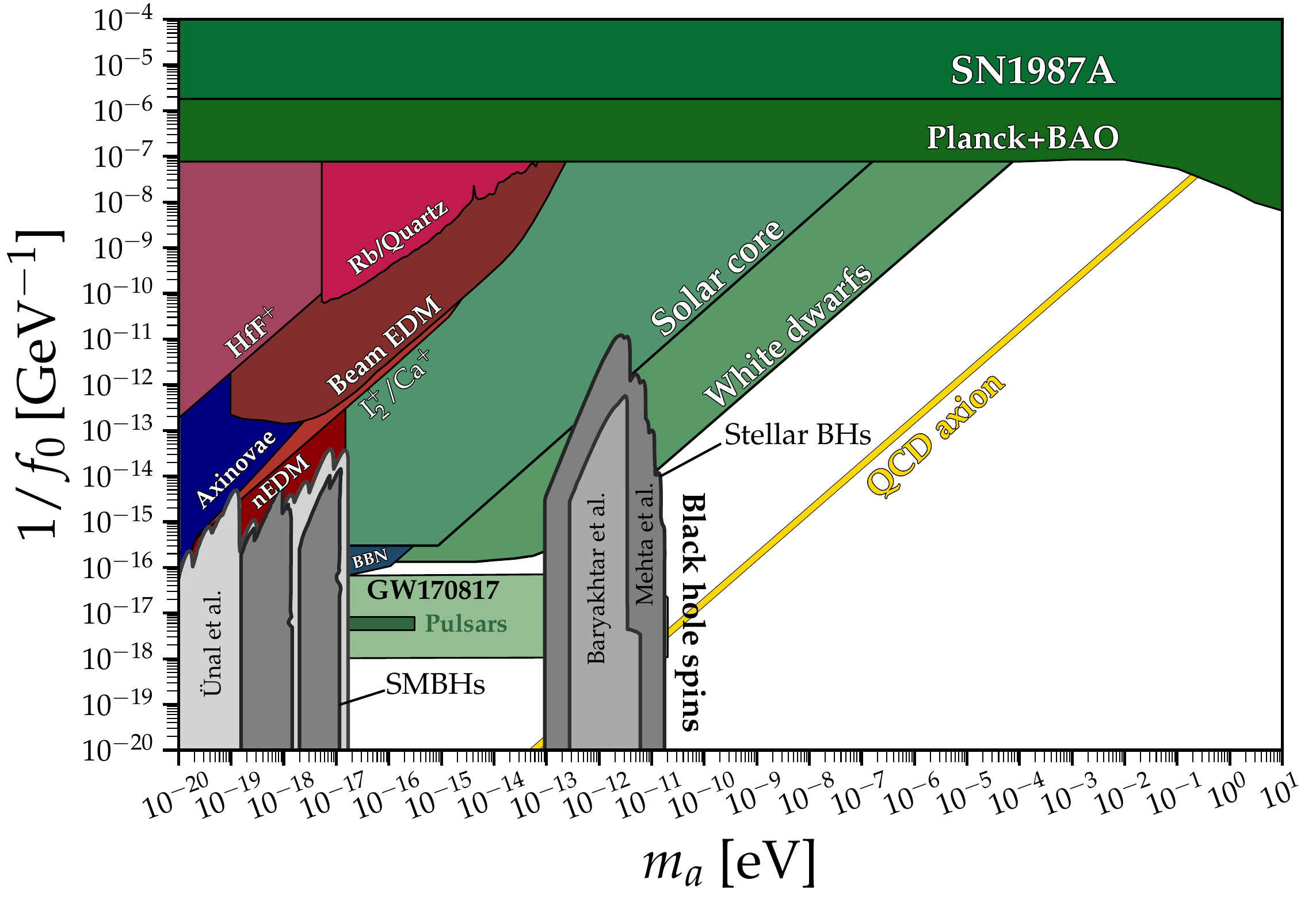}
	\caption{Limits on the effective decay constant $f_0$ of the
          light QCD axion vs its mass. Plot adapted from
          ref.\cite{AxionLimits}.} \label{fig:axilimf0}
\end{figure*}

In order for the light pseudoscalar to be the quintessential QCD
axion, it must offer a solution to the strong CP problem. Although
ref.\cite{Farina:2016tgd} already has a brief discussion in this
context, we outline here, for completeness, how the clockwork
Lagrangian can be consistent with the usual Peccei-Quinn
mechanism. The foremost requirement for the mechanism is the presence
of a global symmetry (at least an accidental one) in the theory. In
our case, this symmetry would correspond to the residual clockwork
symmetry described in the preceding section (see
eq.(\ref{eq:cwgen})). Now, in eq.(\ref{eq:lagpivv}), we can perform a
shift in the $\pi_N$ field so as to exactly cancel the CP violating
topological term $-\Bar{\theta}(\alpha_S / 8 \pi) \Tilde{G}^{\mu \nu}
G_{\mu \nu}$. This, however, is a true cancellation, and a potential
solution to the strong CP problem, only when $\pi_N$ does not acquire
a nonzero vacuum expectation value (VEV) after employing the constant
shift. It is important to note that the explicit breaking of the
$[U(1)]^{N+1}$ symmetry in the clockwork sector, courtesy the
nearest-neighbour interactions, does not generate a nonzero VEV for
the pseudoscalar fields thanks to the residual shift symmetry
$U(1)_{CW}$ \cite{Banerjee:2018grm}. The only other source from where
$\pi_N$ may acquire a nontrivial potential are
nonperturbative effects below the QCD confinement scale, leading
  to an effective potential of the form
\cite{DIVECCHIA1980253,GrillidiCortona:2015jxo}
    \beq \label{eq:axipot}
    V(\pi_N)=-m_{\pi}^2 f_{\pi}^2\sqrt{1-\frac{4 m_u m_d}{(m_u + m_d)^2}\sin^2{\left( \frac{\pi_N}{2 f_{\rm eff}}\right)}}
    \eeq
 where $f_{\pi}$ and $m_{\pi}$ are the pion decay constant and mass,
respectively, and $m_{u,d}$ are the masses of the light SM quarks $u$
and $d$. This shows that $\pi_N$ does not, indeed, acquire a nonzero
VEV\footnote{Naively, the potential would have a degenerate set
    of minima at $\langle \pi_N \rangle = 2 n \pi f_{\rm eff}$ for
    integral values of $n$. However, such nonzero VEVs are unphysical as they can always be shifted to zero by the
    residual symmetry transformation $\pi_N \to \pi_N + 2 n \pi
    f_{\rm eff}$ as a result of the $2\pi$ periodicity of the
    potential.}. Expanding the potential in powers of $\pi_N$, the quadratic term engenders a mass correction to $\pi_N$
\beq
\Delta m_{\pi_N}^2 = \frac{f_{\pi}^2 m_{\pi}^2}{f_{\rm eff}^2}
\frac{m_u m_d}{(m_u + m_d)^2} \, .
\eeq
Going to the physical basis of the CW axions, this generates a small mass for the
lightest state, {\em i.e.}, the KSVZ QCD axion,
given by
\beq \label{eq:laximass}
m_{a_0} \simeq \frac{f_{\pi} m_{\pi}}{f_{\rm eff} \, q^N}
\frac{\sqrt{m_u m_d}}{(m_u + m_d)}, \eeq
and also leads to mass corrections for the heavy modes, albeit orders
of magnitude smaller than $m_{n>0}$.  Clearly, the light axion has a
mass that is exponentially suppressed by the factor $q^N$ as a result
of the clockwork localization. Consequently, one need not assume a
very large scale $f$ in this scenario to accommodate a QCD axion (see
Fig.\ref{fig:axilimf0}). For example, with $q=2$ and $N \gtrsim 15$
one obtains sub-eV masses for the $a_{0}$ even for $f \lapp 1$~TeV
(equivalently, $f_{\rm eff}$ in the few hundred GeVs range). From
eq.(\ref{eq:lagpivv}), it is apparent that such an additional
suppression also appears in the axion's coupling with gluons and the
electroweak bosons above the confinement scale in the form of the
effective decay constant $f_0= {\cal N}_0^{-1} q^N f_{\rm eff} \approx q^N f_{\rm eff}$. Therefore, with a
nominal choice of values for the CW parameters $q$ and $N$, and a
PQ-esque scale $f_{\rm eff}$ which is not too far from the EW scale,
we can have a viable QCD axion that is well within the current
experimental bounds on $f_0$ with respect to the light axion mass
\cite{ParticleDataGroup:2022pth,AxionLimits}. These constraints are
summarised in Fig.\ref{fig:axilimf0} where the yellow line reflects
the mass vs $f_0$ relation as given in eq.(\ref{eq:laximass}) for a
QCD axion.  The plot shows the limits from various experimental
upper bounds on the electric dipole moment of the neutron and the
electron \cite{Schulthess:2022pbp,Abel:2017rtm, Roussy:2020ily} as
well as from atomic and molecular transition experiments
\cite{Madge:2024aot,Zhang:2022ewz}. On the other hand, the
observational limits shown include those obtained from axion star
decays (assuming post-inflationary SSB and axion star formation)
\cite{Fox:2023xgx}, Big Bang Nucleosynthesis (constraining axion DM)
\cite{Blum:2014vsa}, black hole spins
\cite{Mehta:2020kwu,Baryakhtar:2020gao,Unal:2020jiy}, pulsars and
solar core \cite{Hook:2017psm}, binary neutron star gravitational wave
\cite{Zhang:2021mks}, CMB (Planck) and baryon acoustic oscillation
\cite{Caloni:2022uya}, SN1987a \cite{Lucente:2022vuo,Springmann:2024ret} and white dwarfs
\cite{Balkin:2022qer}. Additionally, one also finds constraints from
neutron star cooling as discussed in ref.\cite{Gomez-Banon:2024oux,Kumamoto:2024wjd}.

It is perhaps prudent to remark at this juncture about the possible
nature and implications of the topological defects that may arise in
the multi-axion configuration of the clockwork sector. The spontaneous
breaking of the $U(1)^{N+1}$ symmetry at a temperature $T \sim f \gg
\Lambda_{QCD}$ (post inflation and reheating) is
  expected to lead to the formation of $N+1$ cosmic strings bound to
one another through domain walls~\cite{Long:2018nsl, Higaki:2016jjh,
  Lee:2024toz}. Such a string-wall network would, in principle, tend
to radiate light axions as well as ALPs (as also gravitational waves
\cite{Higaki:2016jjh,Lee:2024toz}) during the course of its
cosmological evolution towards the epoch of the QCD phase
transition\footnote{The ALPs, owing to their relatively small decay
  constant, decay promptly once they are produced.}. In a
  canonical framework, these emissions are estimated to scale as
$f^2$ and, therefore, constitute a large fraction of the axion
abundance in the usual QCD axion cosmology where $f \sim f_0 \gtrsim
10^9 \, \mbox{GeV}$. In the present
  context, though, the clockwork mechanism generates an exponential
separation between the SSB scale $f$ and the effective decay constant
of the light axion $f_0$, such contributions to the axion abundance
is exceedingly suppressed for $f \sim
\mathcal{O}(\mbox{TeV})$.  At the epoch of the QCD phase
transition, the induced potential of eq.(\ref{eq:axipot}) defines a
unique vacuum for the system associated with a domain wall number
$N_{DW}=1$ --- a direct consequence of invoking only a single
generation of the extra $SU(2)$ singlet coloured fermion
$\Psi$. Similar to the case of a single axion
\cite{Barr:1986hs,Kim:1986ax,Chang:1998tb}, such a potential
destabilizes the string-wall network and leads to its eventual
collapse\footnote{Thus, apart from the obvious simplicity, the choice
  of a single species of the singlet heavy quark possesses the
  important feature of alleviating the domain wall problem
  \cite{Sikivie:1982qv} typical to axion models.}, with its energy
being released primarily in the form of axions and ALPs. This
contribution to the axion abundance, too, can be estimated to be
insignificant, with the axion emission being suppressed by the factor
$\sim (m_{a_0}/m)^3$ \cite{Long:2018nsl}.  
  
While a certain abundance of the axion as dark matter is expected to
be generated via radiation and collapse of the topological defects in
the clockwork system, a robust estimate of the same can only be
obtained from a dedicated numerical simulation which is extremely
complicated to perform for a multi-axion setup, especially for $N \gg
\mathcal{O}(1)$. In the absence of such analyses, we admit to a
reasonable assertion, based on the aforementioned arguments, that for
the SSB scale of interest $f\sim \mathcal{O}(\mbox{TeV}) \ll f_{0}$
the dominant contribution to the axion DM abundance emanates from the
misalignment mechanism~\cite{Fox:2004kb}, namely

\beq \label{eq:aximis} \Omega h^2 \simeq 0.2 \left(
\frac{f_0}{10^{11}{\rm GeV}}\right)^{7/6}\left( \frac{
  \theta_0^2}{\pi^2/3}\right)
\eeq
where $\theta_0$ is the average
initial misalignment angle of the QCD axion.
This implies that for the natural choice of
$\theta_0 \sim \mathcal{O}(1)$ in eq.(\ref{eq:aximis}), the constraint
from overclosure demands $f_0 \lesssim 10^{11} \,
  \mbox{GeV}$. The experimental and observational limits demand, on
  the other hand, $ f_0 \gtrsim 10^{9} \, \mbox{GeV}$. The allowed
  window is then wide enough and comparable with the one usually
  encountered for the traditional QCD axion. Indeed, this is why it
is also the target region for several cavity haloscope experiments
such as ADMX, CAPP, HAYSTAC, QUAX, etc. (see \emph{e.g.} figure 90.5
of the PDG review on axions \cite{ParticleDataGroup:2024cfk}).

In the spirit of these arguments, we would be assuming, in the
discussions to follow, benchmarks that are consistent with the above
mentioned bounds on $f_{0}$.

\subsection{Axion physics above the QCD confinement scale}
Concerned with the physics observable at high energy colliders, it
suffices to consider the theory at energy scales below the EW symmetry
breaking scale but significantly above the QCD confinement scale. At
these scales, the axion Lagrangian becomes
\beq \label{eq:lagpivv2}
\begin{split}
\mathcal{L}_{\pi vv} = & - g_{\pi gg} \, \pi_N G^{A \mu \nu} \Tilde{G}^{A}_{\mu \nu}-g_{\pi \gamma \gamma} \,\pi_N  F^{\mu \nu} \Tilde{F}_{\mu \nu} \\
&- g_{\pi \gamma Z} \, \pi_N F^{\mu \nu} \Tilde{Z}_{\mu \nu} - g_{\pi ZZ} \, \pi_N Z^{\mu \nu} \Tilde{Z}_{\mu \nu}.
\end{split}
\eeq
The coefficients in this case are given by
\[
\barr{rclcrcl} \label{eq:coup}
g_{\pi gg} & = & \dis \frac{\alpha_s}{8\pi f_{\rm eff}} \, ,
& \qquad & 
g_{\pi \gamma Z}  & = & \dis
\frac{-4 N_c s^2_w \alpha_{EM} Y^2_{\Psi}}{8 \pi f_{\rm eff} s_w c_w} \, ,
\\[3ex]
g_{\pi \gamma \gamma} & =  & \dis
\frac{2 N_c \alpha_{EM}  Y^2_{\Psi}}{8 \pi f_{\rm eff}} \, ,
& &
g_{\pi ZZ} & = & \dis
\frac{2 N_c s^4_w \alpha_{EM}  Y^2_{\Psi}}{8 \pi f_{\rm eff} s_w^2 c_w^2} \, ,
\earr
\]
where $s_w \equiv \sin\theta_w$ and $c_w \equiv \cos\theta_w$ with
$\theta_w$ being the Weinberg angle.  For the mass-eigenstates, the
  couplings are simply scaled by the mixing parameters,
  \emph{viz.}
  \beq
  g^{(n)}_{a gg} = g_{\pi gg} (f_{\rm eff} \to f_n)
  \eeq
where $f_n \equiv f_{\rm eff}/C_{nN}$ denotes the effective decay
constant of the $n$-th physical axion. Using
  eq.(\ref{eq:eigvec}), it is then readily apparent that the decay
constants of the heavy pseudoscalars are hierarchically smaller than
that of the light axion. For a SSB scale $f \sim
\mathcal{O}(\mbox{TeV})$, this engenders significantly enhanced
couplings of the heavy axions with gluons and the electroweak bosons
as compared with those associated with typical light ALP
candidates. This enhancement, in principle, can facilitate resonant
production of the heavy CW axions at hadron colliders, which is going
to be the main subject of discussion in the ensuing sections.

\section{Clockwork axions at colliders}
\label{sec:colliders}
As is already described above, the (light) axion, when compatible with
the constraints from astrophysical and low-energy
experiments~\cite{Semertzidis:2021rxs}, is nearly invisible at high
energy collider experiments.  What seems promising, instead, is the
prospect of probing the heavy pseudoscalars at the LHC and its
forthcoming high-luminosity upgrade (HL-LHC). In this case, the
production would be dominated by the gluon fusion channel with the
hadronic cross-section given by

\beq \label{eq:partonic_cs}
\sigma(pp \to a_{n})=K^{0}_{\sigma}\frac{d \mathcal{L}_{gg}}{d \Hat{s}}\Bigg|_{\Hat{s}=m_n^2}
\frac{\pi^2}{8 m_n}\Hat{\Gamma}^{(n)}_{gg},
\eeq
where $d \mathcal{L}_{gg}/d \hat{s}$ is the gluon-gluon luminosity and
the $K$-factor $K^{0}_{\sigma}$ encapsulates the higher order QCD
corrections. Using the \texttt{MSTW2008nnlo68} parton densities,
ref.\cite{Mariotti:2017vtv} estimates $K^{0}_\sigma \approx 3.7$.
(including the full NNLO and approximate N$^3$LO corrections
\cite{Ball:2013bra, Bonvini:2014jma, Bonvini:2016frm, Ahmed:2016otz})
over the pseudoscalar mass range $40-125$ GeV . In the absence of a
full computation of the QCD corrections, we make the reasonable
extrapolation that $K^{0}_{\sigma}\sim 3.7$ for the mass range
$10-125$ GeV. For ALP masses beyond $150$ GeV we assume a more
conservative $K^{0}_{\sigma}=2.5$ \cite{Ahmed:2016otz}.

The leading (two-body) decay channels and the corresponding
widths for the ALPs are 
\begin{widetext}
\beq
\begin{split}
  \Gamma(a_n \to gg) &\equiv K_{gg}\Hat{\Gamma}^{(n)}_{gg} = K_{gg}
 |C_{nN} g(x_n)|^2
    \frac{\alpha_S^2}{32 \pi^3} \frac{m_n^3}{f_{\mbox{eff}}^2} \\
    \Gamma(a_n \to \gamma \gamma) &\equiv \Gamma^{(n)}_{\gamma \gamma} = 9 Y_{\Psi}^4
     |C_{nN} g(x_n)|^2
    \frac{\alpha_{EM}^2}{64 \pi^3} \frac{m_n^3}{f_{\mbox{eff}}^2} \\
    \Gamma(a_n \to Z \gamma) &\equiv \Gamma^{(n)}_{Z \gamma} = 9 Y_{\Psi}^4
     |C_{nN} g(x_n)|^2
    \frac{\alpha_{EM}^2 \tan^2{\theta_w}}{8 \pi^3} \frac{(m_n^2-m_Z^2)^3}{m_n^3 f_{\mbox{eff}}^2} \\
    \Gamma(a_n \to Z Z) &\equiv \Gamma^{(n)}_{Z Z} = 9 Y_{\Psi}^4 |C_{nN} g(x_n)|^2
    \frac{\alpha_{EM}^2 \tan^4{\theta_w}}{64 \pi^3} \, \frac{(m_n^2-4m_Z^2)^{3/2}}{f_{\mbox{eff}}^2}  
\end{split},
\eeq
\end{widetext}
where $x_n \equiv 4 m_\Psi^2 / m_n^2$ and the loop-integral $g(x)$ is defined as

\beq
g(x)\equiv
\left\{
\barr{lcl}
\dis    x \left[\sin^{-1}(x^{-1/2})\right]^2\,& \qquad \quad & x \geq 1\\[1.5ex]
\dis \frac{-x}{4} \left[ \ln\frac{1+\sqrt{1 - x}}{1-\sqrt{1-x}} - i \pi\right]^2 \, & & x < 1 \, .
\earr
\right.
\eeq
In the expression for $\Gamma(a_n \to
  gg)$, the QCD correction amounts to $K_{gg}=2.1$
\cite{Djouadi:2005gi}.  Clearly, the ALPs decay dominantly to two
gluons followed by decays to photons (and to $Z\gamma$ and $ZZ$
depending on kinematic feasibility). It is easy to
infer---within the narrow width
approximation--- that the dijet cross-section would exceed the
diphoton one by a factor of \footnote{Note that $K_{\gamma\gamma}
  \approx 1$.} $[2/(9
  Y_{\Psi}^4)]K_{gg}(\alpha_S^2/\alpha_{EM}^2)$. However, as far as
detection at hadron colliders is concerned, the diphoton final state,
understandably, offers better sensitivity as compared to the dijet
channel. We will, therefore, consider only the processes $pp \to a_n
\to \gamma \gamma$ for the case in hand.

The diphoton final state has been extensively studied in the
literature in the context of spin-0 and spin-2 resonances, with
dedicated searches by the ATLAS and CMS
collaborations spanning almost the entire range of masses accessible
with current sensitivities. With measurements thus far reporting, at
best, only marginal excess of events over the SM background, the most
recent analyses from
ATLAS~\cite{ATLAS:2022abz,ATLAS:2021uiz,ATLAS:2024bjr} and
CMS~\cite{CMS-PAS-HIG-20-002, CMS-PAS-EXO-22-024} place the strongest
limits to date on the $pp \to \gamma \gamma$ fiducial cross-section
for spin-0 resonances over a wide mass bracket ranging from nearly $10
\, \mbox{GeV}$ to a few TeVs. Being optimistic that some new physics
indeed exists beyond the SM, this broadly implies two possibilities
---first, that the energy scale of new physics lies
beyond the reach of the LHC and, second,  that new
dynamics exist within the LHC's energy reach, albeit with feeble
interaction strengths with the SM so as to be inaccessible with the
current luminosity reach. In the following, we choose the latter
viewpoint and explore viable scenarios where the diphoton
cross-sections for the CW pseudoscalar spectrum could be very close to
the current exclusion limits at the LHC obtained from the Run-2 data,
such that a sizable enhancement in the signal significance can be
achieved by the end of the ongoing Run-3 phase. This is interesting
because even if a $5\sigma$ discovery looks improbable at the LHC,
such enhancements could hint at a potential discovery at the HL-LHC
taking into account the projected integrated luminosity of $\sim 3000
\, \mbox{fb}^{-1}$, {\em i.e.} nearly 20 times the luminosity achieved
by the end of LHC Run-2 ($\sim 138 \, \mbox{fb}^{-1}$). We consider
for our study three benchmark points categorized by the ALP
masses\footnote{While we do not consider ALP masses below $10$ GeV,
collider searches for lighter axions in general have been studied in
refs.\cite{CidVidal:2018blh,Acanfora:2024spi}.}  in the range $10-30$
GeV, $35-105$ GeV (with two sub-categories in this case) and $150-450$
GeV, respectively, and have as guiding references the ATLAS analyses
carried out in refs.~\cite{ATLAS:2022abz,ATLAS:2021uiz,ATLAS:2024bjr}.

For each of the benchmark points, we choose $\xi = 3$ and $q =
  2$ (as the middle point of the theoretically allowed range) and also
  assume $Y_\Psi = 2/3$ so that the $\Psi$ may mix with the top quark. 
The other particulars are as follows---

\begin{figure}[!ht] 
\includegraphics[scale=0.29,keepaspectratio=true]{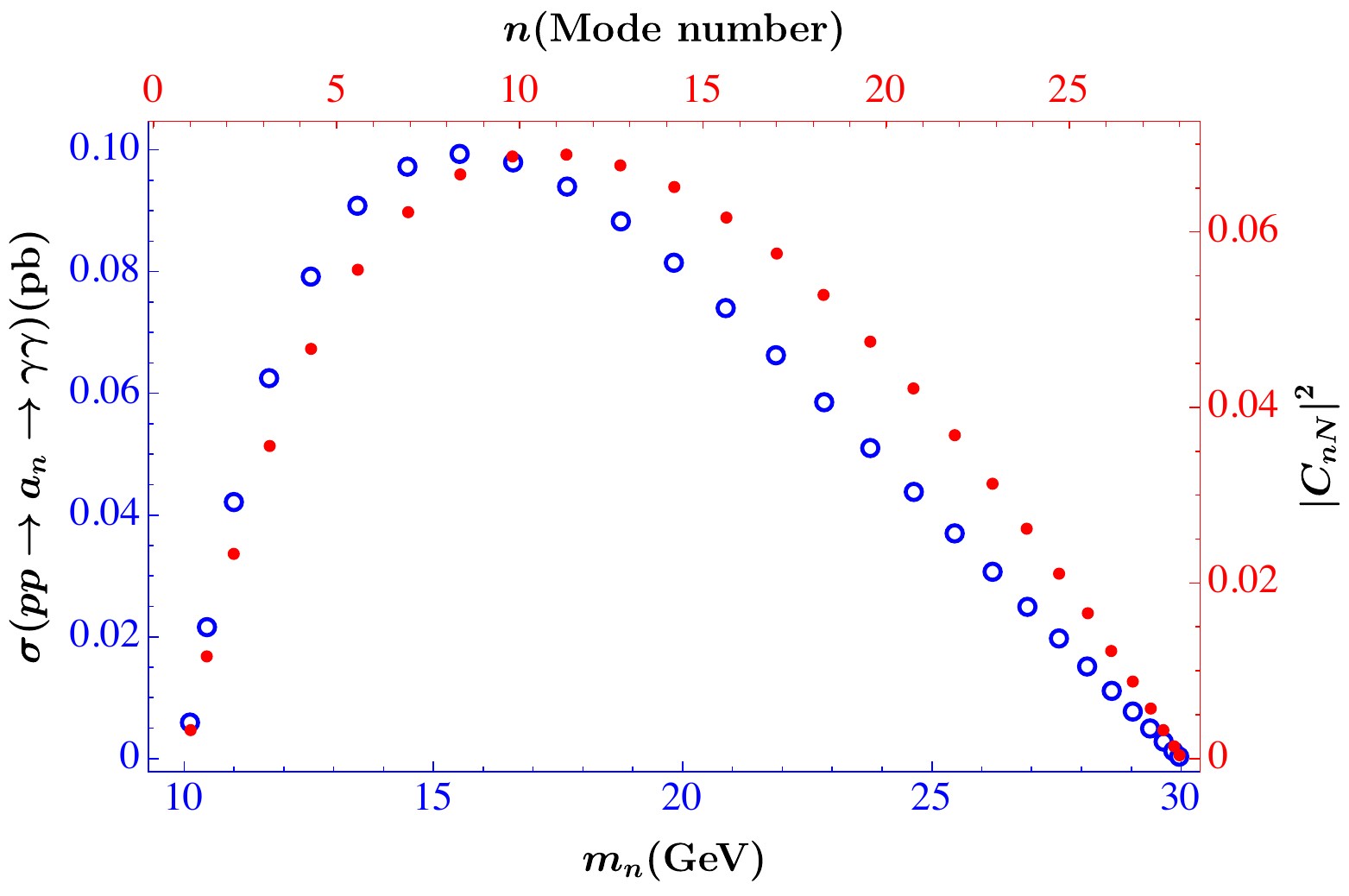}
\includegraphics[scale=0.27,keepaspectratio=true]{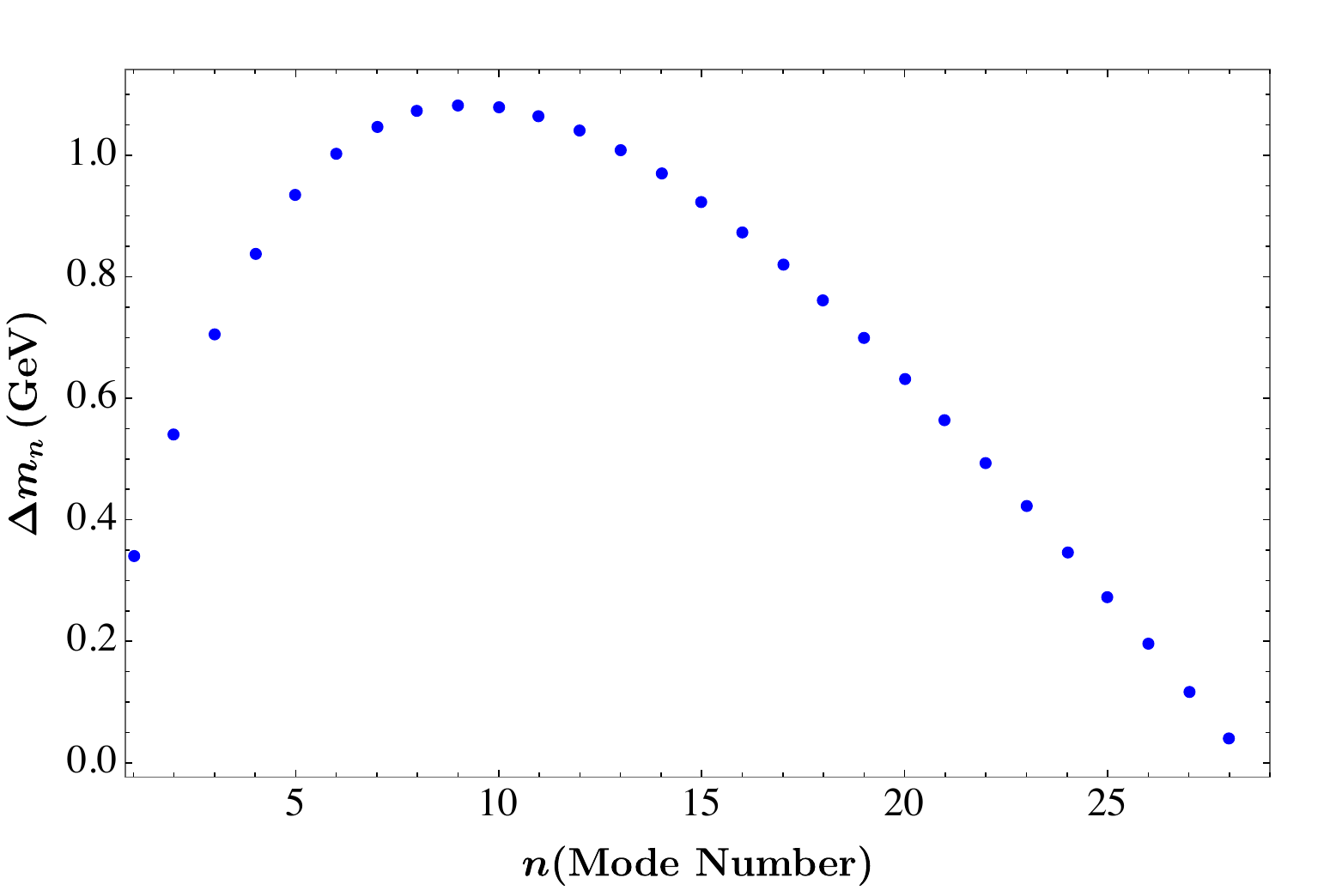}
\caption{\textbf{(top)} Masses and couplings for individual ALPs and corresponding diphoton cross-sections for benchmark I. 
The vertical axis on the right shows the extent of the field $\pi_N$ contained in the mass eigenstates $a_n$. \textbf{(bottom)} Mass-splittings between consecutive ALPs for benchmark I.} \label{fig:10gevf1000analytical}
\end{figure}
\textbf{Benchmark I: } We assume the parameter values $N=28$, $m=10$
GeV and $f=1000$ GeV in the CW sector. The resulting light axion in
this case has a decay constant $f_0 \approx 9 \times 10^{10}$
  GeV and a mass $m_{a_0} \approx 6.4 \times 10^{-5}$ eV. The
corresponding ALP masses span the range $\sim 10-30$ GeV. This
particular choice of the CW parameters, especially the sizable number
of particles $N$, is also motivated by the fact that it is compatible with the light axion being a DM candidate\footnote{The
    notion of the KSVZ axion as dark matter has also been explored in
    ref.\cite{Alonso-Alvarez:2023wig}, albeit in the traditional one
    scalar setup.} (with an acceptable relic density and consistency with cosmological observations)---as discussed in section
  \ref{subsubsec:pqm}---for $\mathcal{O}(1)$ values of the misalignment
  angle\footnote{See figures 4 and 5 in ref.\cite{Long:2018nsl}
      for a summary of the various cosmological constraints on the CW
      parameter space.}. With these choices in place, we can
determine the diphoton cross-section in the narrow-width approximation
(NWA).  Fig.\ref{fig:10gevf1000analytical} (top) shows the cross-sections for
each of the ALPs (integrated over the full final state phase-space)
computed for a center of mass energy $\sqrt{s}=13$ TeV using the
\texttt{NNPDF2.3LO} PDF set \cite{NNPDF:2014otw}. Note that the
cross-section peaks for an intermediate mode of the spectrum, the
exact identity of which depends on the periodic variation of the
couplings due to the clockwork mixing as well as the energy dependence
of the parton-parton luminosity. Fig.\ref{fig:10gevf1000analytical} (bottom),
on the other hand, shows $\Delta m_n$ (the mass-splitting between
consecutive CW modes) along the full spectrum, with the average
mass-splitting being given by $\Delta m \sim 2m/N$. It is interesting
to note that, with the consecutive differences $\Delta m_n \lesssim 1$
GeV over the entire extent of the ALP spectrum, the splittings are
comparable with the prevailing detector resolution. Consequently, the
individual resonances may not be entirely resolved and the events
corresponding to the full spectrum (subject to the assumed bin size)
may even appear as a single broad resonance\footnote{This was also
  hinted at in ref.\cite{Higaki:2015jag}.} within an envelope of mass
width $\sim \, 2 m$ in the diphoton invariant mass distribution.

\begin{figure}[!ht] 
\centering
\includegraphics[scale=0.29,keepaspectratio=true]{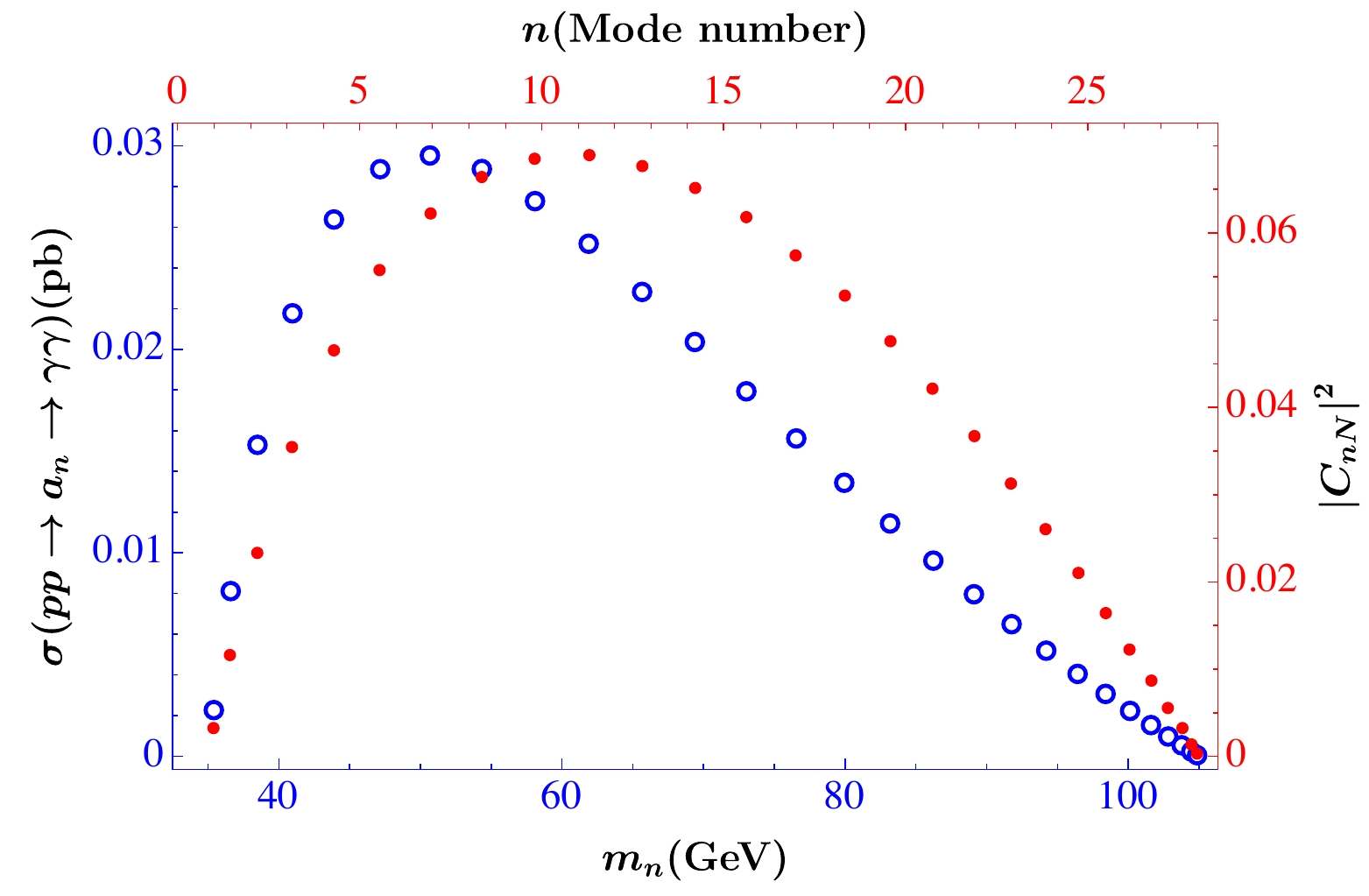}
\vskip 10pt
\includegraphics[scale=0.29,keepaspectratio=true]{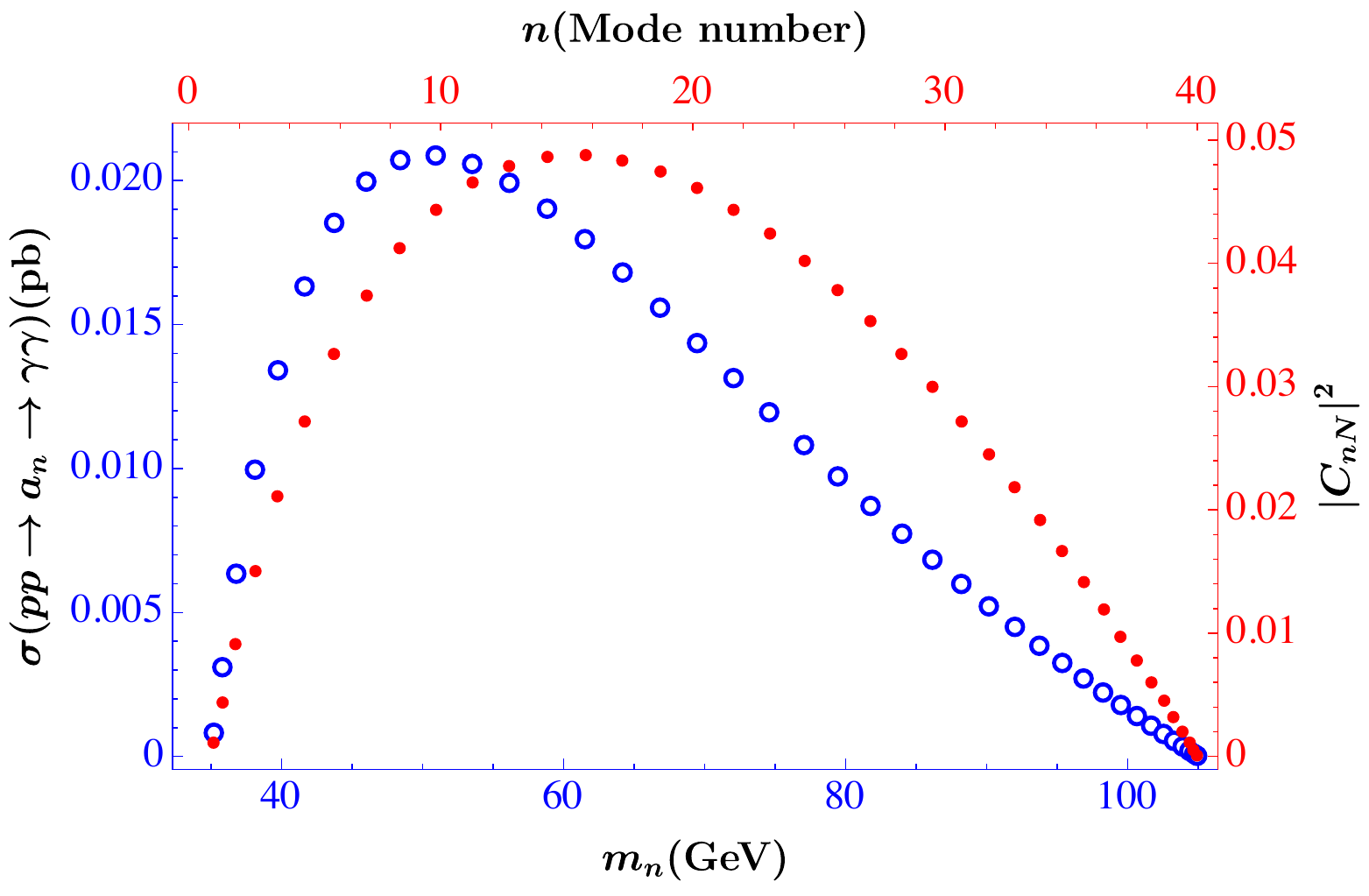}
\caption{Masses and couplings for individual ALPs and corresponding diphoton cross-sections.  \textbf{(top)} benchmark II(a) and
   \textbf{(bottom)} benchmark II(b).}
\label{fig:35gevf1000n28analytical}
\end{figure}

\textbf{Benchmark II(a):} The parameter configuration in the CW
  sector for this case is --- $m=35$ GeV, $N=28$ and $f=1000$ GeV. The
  light axion has the same decay constant and mass as in benchmark I and, thus, is still
    a potential DM candidate, while the ALP masses now range from
  $35$ to $105$ GeV. The corresponding diphoton cross-sections are
  shown in Fig.\ref{fig:35gevf1000n28analytical}.  Due to an increase in
  the value of the parameter $m$ (by a factor of 3.5) the
  mass-splittings in this case, readily inferred from eq.(\ref{eq:
    eigval}), are nearly triple in value compared to that in the
  preceding case. Therefore, in stark contrast with benchmark I, one
  expects to discern at least a few individual peaks in the invariant
  mass distribution of the signal events. We will see that this is
  indeed the case.

\textbf{Benchmark II(b):} Now, we assume a larger set of CW scalars,
namely $N=40$, while keeping the rest of the parameters the same as in
II(a). Then, from eq.(\ref{eq:laximass}) it is clear that the light
axion mass lowers down to a value $m_{a_0} \sim 1.5 \times
10^{-8}$ eV and the effective decay constant increases to $f_0
\approx 4 \times 10^{14}$ GeV. Note that an enhancement in $f_0$ of
this magnitude is incompatible with the DM constraint unless $\theta_0
\ll \mathcal{O}(1)$. Although certain cosmological scenarios could
exist where a small initial misalignment angle seems plausible (see
ref.\cite{Fox:2004kb} for a discussion), we refrain from delving into
those details and, from a strictly phenomenological standpoint of
studying different ALP signatures at the colliders, assume a small
enough value for $\theta_0$ such that the axion relic density is
acceptable. The mass splittings in this case fall in the range $\sim
0.5 - 2.7$ GeV --- smaller than what we obtained in benchmark II(a)
but not small enough to be completely unresolved at the detector.
Consequently, the distribution of signal events, as a function of the
diphoton invariant mass is expected to be an undulating one.

\begin{figure}[!h] 
\includegraphics[scale=0.29,keepaspectratio=true]{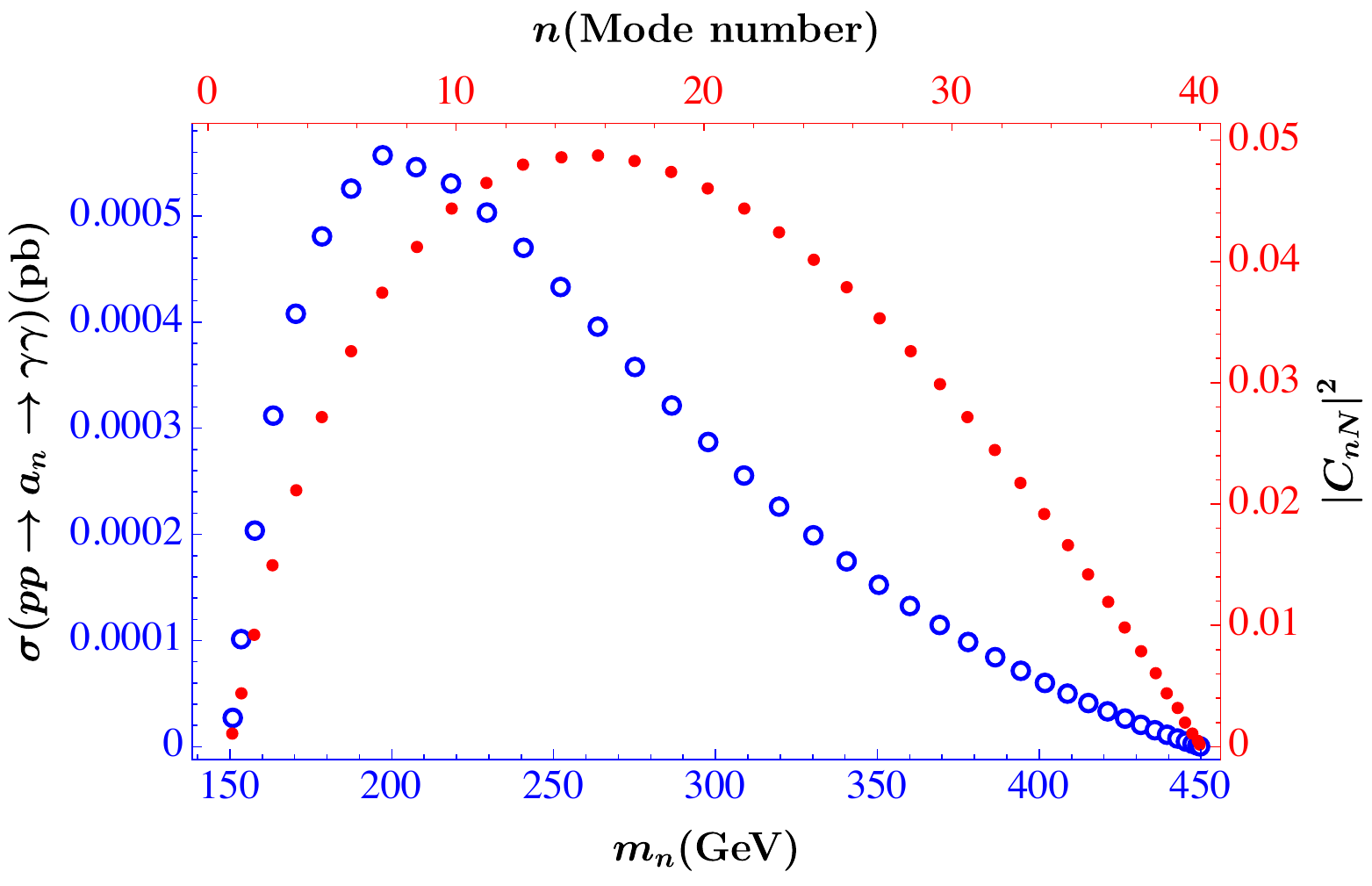}
\includegraphics[scale=0.27,keepaspectratio=true]{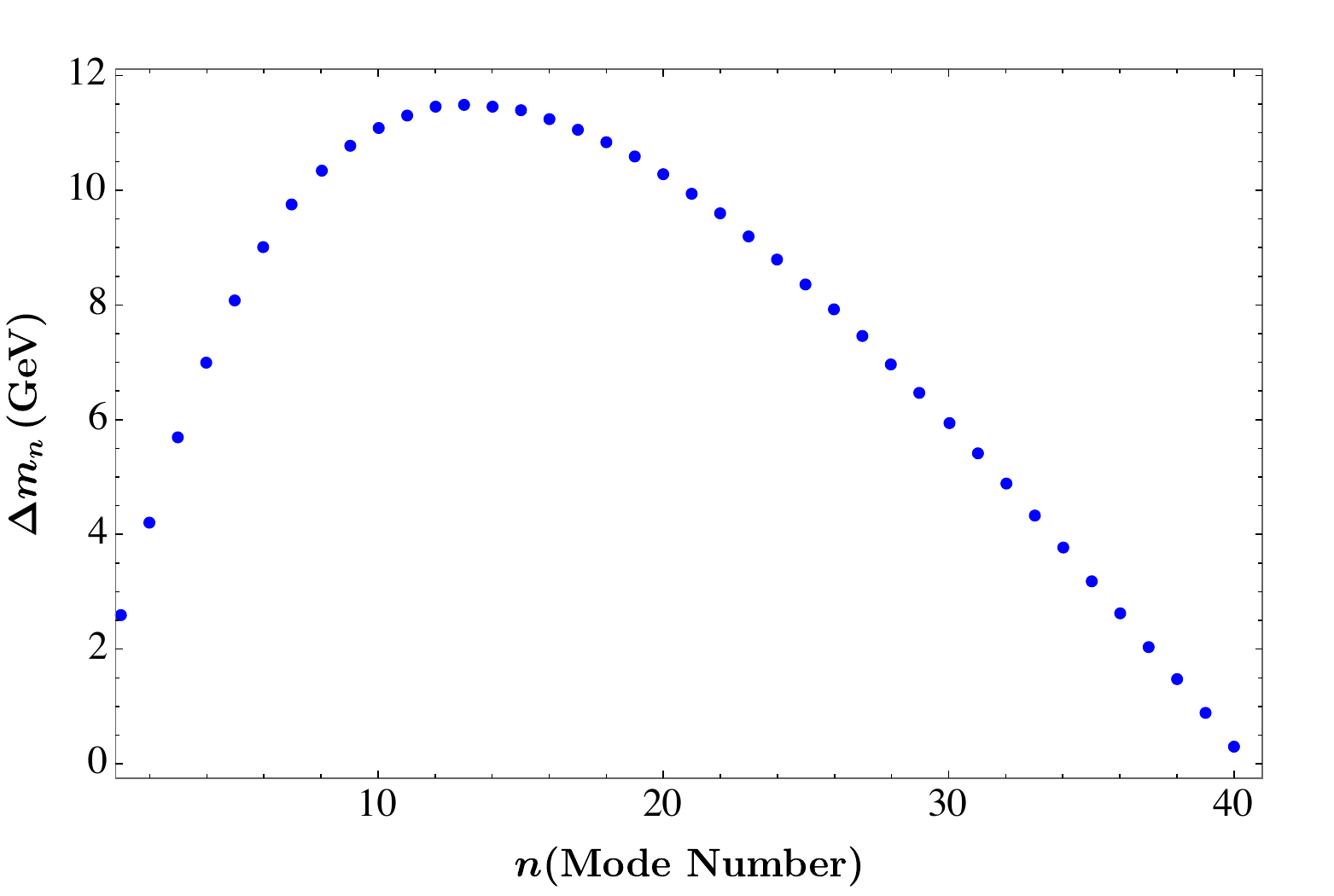}
\caption{Diphoton cross-sections and mass-splittings for benchmark III.} \label{fig:150gevf1600analytical}
\end{figure}

\textbf{Benchmark III: } Exploring ALPs heavier than the SM Higgs, we
consider $m=150$ GeV, $N=40$ and $f=1600$ GeV.  Due to the increase in
the SSB scale $f$, the QCD axion mass is now even slightly
lower than that in benchmark II(b), namely $m_{a_0} \approx 9.7 \times
10^{-9}$ eV with $f_0 \approx 6 \times 10^{14}$ GeV. Naturally, the arguments pertaining to DM abundance as presented for benchmark II(b) are also applicable here. As before, Fig.\ref{fig:150gevf1600analytical} shows
the diphoton cross-sections for the ALP spectrum as well as the
characteristic mass-splittings.

\vskip 10pt
With the model parameters defined, we now simulate the
diphoton signal events and compare with the corresponding SM
backgrounds for the $13$ TeV LHC. We write the Feynman rules of the
models in \texttt{Feynrules}~\cite{Alloul:2013bka} and generate UFO
files that are then imported in the
\texttt{MadGraph5}~\cite{Alwall:2014hca} event generator for a Monte
Carlo simulation of the signal. Further,
\texttt{Pythia8}~\cite{Sjostrand:2014zea,Bierlich:2022pfr} is used for
parton showering and hadronization. Final state objects at the
detector level are reconstructed using the fast simulation tool
\texttt{Delphes}~\cite{deFavereau:2013fsa}.  Within \texttt{Delphes}
we choose to simulate the detector effects using the ATLAS detector
card for the entirety our study.

The SM diphoton (with and without jets) background at the LHC has been
very well-studied by both the ATLAS and CMS collaborations, not only
in the context of the SM Higgs, but also for other exotic resonances,
light and heavy. Indeed, there have been occasional reports of
excesses, only to largely vanish on account of an even more careful
recalibration of the backgrounds. We would be largely using the ATLAS
analyses to estimate the backgrounds and, by extension, use the same
kinematical restrictions {\em etc.} to estimate the signal strength as
well.

\section{Signal and Background profiles}
\label{sec:sig_bkg}
While we would be concentrating on ALPs being produced in gluon fusion
and decaying into a diphoton pair, note that the exact final state
would be dependent on their
  average mass. In particular, a very light ALP, produced sans any
accompanying high-$p_T$ particle, would result in a pair of relatively
soft photons that would, typically, fail trigger requirements. In
other words, a minimal number of high-$p_T$ entities must be present,
and thus, both signal and background estimations would need to be done
accordingly.

\subsection{Light ALPs}

For \textbf{Benchmark Point I}, the ALP masses are in the $10 - 30\,
\mbox{GeV}$ range, and the final state photons need to be boosted to
overcome the detector's trigger level energy threshold. To achieve
this, we follow the strategy employed in the ATLAS search
\cite{ATLAS:2022abz} (and also ref.\cite{Mariotti:2017vtv}) for low
mass diphoton resonances and consider a diphoton final state with upto
two additional jets. To simulate the signal events we first
generate\footnote{Understandably, this is overwhelmingly dominated by
  the subprocesses wherein the final state partons are gluons.} $pp
\to a_n + 0/1/2 \, {\rm partons}$ at the leading order (LO), using
\texttt{MadGraph5}, for $\sqrt{s}=13$ TeV with the \texttt{NNPDF2.3LO}
PDF set, followed by the two-body decay of the on-shell
ALPs\footnote{Given that the ALPs are spinless, there are no
  non-trivial spin correlations. Furthermore, given that they are very
  narrow, there is no real loss of information or accuracy in the
  neglect of possible off-shell effects, as can also be confirmed by
  the use of tools such as the \texttt{Madspin}
  \cite{Artoisenet:2012st} module.}.  This is followed by parton
showering and hadronization via
\texttt{Pythia8}~\cite{Sjostrand:2014zea,Bierlich:2022pfr}.  Matching
and merging of the matrix elements (including the avoidance of
overcounting) for showering is automated using the MLM matching
scheme.

As mentioned previously, for a detector level simulation of the
showered events we use the \texttt{Delphes} tool. The following
summarizes the criteria used to identify and isolate final state objects.

\textbf{Jets:} Within \texttt{Delphes}, jets are reconstructed using
the \texttt{FastJet} package~\cite{Cacciari:2011ma}. Jet clustering is
performed using the anti-$k_T$ algorithm~\cite{Cacciari:2008gp} with
the jet cone radius parameter $R$ chosen to be 0.4. To be consonant
with the ATLAS analyses, we require the minimum transverse momentum of
jets ($p_{Tj}$) to be 20 GeV and its
pseudorapidity to satisfy $|\eta_j|<2.5$. Finally, any two jets must
be separated by $\Delta R(j, j) > 0.7$ where $(\Delta R) \equiv
\left[(\Delta\eta)^2 + (\Delta \phi)^2\right]^{1/2}$ is the separation
in the pseudorapidity-azimuthal angle plane.

\textbf{Photons:} For identification and isolation of photons, a cone
size of $\Delta R =0.2$ around the photon candidate is
considered. Denoting the ratio of the sum of transverse momenta of
isolated objects (tracks, calorimeter towers, etc) to the candidate's
transverse momentum as $p_T^{\text{ratio}}$, it is demanded that
$p_T^{\text{ratio}} < 0.05$. Also incorporated is a $p_T$-dependent
photon identification efficiency (following the ATLAS analysis of
ref.\cite{ATLAS:2022abz}) that ranges from $\approx 70 \%$ at $p_T=22$
GeV to $\approx 90\%$ for $p_T > 50$ GeV.

In addition, we ensure angular separation between photons or jets by
demanding that $\Delta R(\gamma, \gamma) > 0.2$ and $\Delta R(\gamma,
j) > 0.4$. Finally, we demand events with at least two photons and at
least a jet in the final state by imposing $N_\gamma\geq 2 $ and
$N_j\geq 1$.  The above conditions are summarised in Table
\ref{tab:signal_cuts_1_id} for quick reference.

\begin{table}[htbp!]
    \begin{center}
      Channel : $p p \rightarrow a_n + 0/1/2 \,\text{jets}$  \hskip 20pt
      $a_n \rightarrow \gamma \gamma$  \\[1ex]
      \begin{tabular}{c}
        \hline
          Acceptance Cuts \\
       \hline
     \underline{\textbf{Photon identification}} \\
     $\Delta R = 0.2$, $p_T > 0.5$ GeV, $p_T^{\text{ratio}}(\gamma) < 0.05$\\  
       \underline{\textbf{ Jet ident  ification}} \\
      $\Delta R=0.4$ (anti-$k_T$),  $p_T^{j} > 20$ GeV, $|\eta_j| <2.5$\\
       \underline{\textbf{Isolation}} \\
 $\Delta R(\gamma, \gamma) > 0.2$, $\Delta R(\gamma, j)>0.4$, $\Delta R(j, j)>0.7$ \\
 $N_\gamma\geq 2 $, $N_j\geq 1$ \\
       \hline
        \end{tabular}
    \caption{Acceptance cuts for the final state
      objects in benchmarks I and II \cite{ATLAS:2022abz}.}

    \label{tab:signal_cuts_1_id}
        
    \end{center}
    
\end{table}



As for the QCD corrections to the process, within the narrow width
approximation, we may factorise these separately for the production
and the subsequent decay. For such masses of the ALPs, the QCD
corrections to the production cross section can largely be summarised
in terms of $K$-factors, with $K^{1}_{\sigma}\sim K^{2}_{\sigma}
\simeq 2$ \cite{Gershtein:2020mwi} where the superscript indicates the
number of jets in the final state. As for the decay, while the correction
to the photonic branching fraction is small, that to the gluonic one
is substantial, {\em viz.} $K_{gg}=2.1$ \cite{Djouadi:2005gi},
and serves to scale the diphoton branching
fraction (since $\mbox{BR}_{\gamma \gamma} \approx \Gamma_{\gamma
  \gamma}/\Gamma_{gg}$). For the effective $K$-factor, then, 
$K^{1,2}_{\sigma}/K_{gg} \approx 1$ for the two 
benchmark points with light ALPs.

\begin{figure}
    \centering
 \includegraphics[width=0.46\textwidth,height=0.30\textwidth]{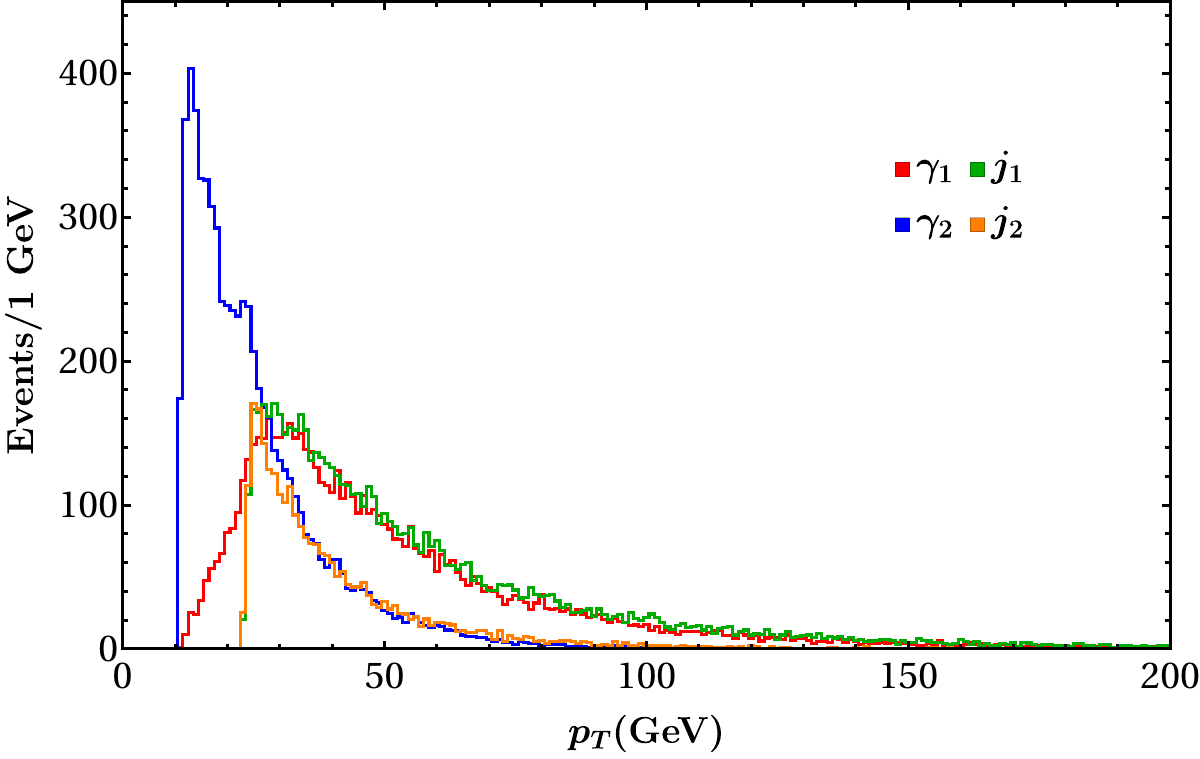}
\includegraphics[width=0.46\textwidth,height=0.30\textwidth]{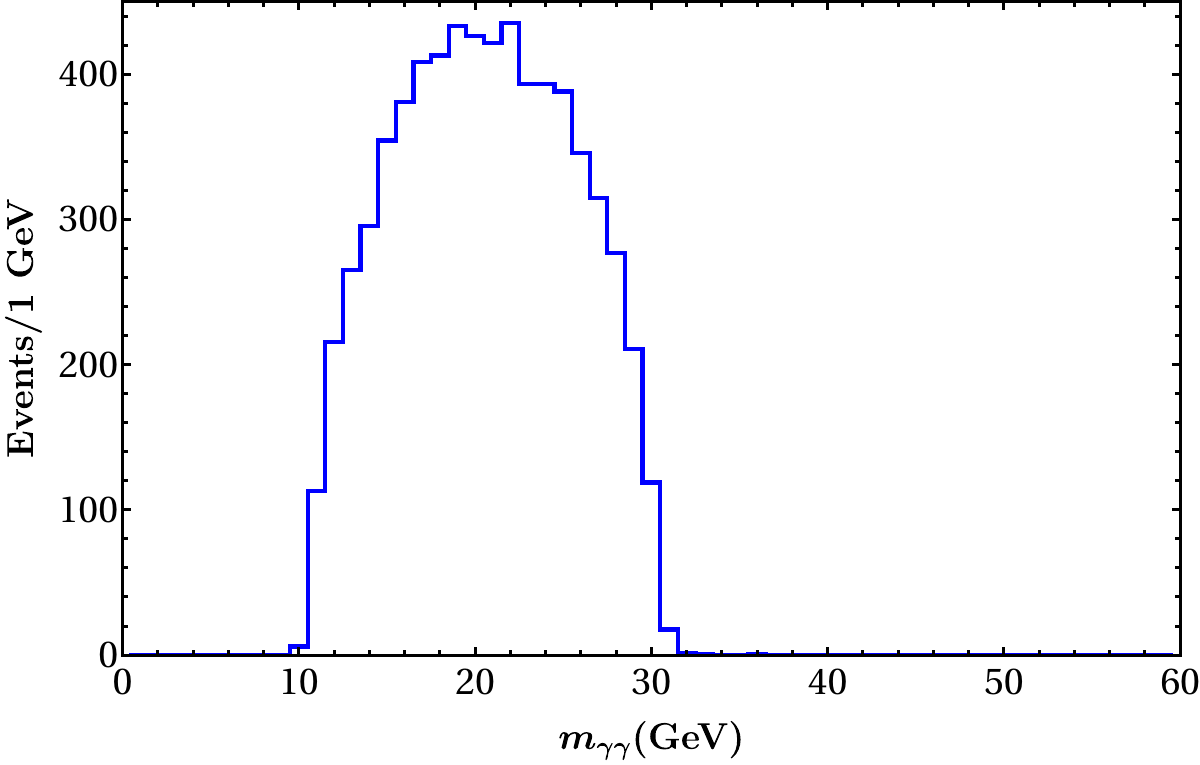}

      \caption{BP-I: event distributions,
        after applying acceptance cuts alone, {\em vide}.
        Table~\ref{tab:signal_cuts_1_id}.  {\bf (top)} $p_T$
          distributions for the leading photon ($\gamma_1$),
        sub-leading photon ($\gamma_2$), leading jet ($j_1$) and the
        sub-leading jet ($j_2$) and {\bf (bottom)} the two-photon
          invariant mass.}
    \label{fig:bp1_ptnocut}
\end{figure}

Using \texttt{MadAnalysis5} \cite{Conte:2012fm} to analyze the signal
events following the acceptance cuts, we order the photons and the
jets (wherever applicable) in terms of their $p_T$ and present, in
Fig.\ref{fig:bp1_ptnocut} (top), the corresponding
distributions, for an integrated luminosity of $\mathcal{L}=138 \,
\mbox{fb}^{-1}$. In this, we use a bin size of $1$ GeV, as in the
ATLAS analysis~\cite{ATLAS:2022abz}. While the sharp edges (at 20 GeV)
in the jet\footnote{Note that the events with 2 jets are only a
  subset of the accepted set of events.} $p_T$  distributions are but
reflections of the acceptance cut imposed, the fast fall off at higher
$p_T$s is char acteristic of QCD radiation. The positions of the peaks
are just caused by an interplay of the two effects. As for the
photons, the decay of an ALP, in its rest-frame, is isotropic with
each daughter photon having an energy equalling half the ALP
mass. This distribution, of course, has to be convoluted with the
$p_T$ of the ALP itself (as counterbalancing the jet $p_T$s). And,
finally, there are additional convolutions with both the ALP
mass-spectrum and the corresponding production cross sections.
  
Of particular interest is the diphoton invariant mass distribution
(Fig.\ref{fig:bp1_ptnocut} (bottom)) which encompasses the
contributions of the entire spectrum of the $28$ ALPs. It is
instructive to note the difference with the total production cross
section (as in Fig.\ref{fig:10gevf1000analytical} (top)) which shows
a maximum for $m_a \approx 15$ GeV, owing, as we have discussed, to a
combination of the variation in the coupling strength as well as the
gluon-gluon flux.  However, once the extra jets are required to be
radiated, the kinematics (and the mass-dependence of the flux) does
change considerably. Even more importantly, the
requirements\footnote{Further corrections arise from the
$p_T$-dependence of the detector efficiencies.} on $p_T(\gamma)$ serve
to suppress the contribution from the very low mass ALPs, resulting in
the significantly shifted maximum (now around $m_{\gamma
  \gamma}\approx 20$ GeV) as in Fig.\ref{fig:bp1_ptnocut} (bottom).

 For the background, diphoton (associated with additional jets)---
 initiated by $q\bar q$ as well as $gg$--- within the SM are considered.
 However, since there is a finite probability for a putative jet
 masquerading as a photon in the detector (this is especially true of
 a low energy jet depositing a large fraction of its energy in the
 electromagnetic calorimeter), one needs to consider such
 contributions as well. Thus, the backgrounds emanate primarily from
\begin{itemize}
\item $p p \to \gamma \gamma$, 
\item $p p \to \gamma j$,
\item $p p \to j j$.
\end{itemize}
in each case, accompanied by upto two additional jets.

Other backgrounds emanate from $e^\pm$ not leaving discernible tracks,
thereby faking photons. These, though, contribute negligibly and can,
therefore, be largely ignored. For an estimate of the relevant
background distribution in the diphoton spectra, we again refer to the
ATLAS analysis \cite{ATLAS:2022abz} wherein the fiducial region in the
phase-space is defined using the selection cuts on $E^{\gamma}_T$,
$p^{\gamma \gamma}_T$ and $\eta_{\gamma}$ as mentioned in Table
\ref{tab:signal_cuts_1_sel}. The diphoton background\footnote{The
background distribution as estimated in the ATLAS analysis
\cite{ATLAS:2022abz} for a bin size of 1 GeV is available at
\url{https://atlas.web.cern.ch/Atlas/GROUPS/PHYSICS/PAPERS/HIGG-2019-23/}.}
thus determined in the analysis has an uncertainty which is
predominantly statistical.
 \begin{table*}[htbp!]
    \begin{center}
    \begin{tabular}{c|c}
        \hline
          Channel & Event Selection Criteria \\
       \hline
&  $N_{\gamma}=2$, $1 \leq N_j \leq 2$,   \\
$p p \rightarrow a_n + 0/1/2 \,\text{jets}$& $|\eta_\gamma|<2.37$ (excluding barrel-to-endcap region $1.37<|\eta_\gamma|< 1.52$),\\
$a_n \rightarrow \gamma \gamma$ & $E_T (\gamma) > $22 GeV, $p_T^{\gamma\gamma} > 50$ GeV \\
       \hline
       \end{tabular}
    \caption{Selection cuts applied to form the fiducial signal regions for
      benchmarks I and II \cite{ATLAS:2022abz}.}

    \label{tab:signal_cuts_1_sel}
        
     \end{center}
    
\end{table*}
 
\begin{figure}[!ht]
    \centering
      \includegraphics[width=0.46\textwidth]{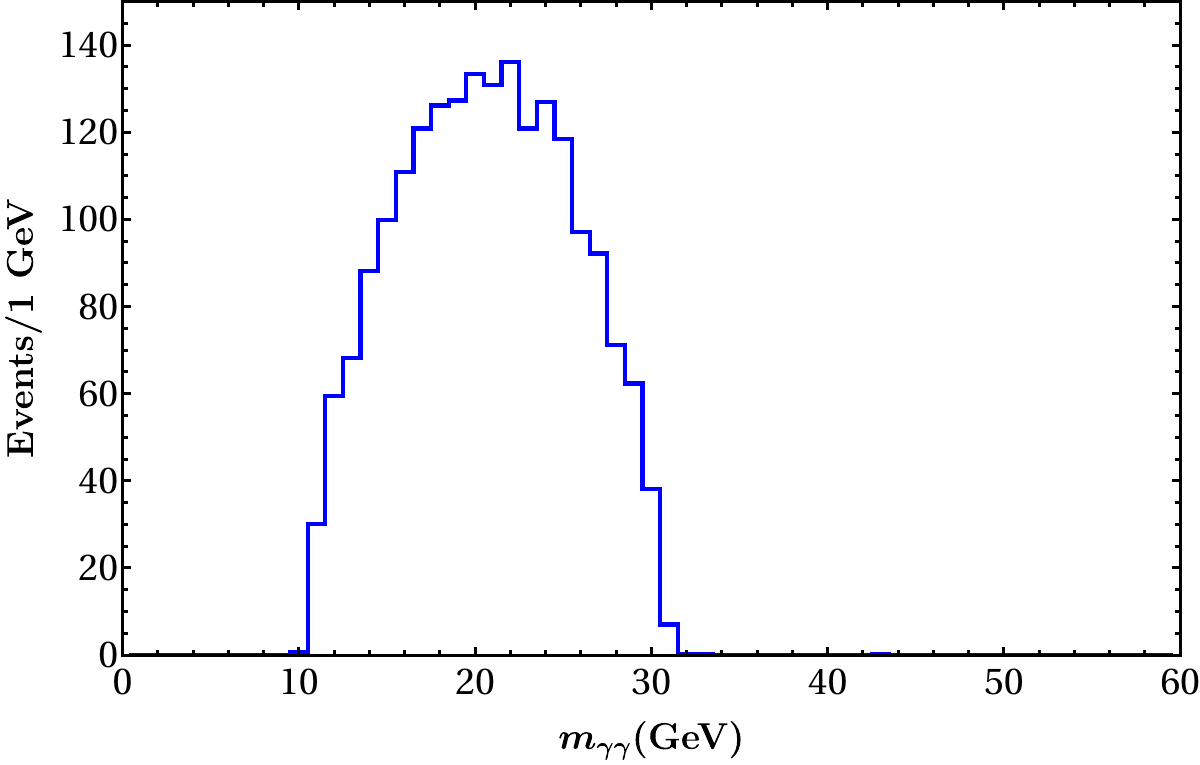}
    \caption{BP-I: Diphoton invariant mass distribution after applying selection cuts.}
    \label{fig:bp1maacut}
\end{figure}

 To examine the viability of the diphoton signal in the light of the
 ATLAS analysis, we must subject the former to the same selection
 criteria as the latter. These are listed in
 Table~\ref{tab:signal_cuts_1_sel}. As an examination of
 Fig.\ref{fig:bp1_ptnocut} (top) shows, the $E_T(\gamma)$ cut, especially
 when applied to the subleading photon, does eliminate a
 non-negligible fraction of the signal events. Even harder is the cut
 on $p_T(\gamma\gamma)$, for this requires that the ALP (and, hence,
 the radiated off jets) must carry a substantial $p_T$. The
   consequent reduction of events is reflected in the $m_{\gamma
     \gamma}$ distribution as shown in Fig.\ref{fig:bp1maacut}.
   
A striking aspect to note here (this was already evident in
Fig.\ref{fig:bp1_ptnocut} (bottom)) is that although the diphoton
distribution receives contributions from all the ALPs in the spectrum,
the separation between the individual resonances is smeared by the
detector resolution\footnote{We have made use of the energy resolution
  function for the ATLAS detector's ECAL \cite{AHARROUCHE2006601} as
  implemented within \texttt{Delphes}. For the phase-space region of
  interest throughout the presented analysis (i.e. for BP-I, II and
  III), the energy resolution of the individual photons varies roughly
  in the range $\sim[0.32, 2.3]$ GeV.} to a degree that the entire
distribution appears as a single broad resonance --- an
\textit{iceberg} of axions, so to say --- with a FWHM $\sim 16$
GeV. This feature, as previously mentioned, is peculiar to the
clockwork spectrum for any mass scale $m$ as long as $N$ is adequately
large.

\begin{figure}
    \centering
    \includegraphics[width=0.48\textwidth]{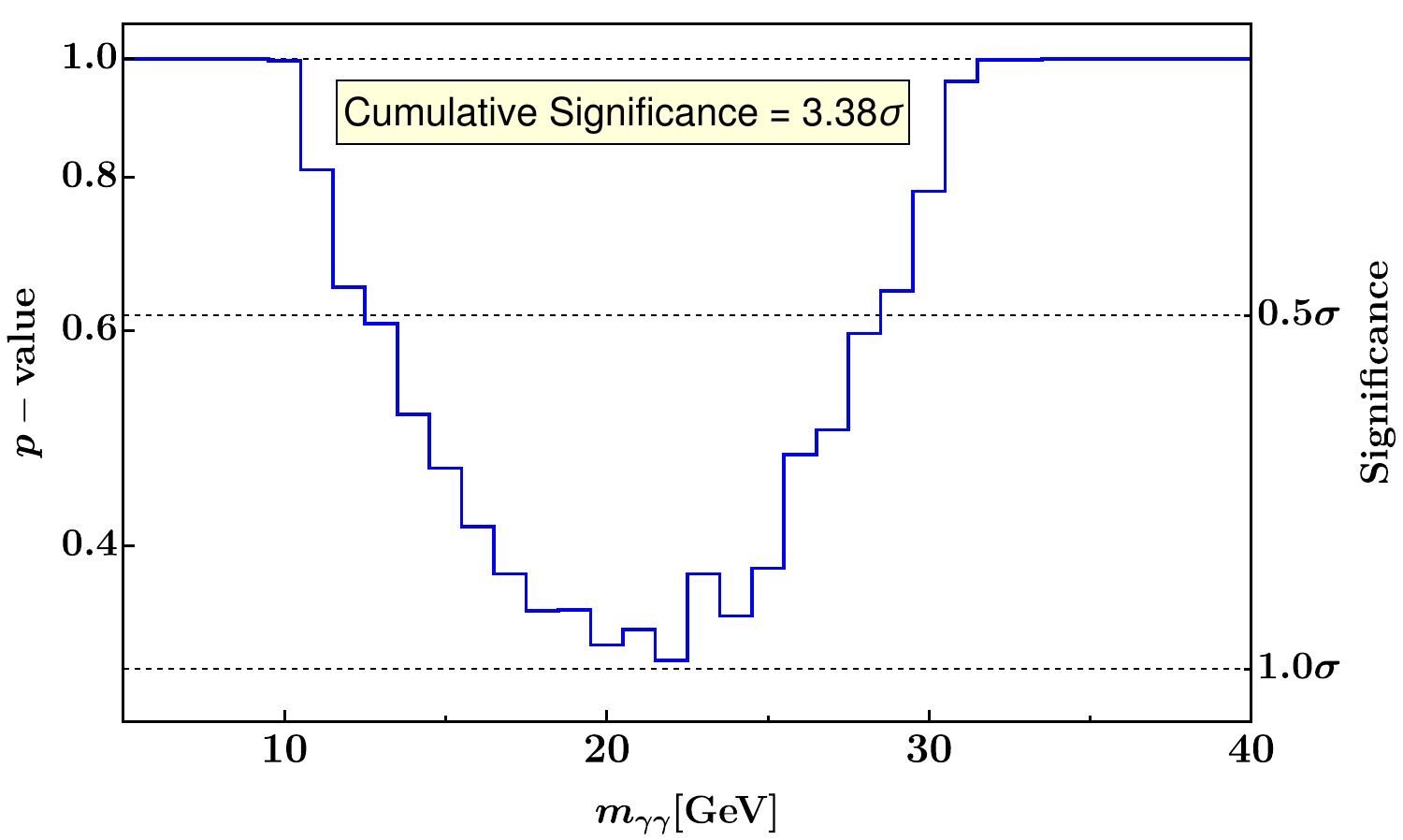}
    \caption{Bin-wise significance for benchmark I at the LHC for $\mathcal{L}=138 \, \rm fb^{-1} $.}
    \label{fig:chisq_model1}
\end{figure}

We can now estimate the significance of the signal events over the
background bin-by-bin using $S=N_{sig}/\sqrt{N_{sig}+N_{bkg}}$, where
$N_{sig}$ and $N_{bkg}$ denote the number of signal and background
events, respectively, in a particular bin. Since the background
uncertainty is statistics dominated, this gives quite a robust
estimate of the sensitivity and can be bettered only by a dedicated
search. The bin-wise significances (and the corresponding $p$-values)
for $\mathcal{L}=138 \, \mbox{fb}^{-1}$ with a bin width of $1$ GeV
are depicted in Fig.\ref{fig:chisq_model1}, and the maximum individual
significance of $S^{1 \, \rm GeV}_{138}\sim 1 \sigma$ is obtained in
the bin corresponding to $m_{\gamma \gamma} \sim 21-22$ GeV. It is
tempting to increase the bin widths as this would be expected to
substantially increase the per-bin significance owing to the
individual peaks being closely packed within the invariant mass range
$10-30$ GeV. However, since the existence of a wide (and finite) band
of resonances is characteristic to the clockwork scenario, the shape
of the invariant mass distribution could itself behave as a
discriminator--- more so in the case of small mass-splittings--- and
by enlarging the bin width one would, obviously, lose information
about the underlying profile of the spectrum. In particular, this
would help discriminate between a broad resonance and multiple sharp
ones\footnote{A further discriminant here would be the comparison
between the width of the resonance and the total signal size. For a
resonance, production cross section is simply related to the total
width, whereas the size of the signal is just the product of the cross
section, the branching fraction and the overall detector efficiency
(the last being a known quantity).}. A more useful method of
estimating the total signal significance would be to consider the sum
$S_c \equiv \sum_i S_i^2$ which, of course, is $\chi^2$-distributed
for the appropriate degrees of freedom. Using this, we find for this
benchmark point, the expected cumulative significance to be $S_{c,
  138} = 3.38 \sigma$ for $\mathcal{L}=138 \, \rm fb^{-1}$.  On the
same note, it is worth alluding to the prospect that the significance
can further build up by a sizable amount with an increase in the
luminosity, even by as early as the end of the ongoing Run-3 phase of
the LHC with a projected luminosity reach of $\sim 300 \, \rm
fb^{-1}$. A simple estimate of this can be obtained by scaling the
current significance with the luminosity, i.e.  \beq S_{c, 300}=S_{c,
  138} \times \sqrt{\frac{300}{138}} \approx 4.98 \sigma \, , \eeq
which makes this benchmark point imminently testable by the end of Run
3.

\begin{figure}
    \centering
    \includegraphics[width=0.46\textwidth]{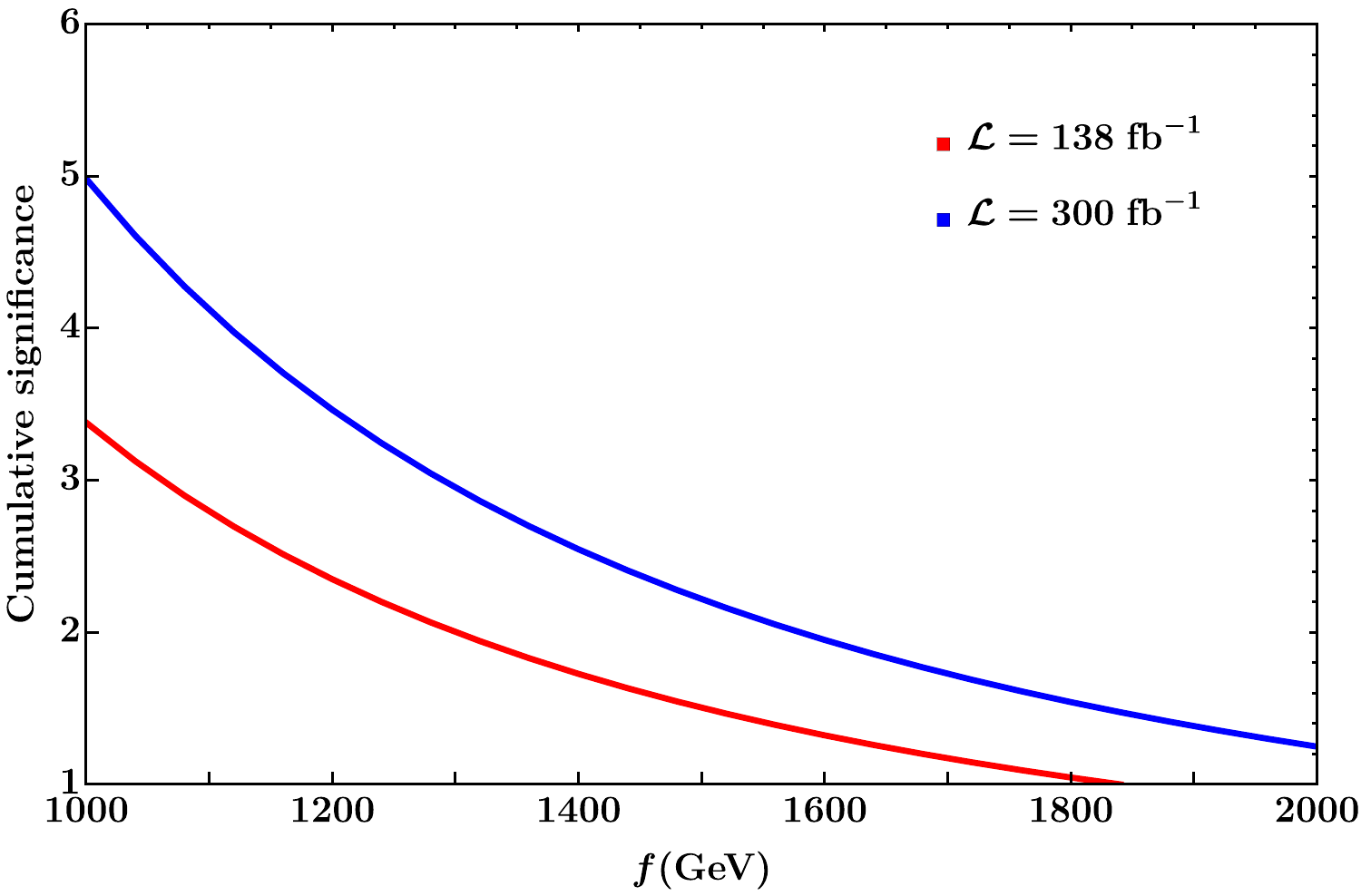}
    \caption{Cumulative significance as a function of the SSB scale $f$ for benchmark-I at the Run 2 (red curve) and Run 3 LHC (blue curve).}
    \label{fig:fplot_model1}
\end{figure}

Apropos of the current benchmark point, Fig.\ref{fig:fplot_model1}
shows the variation of $S_c$ with the SSB scale $f$ for the $138$ $\rm
fb^{-1}$ luminosity case alongwith a projection for the $300$ $\rm
fb^{-1}$ reach of the LHC. The variation follows the approximate relation $S_c \propto f^{-2}$ which can be
understood from the fact that, for a particular choice of the CW
parameters, the pertinent cross-sections scale roughly as $f^{-2}$
with the pole masses being quite insensitive to it.


At this stage, it is worth examining if such a signal could have
emanated from any other theory. First and foremost, the decay of the
resonance into a diphoton pair rules out a spin-one particle. While
spin-two or higher are possible, such fundamental particles are
disfavoured on theoretical grounds ({\em e.g.}, a graviton, while
well-motivated, would not only be much narrower, but would also decay
democratically into other channels). Such arguments narrow down the
possibilities to only a scalar or a pseudoscalar. This apart, a width
comparable to the mass (as is the case here), would require the
couplings to be in the nonperturbative
regime~\cite{Salvio:2016hnf,Franceschini:2015kwy}. (For
instance, a single ALP resonance with a total decay width and mass
$\Gamma(a) \sim m_a \sim \mathcal{O}(10 {\rm \, GeV})$ would require
its decay constant\footnote{The inequality takes into account the
dependence on the constituents of the anomaly loop.} to be $f \lesssim
100$ MeV!). Irrespective of the underlying theory, such a large width
would have translated, {\em vide} eq.(\ref{eq:partonic_cs}), to a
total production cross section several orders of magnitude larger than
what is the case here. This would, immediately, raise question as to
the reason for the suppression of the diphoton branching ratio that
such a production cross section (alongwith the observed rates) would
imply.  Furthermore, an entity that couples so strongly would leave
tell-tale signatures in myriad other processes (a trivial example
being double diffractive processes leading to a pair of central jets
with large rapidity gaps) and, hence, would have been long discovered.

In other words, if such an excess were to be observed, a single
resonance hypothesis would be immediately ruled out. If it were to be
then interpreted as an overlap of multiple resonances, two issues
would need to be considered. The size of the excess imposes
restrictions in the plane of the individual total widths and the
diphoton branching fraction. On the other hand, a conservative
estimate of the number of peaks can be obtained by comparing the width
of the excess to the invariant mass resolution operative at such
masses.  And, finally, the very shape of the excess would dictate that
the couplings of the said resonances must conform to a very specific
pattern. To reproduce the features in a generic ({\em i.e.},
non-clockwork) model (or even in multi-axion scenarios involving a new
$N$-flavor confining sector
\cite{Alexander:2023wgk,Alexander:2024nvi}) would, thus, require many
apparently {\em ad hoc} assumptions to be made. On similar grounds, it
can also be argued that a broad resonance signal --- with a finite
bandwidth that is comparable with the characteristic mass scale of the
resonance $\sim m$---is incompatible with, say, the Kaluza-Klein
spectra of extra-dimensional scenarios consisting, in principle, of an
infinite number of states (upto a natural cutoff of the theory), for
then the distribution of resonances would extend over a large mass
range and truncate only when the cross-sections start to fall near the
kinematic limits of the experiment.

It is perhaps worth reiterating at this point that the analysis
strategy employed here is complementary to the collider study of the
clockwork graviton model carried out in
refs.\cite{Giudice:2017fmj,Beauchesne:2019tpx}. Of the different
strategies investigated therein, a particularly interesting approach
involves searching for periodic signals in various kinematic
distributions. However, the said technique, while well-suited for the
tower-like spectra typically encountered in clockwork models, seems to
be more efficient when the resonances are experimentally
resolvable. The analysis above, therefore,
presents a categorically new perspective to look at clockwork spectra
at colliders, especially for the case of light ALPs with tiny
mass-splittings.

\subsection{Intermediate mass ALPs}

With an intermediate mass range set by the
parameters $m=35$ GeV and $f=1000$ GeV, the analysis particulars in
this case are largely the same as described for benchmark I. With the
ALP masses in this case spanning the range $35-105$ GeV, a sizable
fraction of the spectrum still falls in the low mass region for which
we require the final state photons to be boosted due to reasons
delineated for benchmark I. We, therefore, choose to adopt the same
strategy as in benchmark I for the collider simulation and consider a
signal consisting of CW axions produced in association with upto two
jets (and the axions further decaying to two photons). Photons and
jets are reconstructed using the same criteria as in the case of
Benchmark point I (see Table \ref{tab:signal_cuts_1_id}).

\subsubsection{Benchmark Point II(a)}
\begin{figure}
    \centering
    \includegraphics[width=0.46\textwidth,height=0.30\textwidth]{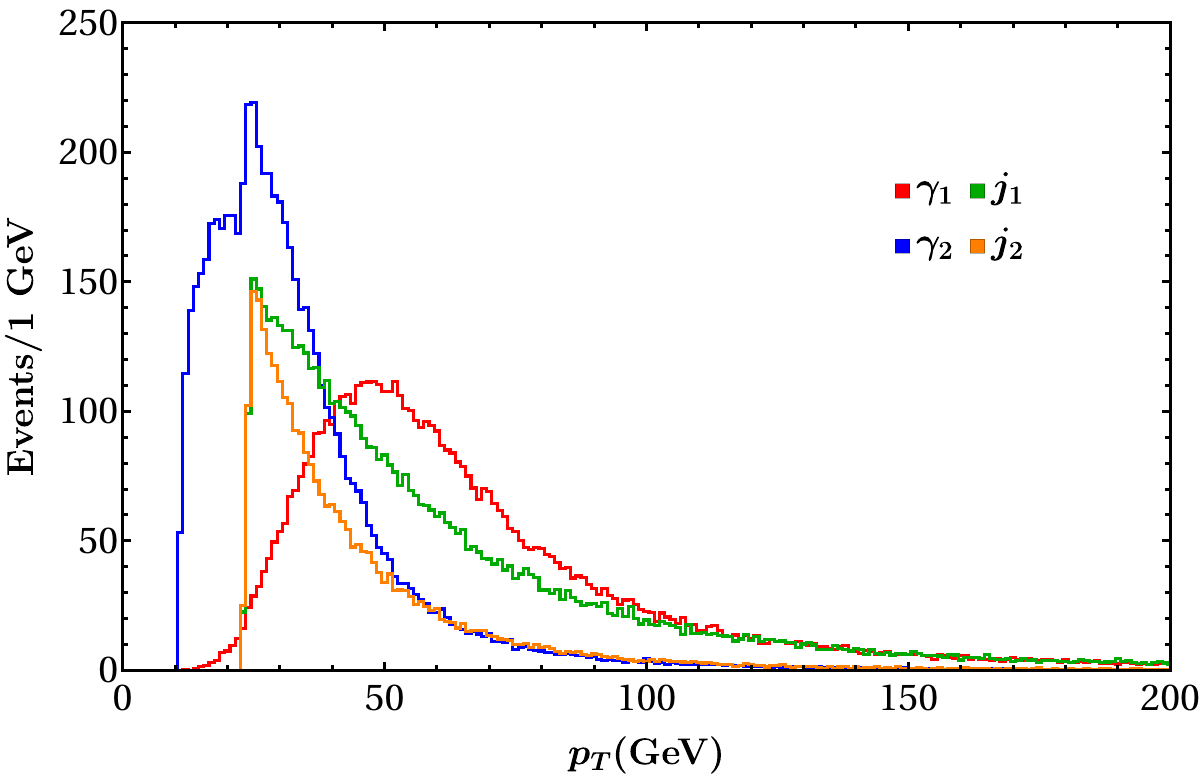}
      \includegraphics[width=0.46\textwidth, height=0.30\textwidth]{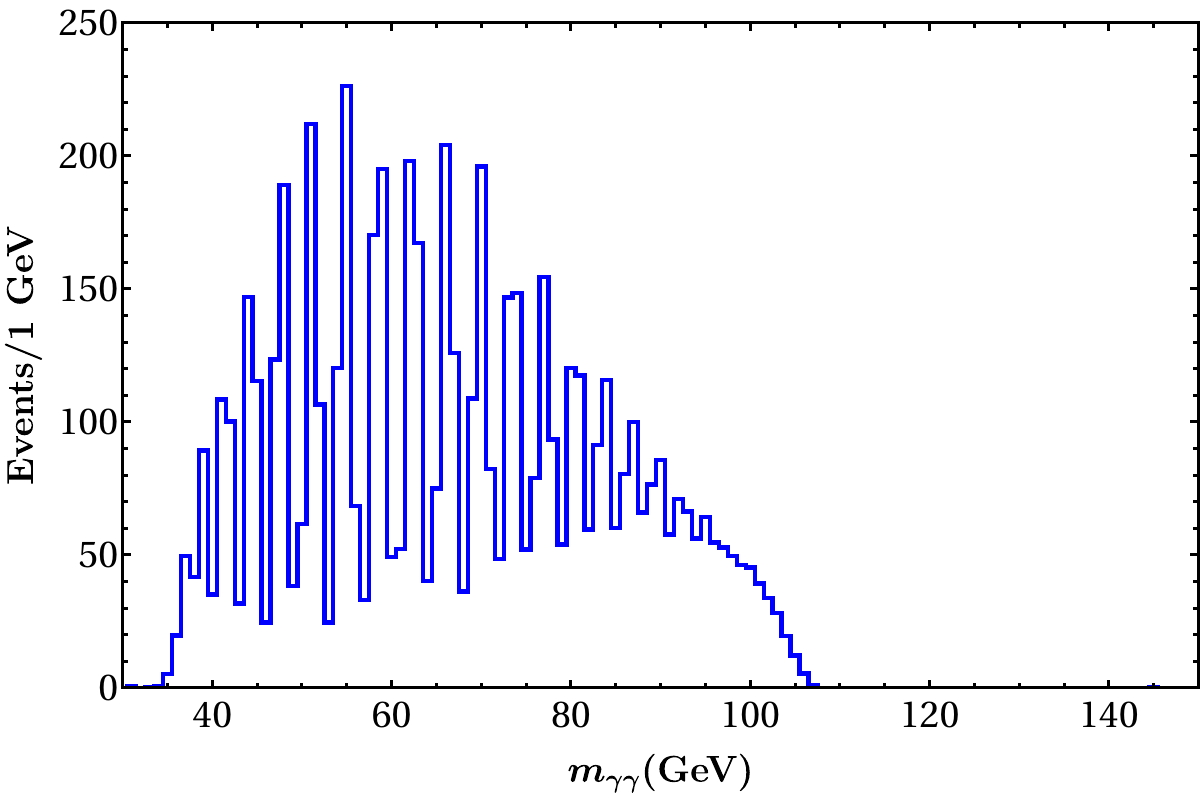}
      
    \caption{BP-II(a): event distributions,
        after applying acceptance cuts alone, {\em vide}.
        Table~\ref{tab:signal_cuts_1_id}.  {\bf (top)} $p_T$
          distributions for the leading photon ($\gamma_1$),
        sub-leading photon ($\gamma_2$), leading jet ($j_1$) and the
        sub-leading jet ($j_2$) and {\bf (bottom)} the two-photon
          invariant mass.}
    \label{fig:bp2a_ptnocut}
\end{figure}

The $p_T$ distributions for the leading and the sub-leading photons
thus obtained for the signal events alongwith the corresponding
distributions for the jet $p_T$'s are shown in
Fig.\ref{fig:bp2a_ptnocut} (top), assuming a bin
size of 1 GeV. The distributions are similar in nature to what was
obtained for benchmark point I, albeit with a shift in the photon
distributions towards higher $p_T$ values due to the ALPs being
heavier in this case. The attendant diphoton invariant mass
distribution is shown in Fig.\ref{fig:bp2a_ptnocut}
(bottom). As for the SM backgrounds, for the mass
range $m_{\gamma\gamma} \in [10,80]$~GeV, once again we adopt the
ATLAS analysis \cite{ATLAS:2022abz}. For the $[80,110]$~GeV window, in
the absence of an appropriate analysis\footnote{The ATLAS search for
  diphoton resonances in the mass range $m_{\gamma \gamma}=66-110$ GeV
  \cite{ATLAS:2024bjr} does provide a background estimation in the
  region of interest. However, the event selection criteria employed
  therein is different than that in the low mass analysis that we
  refer to, which makes it difficult to match the background events
  corresponding to the low and intermediate mass ranges in a
  meaningful way.}, we choose to {\em overestimate} the background, by
holding it at the level of $m_{\gamma \gamma}=80$ GeV. Thus, our
significance estimate would be somewhat conservative.

\begin{figure}
    \centering
      \includegraphics[width=0.46\textwidth]{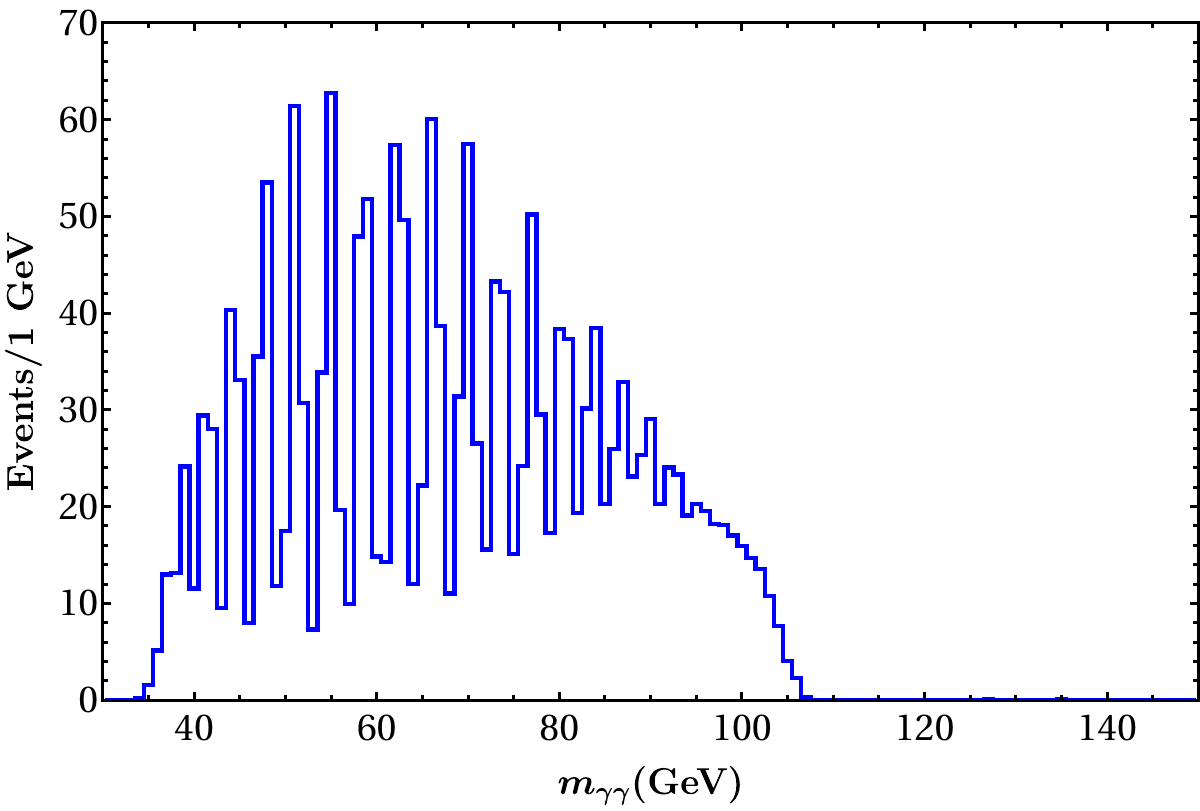}
    \caption{BP-II(a): Diphoton invariant mass distribution after applying selection cuts.}
    \label{fig:bp2amaacut}
\end{figure}

\begin{figure}
    \centering
    \includegraphics[width=0.48\textwidth]{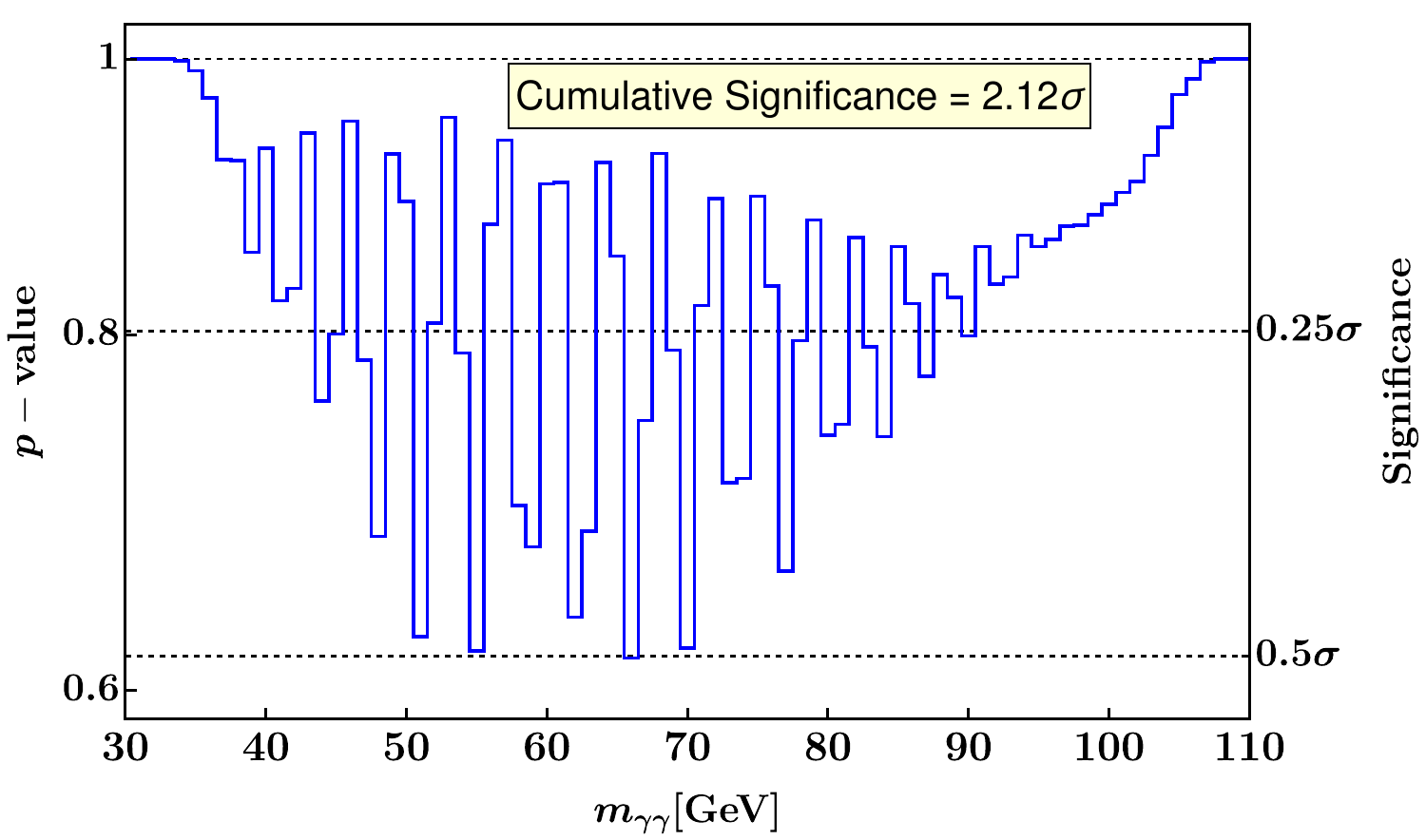}
    \caption{Bin-wise significance for benchmark II(a) at the LHC for $\mathcal{L}=138 \, \rm fb^{-1} $.}
    \label{fig:chisq_model2a}
\end{figure}
     
To obtain the signal diphoton distribution in the fiducial region, we
employ the selection cuts mentioned in Table
\ref{tab:signal_cuts_1_sel}. As for benchmark I, the kinematic cuts
substantially reduce the number of events in the signal distribution,
shown in Fig.\ref{fig:bp2amaacut}. Given the characteristic
mass-splittings of this benchmark, we see that a large portion of the
spectrum (consisting of 28 ALPs) in the diphoton distribution could
potentially be resolved at the detector. Now, a bin-wise significance
estimation, under the conservative assumption of the background events
beyond $m_{\gamma \gamma} \approx 80$ GeV, leads to a distribution as
shown in Fig.\ref{fig:chisq_model2a} with a cumulative significance
$S_{c,138} = 2.12\sigma$ over the signal region. While spectral
techniques~\cite{Giudice:2017fmj,Beauchesne:2019tpx} may enhance
sensitivity in scenarios where multiple resonances are moderately
separated and experimentally resolvable, the present analysis adopts a
more conservative, profile-based strategy. This approach focuses on
the overall signal shape, which effectively manifests as a single
broad structure due to small mass splittings and finite detector
resolution.

\subsubsection{Benchmark Point II(b)}
\begin{figure}
    \centering
      \includegraphics[width=0.46\textwidth]{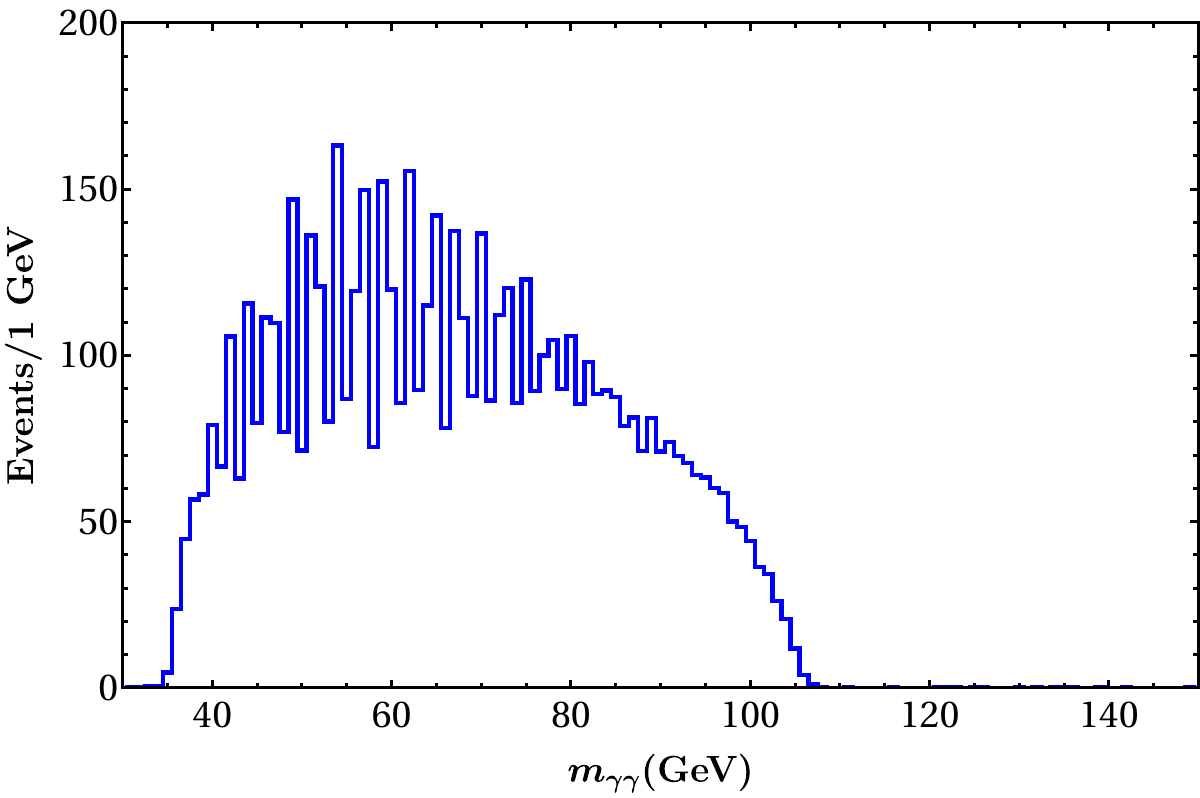}
    \caption{BP-II(b): Diphoton invariant mass distribution after applying acceptance cuts alone.}
    \label{fig:bp2nocutmaabin1}
\end{figure}

\begin{figure}
    \centering
      \includegraphics[width=0.46\textwidth]{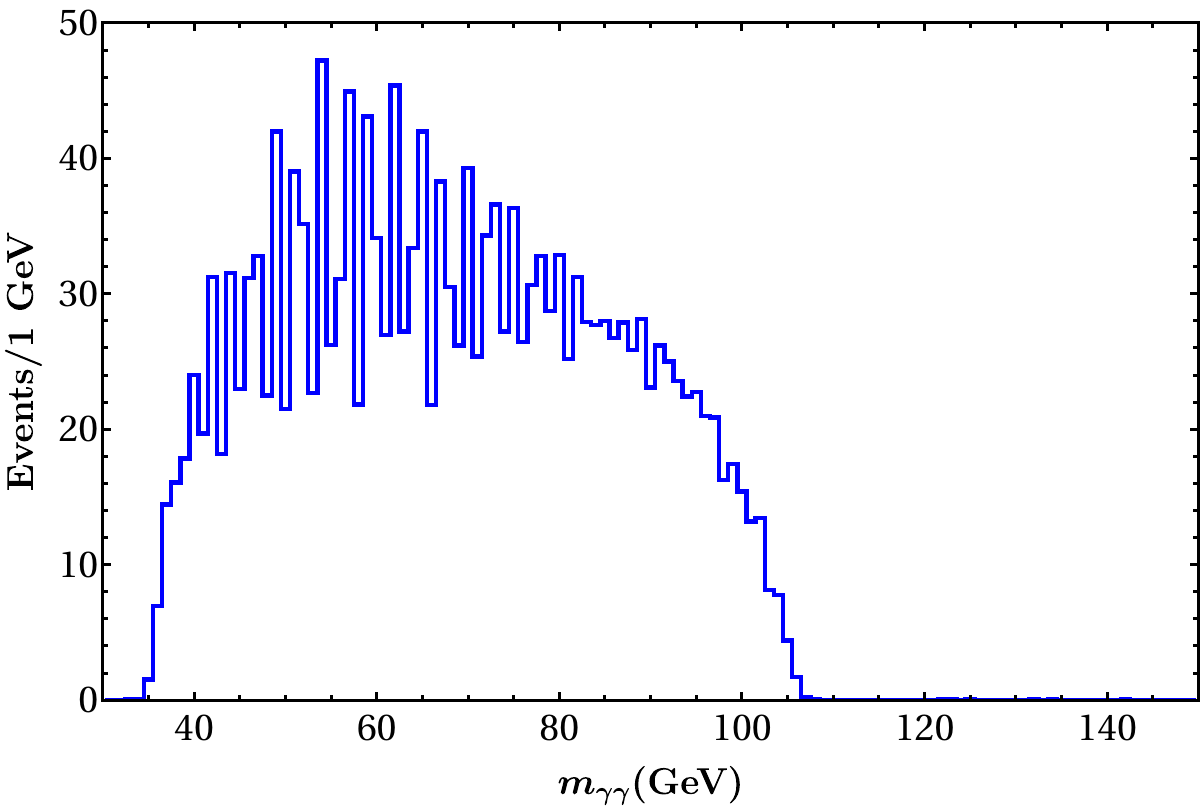}
    \caption{BP-II(b): Diphoton invariant mass distribution after applying selection cuts.}
    \label{fig:bp2maacut}
\end{figure}

\begin{figure}
    \centering
    \includegraphics[width=0.48\textwidth]{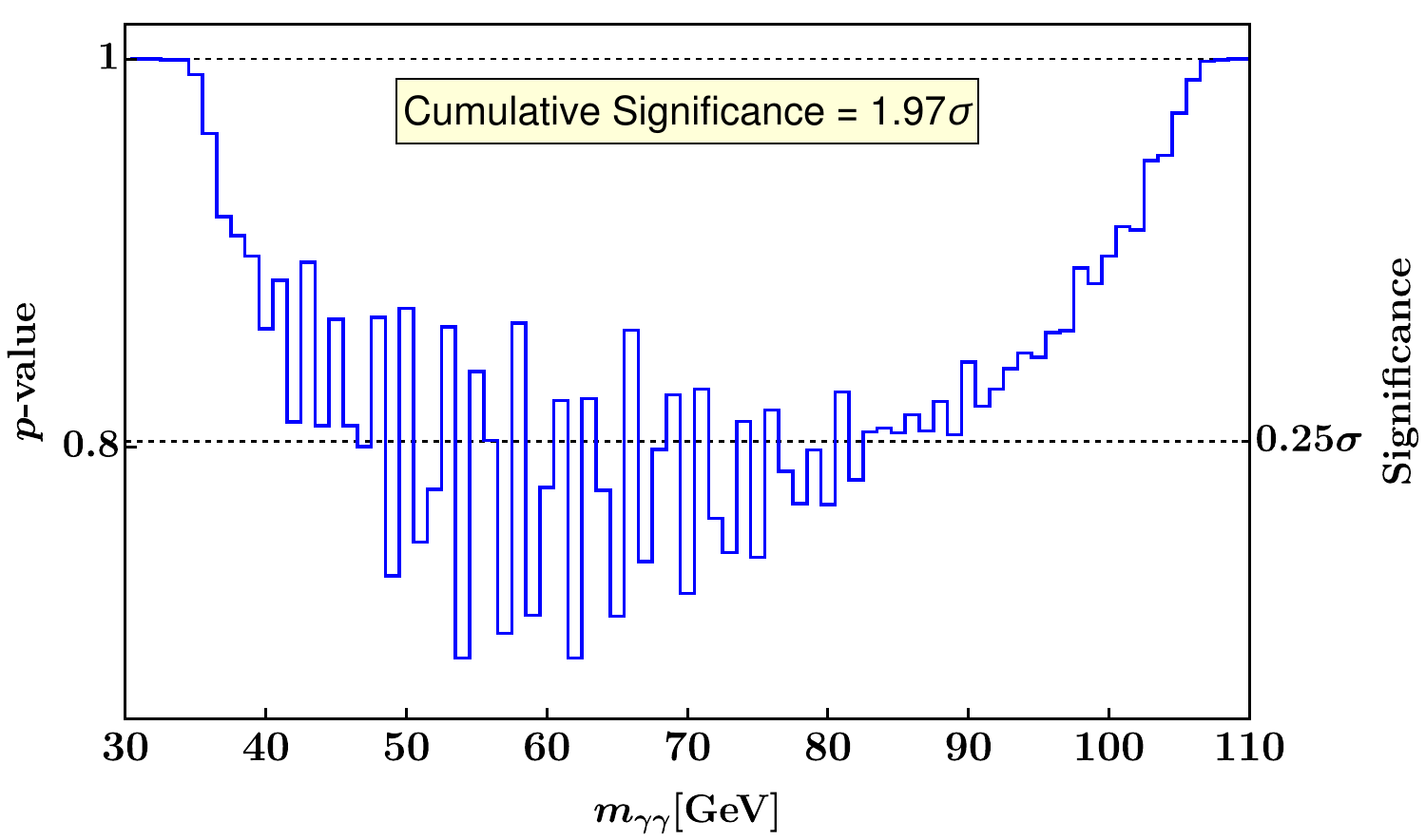}
    \caption{Bin-wise significance for benchmark II(b) at the LHC for $\mathcal{L}=138 \, \rm fb^{-1} $.}
    \label{fig:chisq_model2}
\end{figure}

As the only modification we have in this case is the increase in the
number of ALPs to $N=40$, we follow the same strategy for the analysis
as before. The signal's diphoton invariant mass distribution post
acceptance cuts is shown in Fig.\ref{fig:bp2nocutmaabin1}. In contrast
with the outcome for benchmark II(a), the sizes of the individual
peaks have decreased, with the cross-sections scaling as $\sim |C_{n
  N}|^2 / N$, and their smearing is now relatively more pronounced as
a result of the smaller mass-splittings, both effected by the increase
in the number of ALPs. The corresponding distribution in the fiducial
region is presented in Fig.\ref{fig:bp2maacut}, exhibiting,
expectedly, the same signal shape as in Fig.\ref{fig:bp2nocutmaabin1},
although with a reduced number of events. The bin-wise significance
estimate is shown in Fig.\ref{fig:chisq_model2} with the corresponding
cumulative significance being $S_{c,138} = 1.97\sigma$.

Clearly, the two cases lead to slightly different projections for the
$300 \, \rm fb^{-1}$ luminosity reach, viz. $S^{(a)}_{c,300} \approx
3.12\sigma$ and $S^{(b)}_{c,300} \approx 2.9\sigma$. Thus, it may not
be an overestimate to say that, upon a proper treatment, if the
background distribution is found to saturate or slowly fall beyond
$m_{\gamma \gamma} \sim 80$ GeV, then benchmark II could also be
potentially probed by the end of LHC's Run 3 phase.

For either the intermediate or the heavy-mass cases, the individual
peaks are, thus, expected to be visible. The presence of such a bunch
of closely packed resonances, with the specific envelope, would
immediately point towards a clockwork-like
scenario\footnote{Naively, an alternative could be a composite
    sector with multiple excitations. However, not only would such a
    compositeness scale be very surprising, the abrupt ending of the
    excitations as well as the shape of the envelope would be hard to
    explain.}.


\subsection{Heavy ALPs}

For ALPs that are significantly heavier than considered hitherto, the
decay photons would be expected to carry sufficient energy (and $p_T$)
for the event to be triggered even without any additional jet. An
example is afforded by the aforementioned {\bf Benchmark III}, wherein
the ALP masses span the range $\sim 150 - 450$ GeV.  However, although
extra jets are not needed for triggering, events with such jets do
contribiute significantly to the signal cross sections. Hence, we
define the signal as a semi-inclusive one composed of a pair of
energetic photons with upto two additional jets\footnote{The inclusion
  of events with three or more jets does not improve the signal to
  noise ratio.}. The corresponding $K$-factors, for this mass range,
are taken from refs.\cite{Ahmed:2016otz,Boughezal:2015dra}.

\begin{table}[!ht]
    \begin{center}
    \begin{tabular}{c|c}
        \hline
          Channel & Acceptance Cuts \\
       \hline
&\underline{\textbf{Photon identification}} \\
 $p p \rightarrow a_n  \,$ & $\Delta R=0.4$, $p_T > 0.4$ GeV, $p_T^{\text{ratio}}(\gamma) < 0.12$\\
       $a_n \rightarrow \gamma \gamma$     &\underline{\textbf{Jet identification}} \\
       & $\Delta R=0.4$ (anti-$k_T$),  $p_T^{j} > 20$ GeV, $|\eta_j| <2.5$\\
       & \underline{\textbf{  Isolation}}\\ 
       & $\Delta R(\gamma, \gamma) > 0.4$, $\Delta R(\gamma, j)>0.4$, $\Delta R(j, j)>0.7$ \\
       & $N_\gamma \geq 2$ \\
       \hline
    \end{tabular}
    \caption{Acceptance cuts for the final state objects in benchmark III \cite{ATLAS:2021uiz}.}
    \label{tab:signal_cuts_2_id}
    \end{center}
  
\end{table}

\begin{figure}[!h]
    \centering
    \includegraphics[width=0.46\textwidth, height=0.30\textwidth]{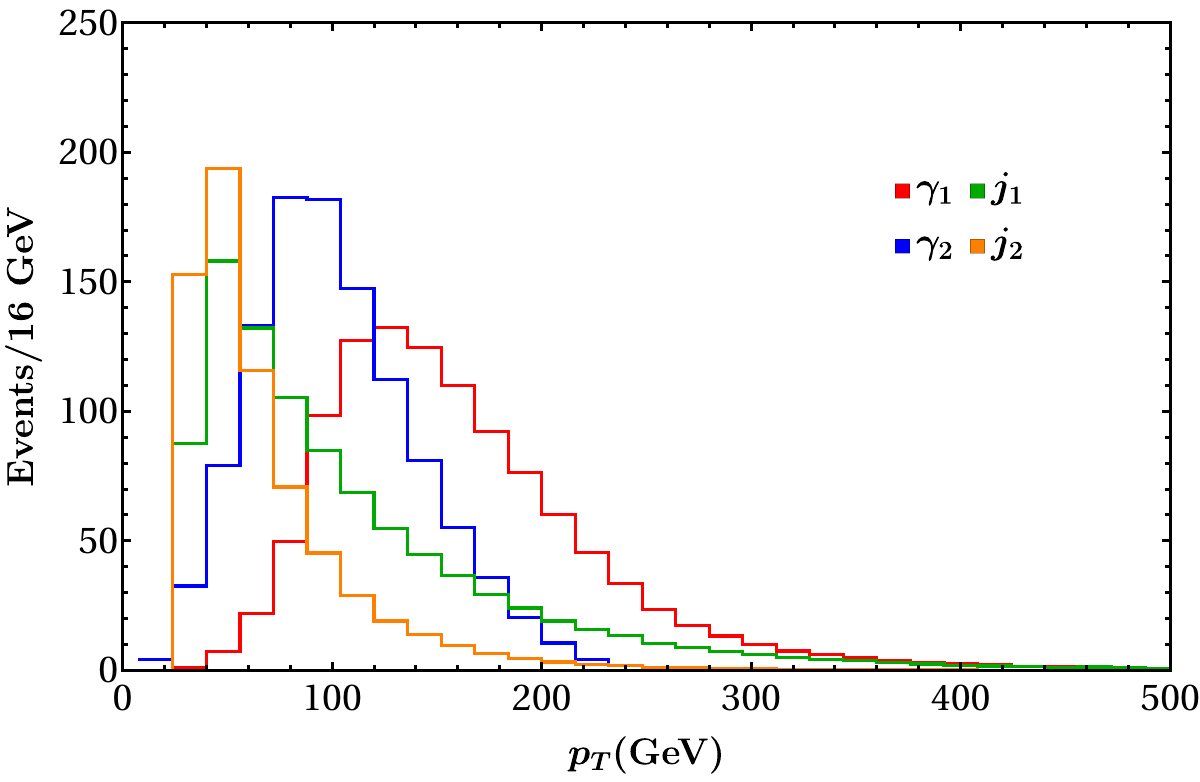}
    \includegraphics[width=0.46\textwidth, height=0.30\textwidth]{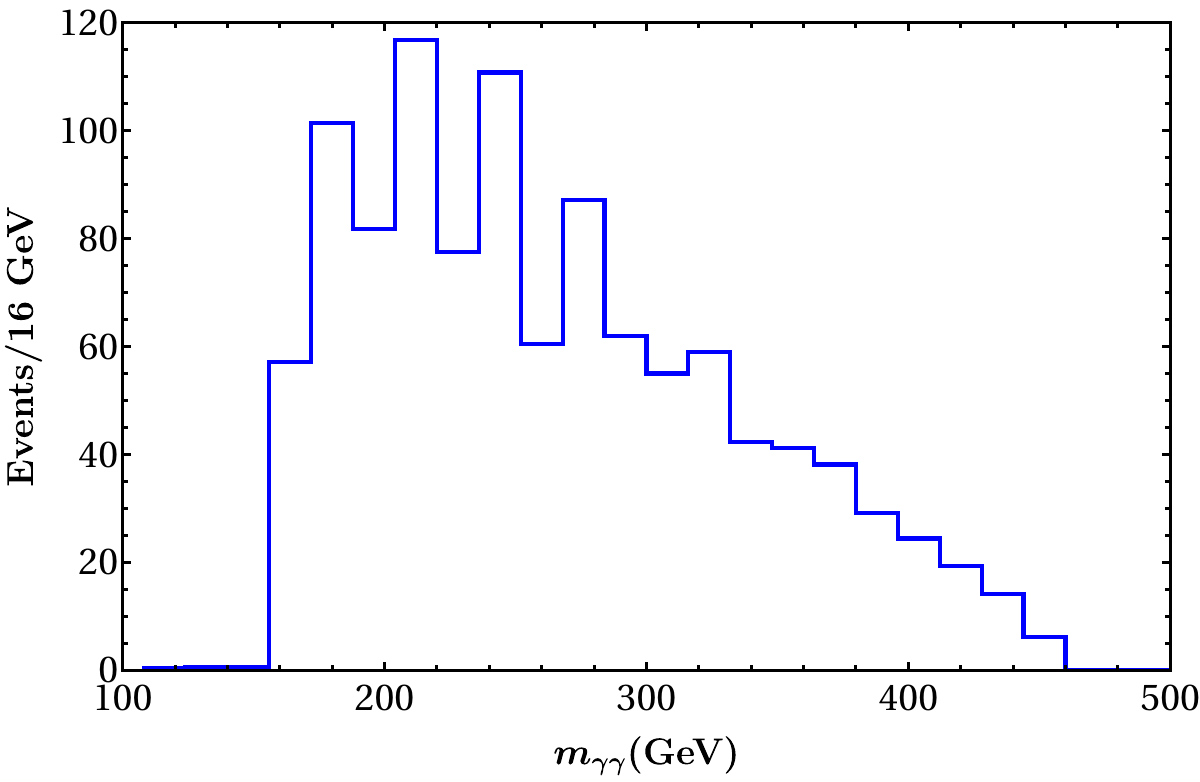}
      
    \caption{ BP-III: event distributions,
        after applying acceptance cuts alone, {\em vide}.
        Table~\ref{tab:signal_cuts_2_id}.  {\bf (top)} $p_T$
          distributions for the leading photon ($\gamma_1$),
        sub-leading photon ($\gamma_2$), leading jet ($j_1$) and the
        sub-leading jet ($j_2$) and {\bf (bottom)} the two-photon
          invariant mass.}
    \label{fig:bp3_ptnocut}
\end{figure}

The background, thus, would receive contributions from essentially the
same channels as assumed in the low mass case. The signal profile
being different, we, though, would need to adequately tune the
assessment, and, in this, we are guided by the ATLAS diphoton analysis
\cite{ATLAS:2021uiz} germane to this mass range. Post showering and
hadronization, using \texttt{Pythia8}, we take for the identification
and isolation of photons in the \texttt{Delphes} detector simulation a
cone size of $\Delta R =0.4$ around the photon candidate. The
requirement on the parameter $p_T^{\rm ratio}$, as defined earlier, is
now reset to $p_T^{\rm ratio} < 0.12$. The complete photon and jet
identification criteria, as adapted from the ATLAS analysis
ref.\cite{ATLAS:2021uiz}, are displayed in Table
\ref{tab:signal_cuts_2_id}.  The $p_T$ distributions of the photons
and the jets for the generated signal\footnote{In contrast with
  benchmarks I and II, events without jets in the final state also
  contribute significantly in this case and, therefore, $N(2 j) < N(1
  j) < N_{Tot}$.}  events are shown in Fig.\ref{fig:bp3_ptnocut} (top),
adopting a bin size of $16$ GeV as in the ATLAS
search\cite{ATLAS:2021uiz}.

Fig.\ref{fig:bp3_ptnocut} (bottom) shows the corresponding diphoton
invariant mass distribution. Taking cue from ref.\cite{ATLAS:2021uiz},
we then apply the selection cuts mentioned in Table
\ref{tab:signal_cuts_2_sel} in order to define the fiducial volume in
the phase space.

\begin{figure}
    \centering
      \includegraphics[width=0.46\textwidth]{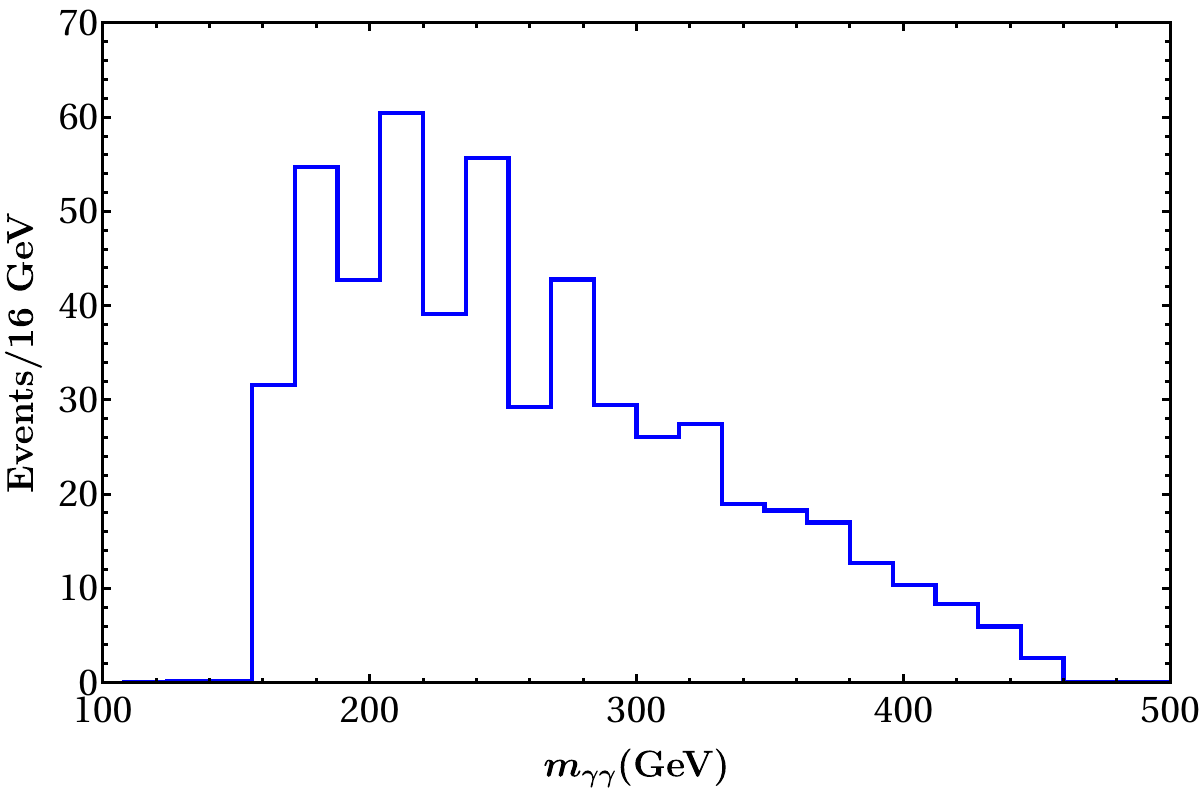}
    \caption{BP-III---Diphoton invariant mass distribution after applying selection cuts.}
    \label{fig:bp3maacut}
\end{figure}

\begin{figure}
    \centering
    \includegraphics[width=0.48\textwidth]{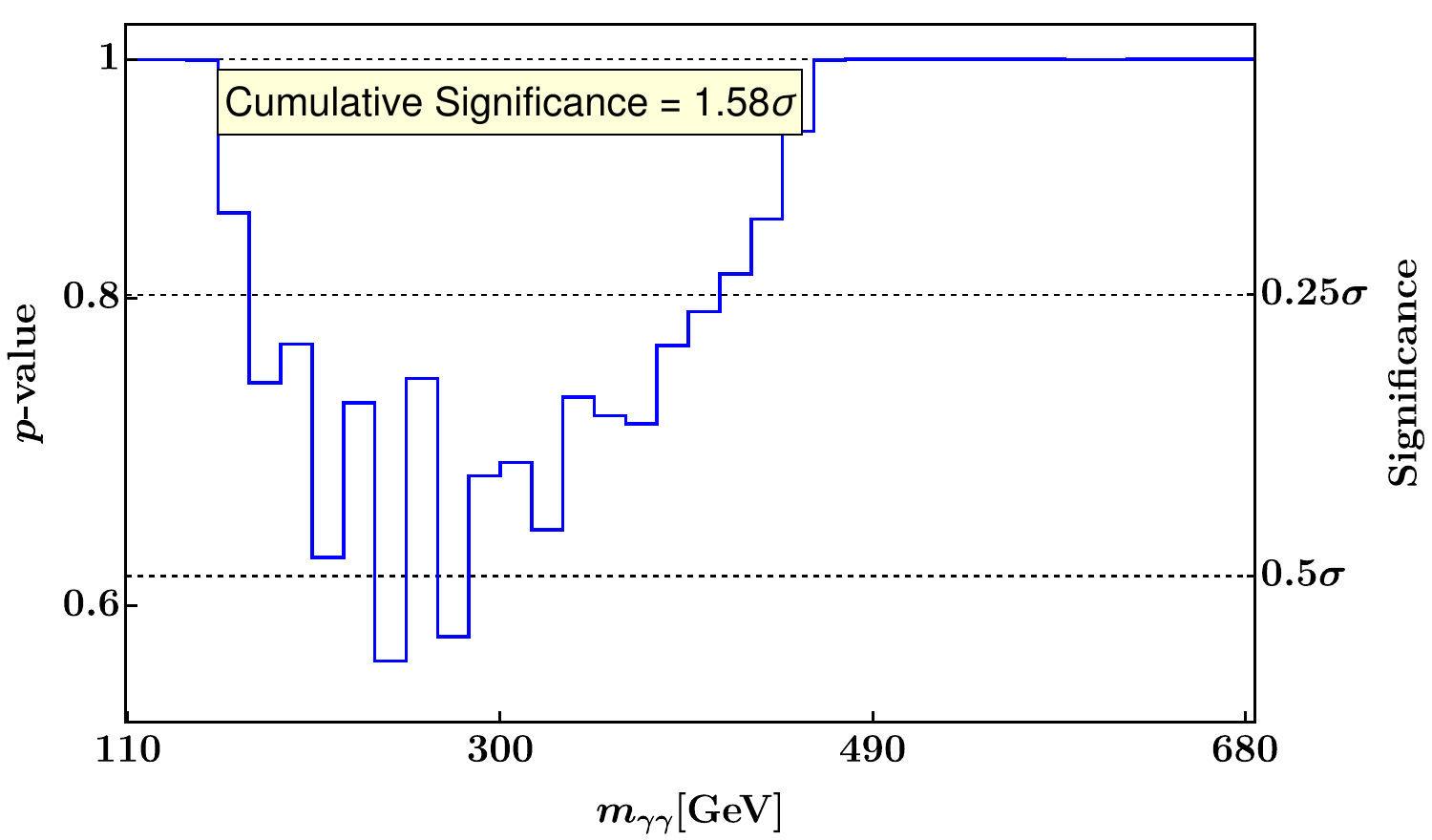}
    \caption{Bin-wise significance for benchmark III at the LHC for $\mathcal{L}=138 \, \rm fb^{-1} $.}
    \label{fig:chisq_model3}
\end{figure}

  \begin{table*}[htbp!]
    \begin{center}
    \begin{tabular}{c|c}
        \hline
          Channel & Event Selection Criteria \\
       \hline
    &  $N_\gamma = 2$, $N_j\leq 2$   \\
           $p p \rightarrow a_n  \,$   & $|\eta_\gamma|<2.37$ (excluding barrel-to-endcap region $1.37<|\eta_\gamma|< 1.52$),\\
           $a_n \rightarrow \gamma \gamma$     & $E_T (\gamma_1) > 0.3 \, m_{\gamma \gamma},\,E_T (\gamma_2) > 0.25 \,  m_{\gamma \gamma}$\\
                                       &$p_T^{j} > 20 $ GeV, $|\eta_j| <2.5$\\
                                       
       \hline
    \end{tabular}
    \caption{Event selection cuts applied to form the fiducial signal regions for benchmark III \cite{ATLAS:2021uiz}.}
    \label{tab:signal_cuts_2_sel}
        
    \end{center}
\end{table*}
The resulting diphoton invariant mass distribution is shown in
Fig.\ref{fig:bp3maacut}, whereas Fig.\ref{fig:chisq_model3} shows the
corresponding bin-wise significance (with the background profile
  adopted from ref.\cite{ATLAS:2021uiz}) over the diphoton signal
region with a cumulative significance $S_{c,138} \approx 1.58
\sigma$. For this case, the projected enhancement for a $300 \, \rm
fb^{-1}$ luminosity reach is $S_{c,300} \approx 2.33 \sigma$. Thus, in
contrast to benchmarks I and II, the high mass scenario will not
be readily accessible in the LHC's Run 3. It is perhaps
worth speculating, though, that benchmark III, with heavy ALPs, could
potentially reach the discovery threshold during the forthcoming high
luminosity phase of the LHC with a projected luminosity enhancement by
a factor of nearly 20 times the current value.

\section{The Vector-like quarks and the heavy scalars} \label{sec:VLQ}
Beyond the low energy spectrum of the pseudoscalars, the model, as
described in section \ref{sec:model}, also contains a heavy quark with
the $SU(3)_c \times SU(2)_L \times U(1)_Y$ charge assignment
$(3,1,2/3)$ as well as $N+1$ heavy radial scalar singlets (the
partners of the pseudoscalars). Although not germane to the main
objectives
of our analysis, it is worth outlining here the dynamics of both the
VLQ and the heavy scalars so as to establish the consistency of the
model, especially in view of the fact that the assumed mass scales
(characterised by the SSB scale $f$) of these heavy particles are, in
principle, accessible at the LHC.

Considering that $m^2 \ll f^2$ in all the benchmarks, the off-diagonal quadratic terms for the radial scalars in the Lagrangian (eq.(\ref{eq:lagscal})) are hierarchically smaller than the diagonal mass terms
governed by the coupling $\lambda$, which is presumed to assume $\mathcal{O}(1)$ values. Consequently, all the radial scalars are
nearly degenerate with masses $\sim \sqrt{2 \lambda}f$. To simplify
matters, we assume the other new quartic coupling, {\em viz.}
$\lambda_{\Phi H}$, is also small, thereby automatically relaxing constraints from the stability of the SM Higgs potential, triviality, etc. Similar to the pseudoscalars, these
heavy scalars have effective interactions with the gluons, photons and
the EW vector bosons with the respective couplings given by $g_{\pi V
  V}/\xi$, where $g_{\pi V V}$ represent the pseudoscalar couplings as
shown in eq.(\ref{eq:lagpivv2}).

As stated in the discussion following eq.(\ref{eq:yukawa}), we assume,
for the sake of simplicity, that the singlet VLQ mixes predominantly
with the SM top quark, thereby automatically suppressing FCNCs
involving the first two generations of quarks.  Post-EWSB, the two
quarks mix and the corresponding transformations can be expressed (in
the limit of neglecting the very small mixings with the first two
generations) as

\beq
\begin{split}
\left(\Psi_L \quad u^{(3)}_L\right)^T &= U \, \left(T_L \quad t_L\right)^T \ ,
\qquad {\rm and} \\
\left(\Psi_R \quad u^{(3)}_R\right)^T &= V \, \left(T_R \quad t_R\right)^T \ ,
\end{split}
\eeq
where $T, t$ represent the mass eigenstates and $U, V$ are the
(special unitary) mixing matrices. Since $\Psi_R$ and
  $u^{(3)}_R$ have identical quantum numbers, there is no
  gauge-mediated FCNC involving $T_R$ and $t_R$ and the weak gauge
couplings of the VLQ-like state $T$ can be expressed as

\begin{widetext}
\beq
\begin{split}
{\cal L}_{T-V}  \dis = \left( \frac{g}{\sqrt{2}}U_{tT}\right)W_{\mu}\,\Bar{b}_L \gamma^{\mu}T_L
 \; + \left(\frac{g}{2 c_w} U^{*}_{t t}U_{t T}\right)Z_{\mu}\, \Bar{t}_{L}\gamma^{\mu}T_{L} + \dis \frac{g}{c_w} \, 
Z_{\mu}\, \Bar{T}\gamma^{\mu} \left[- \, \frac{2}{3}s_w^2+\frac{g}{2c_w}U^{*}_{T t}U_{t T} P_L  \right] T + \mbox{h.c.} 
\end{split}
    \label{VLQ_gauge}
\eeq
\end{widetext}
where $P_{L,R}$ are the usual chiral projection operators. On the
other hand, the FCNCs involving the scalars, as derived from
eq.(\ref{eq:lagpsi2}), are given by

\begin{widetext}
\beq
\begin{split}
\mathcal{L}_{T-\phi} &= -\frac{1}{\sqrt{2}} \Bar{t}_{L}\left[ \left(\lambda_{\Psi} U^*_{tT}V_{TT} + y'_{\Psi} U^*_{tT}V_{tT}\right) \phi_N + \left(\lambda_h U^*_{tt}V_{tT}+y_{\Psi} U^*_{tt}V_{TT}\right)h \right]T_{R}\\
& -\frac{1}{\sqrt{2}} \Bar{T}_{L}\left[ \left(\lambda_{\Psi} U^*_{TT}V_{Tt} + y'_{\Psi} U^*_{TT}V_{tt}\right) \phi_N + \left(\lambda_h U^*_{Tt}V_{tt}+y_{\Psi} U^*_{Tt}V_{Tt}\right)h \right]t_{R} + \mbox{h.c.} \ ,
\end{split}
    \label{VLQ_scalar}
\eeq
\end{widetext}
where $\lambda_h$ denotes the top-Higgs Yukawa coupling.
To obtain the pseudoscalar FCNCs, it is perhaps the most convenient to
make use of the shift symmetry and rescale the field $\Psi_L \to
e^{i\xi \pi_N/f} \Psi_L$ in eq.(\ref{eq:lagpsi2}). This, in turn,
results in a pseudovector FCNC term in the full Lagrangian through the
kinetic term of the field $\Psi_L$, namely,
\beq
\mathcal{L}_{T-\pi} = \frac{\partial_{\mu} \pi_N}{f} \, \Bar{T}_L \gamma^{\mu} \left( U^{*}_{TT} U_{Tt}\right)t_L + \mbox{h.c.},
\eeq
with $\pi_N$ being expressed in terms of the mass eigenstates
  $a_n$ through $\pi_N = \sum_k C^{-1}_{Nk} a_k$ with the matrix $C$ as
  defined in eq.(\ref{eq:eigvec}).  For $f \gtrsim 1000$ GeV and
$m_{T} \sim \lambda_{\Psi} f / \sqrt{2} \, > (m_{W,Z,h,\phi}+m_t)$,
the primary decay channels for the VLQ are clearly $T \to b \, W$, $T
\to t \, Z$, $T \to t \, h$, $T \to t \, \phi_N$ (if kinematically
allowed) and $T \to t \, a_n$ (see appendix \ref{sec:appdecay} for the
decay width expressions \cite{Atre:2011ae}). Note that the decays to
the new scalars and pseudoscalars appear in addition to the
conventional VLQ decay modes to the SM particles and, therefore, it is
worth comparing the branching fractions pertaining to the different
channels. As this needs
  specifying a few further parameters, we begin by doing so for each
  of the benchmark points:

\textbf{BP-I and II:} We assume $\lambda_{\Psi}
= 2.2$, $y_{\Psi}=\epsilon y'_{\Psi}$, $\lambda=1.8$ and $y'_{\Psi}
\lesssim 0.1$ with $\epsilon=0.1$. This choice of
$\lambda_{\Psi}$ is motivated by the experimental
lower bound on the VLQ mass ($m_T \gtrsim 1500$ GeV) to be discussed
below.  However, as can be expected from such a large value of
  the coupling, the evolution is fast and an examination of the
  three-loop renormalization group
  equations \cite{Buttazzo:2013uya} leads to the conclusion that the
  theory becomes a very strongly coupled one at a scale $\mu \sim 10$ TeV, signalling that new
  physics must take over well before this scale\footnote{A cutoff near $10$ TeV also serves to mitigate probable issues related to naturalness in the extended scalar sector, exhibiting only a \emph{little} hierarchy problem.}. Given that the clockwork
  model is {\em not} a UV-complete one, this, {\em per se}, may still
  be overlooked. However, we return to this point in the next
  section.

Note that the aforementioned collider limits assume that the VLQ
decays within the detector. If, on the other hand, its decay length was large enough that it
does not decay within the LHC detector\footnote{This could happen if
    both $y'_\Psi$ and $y_\Psi$ are tiny. The VLQ, nonetheless,
    has to decay on cosmological time scales so as not to violate the
    strong constraints on coloured and electrically charged dark
    matter candidates. Possibility of a bound state, made of stable
    neutral VLQs, being a DM candidate is discussed in
    \cite{DeLuca:2018mzn}.}, it would hadronize and leave very
distinctive signals. While such ``R-hadrons" have been considered in
the literature \cite{Baer:1998pg,Arkani-Hamed:2004ymt,Giudice:2004tc,
  Kilian:2004uj,Hewett:2004nw,
  Arvanitaki:2005nq,Fairbairn:2006gg,Diaz-Cruz:2007ewo,Choudhury:2008gb,
  DiLuzio:2015oha}, the corresponding experimental studies are not as
exhaustive \cite{ATL-PHYS-PUB-2019-019}. The limits (typically ranging from
 $\sim 900$~GeV to $\sim 1250$~GeV) depend on the details of the hadronization model assumed. We refrain from
considering this possibility in our analysis.

\textbf{BP-III:} The considerably larger value of the SSB scale $f$
allows us the luxury of choosing a relatively smaller value for the
Yukawa coupling $\lambda_{\Psi}$, and we consider, instead,
$\lambda_{\Psi} = 1.5$, $\lambda=0.7$ and $y_{\Psi}=\epsilon
y'_{\Psi}$, $y'_{\Psi} \lesssim 0.1$ with $\epsilon=0.1$. For such a
choice, the running of $\lambda_{\Psi}$ is significantly slower and
the strong coupling phase of the theory lies near $10^8$ GeV.

While in eqs.(\ref{VLQ_gauge} \& \ref{VLQ_scalar}) we have not
  listed any alterations in the SM couplings of the top-quark, it is
  obvious that certain changes would be wrought. However, these
  changes are only higher-order in the $T$--$u_3$ mixing $U_{tT}$, and
  with the latter not being large, are well below the current
  sensitivity limits (the strongest being that for the SM CKM element
  $V_{tb}$ \cite{ParticleDataGroup:2024cfk}), whether from flavour physics or from top-decay.

\begin{table*}[ht]
\centering
\begin{tabular}{||l|lll||}
\hline
 &                       & \textbf{Branching Ratios}                      & \\ \cline{2-4} 
 \textbf{Channel} & \multicolumn{1}{l|}{SM} & \multicolumn{1}{l|}{BP-I \& II} &  BP-III\\ \hline \hline
 $T \to b \, W$& \multicolumn{1}{l|}{0.5} & \multicolumn{1}{l|}{0.44} & 0.47 \\ \hline
 $T \to t \, Z$ & \multicolumn{1}{l|}{0.25} & \multicolumn{1}{l|}{0.21} & 0.23  \\ \hline
 $T \to t \, h$& \multicolumn{1}{l|}{0.25} & \multicolumn{1}{l|}{0.23} &  0.25 \\ \hline
$T \to t \, a_{(\rm all)}$ & \multicolumn{1}{l|}{---} & \multicolumn{1}{l|}{0.12} & 0.05 \\ \hline \hline
 $m_T$ \textbf{lower limit} & \multicolumn{1}{l|}{1540 GeV \cite{CMS:2022fck,Alves:2023ufm}} & \multicolumn{1}{l|}{1500 GeV} &  $\approx$ 1540 GeV \\ \hline
\end{tabular}
\caption{VLQ branching fractions for BP-I,II and III (The SM column
  refers to the assumed branching fractions for VLQ searches
  \cite{ATLAS:2023pja,CMS:2022fck}). The bottom row displays the
  resulting lower bounds on $m_T$, as derived from the limits on the
  VLQ pair-production cross-section presented in
  ref.\cite{CMS:2022fck}}.
\label{tab:br_vlq}
\end{table*}

The constraints on the VLQ sector would, thus, come from direct
observations at colliders\footnote{Collider phenomenology of such KSVZ-VLQs have also been studied in refs.\cite{Ghosh:2022rta,Ghosh:2023xhs,Ghosh:2024boo}.}. At the large hadron collider, the
overwhelmingly leading production mechanism is the QCD-driven
one\footnote{With $U_{tT}$ not being large, single production is
  suppressed and the consequent bounds~\cite{ATLAS:2023pja} are
  relaxed.}.  Once pair-produced, the $T$s would decay promptly.
Table \ref{tab:br_vlq} lists the leading VLQ branching ratios for the
three benchmark points\footnote{For the chosen set of couplings,
    a decay of $T$ to radial scalars is kinematically forbidden. For
    any phenomenologically viable set of couplings, the branching
    fractions remain much smaller than those listed in the table.}.
It can be readily ascertained that the branching fractions have a very
small dependence on the free parameter $y'_{\Psi}$.

As the table shows, the branching fractions relevant to
  the standard search algorithms are not overly affected. Consequently,
  the derived limits are only slightly relaxed at best. On the other hand, 
  it might be interesting to consider 
  exotic channels such as $T \to t + a_n \to t + \gamma\gamma$. Hitherto
  (largely) unexplored, these might be of interest at future runs of the LHC.

\section{Summary and Conclusion}
\label{sec:conclusion}
We have examined the minimal QCD axion model within the
\emph{clockwork} paradigm and investigated the prospects of observing
the massive ALPs that the model engenders at current and future hadron
colliders. To this end, we employ a KSVZ-like setup with a heavy
coloured $SU(2)_L$ singlet quark (as a top-partner) which couples to
one end of a CW chain of $(N+1)$ complex scalars. Being chirally
charged under the site-specific $U(1)$---which acts as an analogue of
the PQ symmetry---this quark generates the requisite
chiral anomaly for addressing the strong CP problem in a natural
manner. The lightest pNGB in the CW sector identifies as the QCD axion
while the accompanying heavier pNGBs in the spectrum behave as ALPs
with a characteristic mass scale $\sim m q$. Through the chiral
anomaly, these axions then couple to the gluons, the photon and the
$Z$ boson, albeit with hierarchically\footnote{The hierarchy is generated dynamically.} different decay constants for
the QCD axion and the ALPs, namely $f_{0} \sim q^{N}f$ and $f_{n>0}
\sim f$, thanks to the CW mechanism. Being naturally consistent with
the current experimental and observational constraints, the large
suppression in the effective couplings of the light axion renders it
practically invisible at high energy colliders (as is the case with
all traditional QCD axions).

 For a reasonable parameter configuration such as
 $q>1$, $N\sim \mathcal{O}(10)$, $m \gtrsim 10$ GeV and $f\sim 1$ TeV,
 what is {\em more attractive}
 instead is the production and detection of the ALPs, especially at the currently operating LHC and its
 future derivatives. To perform a quantitative analysis, we classify
 three benchmark scenarios according to the ALP masses in the range
 $10-30$ GeV (BP-I), $35-105$ GeV (BP-II) and $150-450$ GeV
 (BP-III). Such varied mass scales for the ALPs are enabled by the fact
 that the CW mechanism is based on the premise of localization in the
 theory space and, thus, is practically independent of the mass scales
 assumed in the theory. We performed an analysis of the expected
 signal profiles pertaining to the three benchmark points for the
 production of ALPs (with upto two additional jets) via gluon fusion
 and their subsequent decay to two photons at the LHC for an
 integrated luminosity of $138{\rm \, fb}^{-1}$. As anchor points for
 the analysis and for estimates of the pertinent background
 distributions, we referred to the ATLAS searches for diphoton
 resonances in the relevant invariant mass regions
 \cite{ATLAS:2022abz,ATLAS:2021uiz}.

What stands out as particularly interesting is the case of the light
ALPs where the number of particles $N$ that is consistent with the
allowed window for the QCD axion's decay constant also corresponds to very small
mass-splittings among the ALPs. The resulting diphoton invariant mass
distribution for the signal events turns out to be such that the
individual resonances are smeared by the detector's resolution
(implemented in the form of a fast simulation of the ATLAS detector)
and, thereby, overlap with each other with the full CW spectrum of
light ALPs manifesting itself, instead, as a single but very broad
resonance. It becomes immediately apparent, though, that a single such
resonance is entirely untenable, and that such an excess must have
resulted from the merging of several individual narrow resonances. On
the other hand, treating the spectrum as a whole also offers the
possibility of a better, more inclusive, estimation of the signal's
statistical significance over the background. For instance, the
simplistic \emph{cumulative} estimation defined in this work gives a
sizable significance value of $S_c \sim 3.4 \sigma$ (for an integrated luminosity of only $138~{\rm fb}^{-1}$) in BP-I, even
though the significance estimates of the individual bins in the signal
distribution are comparatively small. Moreover, when scaled by the
projected luminosity reach of the ongoing Run 3 phase of the LHC,
\emph{viz.} $\sim 300{\rm \, fb}^{-1}$, the cumulative significance
gets enhanced to $S_c \sim 5 \sigma$. Of course, a dedicated search
for such signals would call for a more detailed statistical analysis
that is tailored to the fact that the individual resonances of the CW
spectrum are characteristically correlated.

 As for the heavy ALPs, e.g. those encountered in BP-III, the
 cumulative significance hovers just below $3 \sigma$ even with
 $300{\rm \, fb}^{-1}$ of data. It should be borne in mind, though,
 that our analysis and projections are strictly based on the published
 ATLAS data. It is conceivable that a more refined analysis would
 serve to improve the sensitivity.  What is exciting, though, is the
 prospect of probing a wide window of the CW ALPs at the HL-LHC with a
 planned luminosity reach of $\sim 3000{\rm \, fb}^{-1}$, where larger
 portions of the diphoton invariant mass profile could start to become
 apparent --- the structure of the axion \emph{iceberg} revealing
 itself --- and even scenarios like BP-III could possibly surpass the
 discovery threshold. We note that the kind of multi-ALP signals
 discussed here possess a very broad and distinct quasi-periodic
 profile which cannot be reproduced in models with a single heavy ALP
 or in generic multi-scalar scenarios.

A key issue in our analysis has been the opposing pulls on the SSB scale $f$ from the need to increase the
effective ALP--gluon--gluon coupling (scaling as $f^{-1}$) on the one
hand and the lower bound on the VLQ mass (scaling as $f$) on the
other. While modifying the number of sites $N$, or the exponent
parameter $q$ does make a difference, the effect is a muted one
(especially given that renormalizability of the low energy theory, as
well as vacuum stability, requires that $q \leq 3$). However, given
that the light ALP solutions presented here (benchmark points I \& II)
indicate that further new physics must exist by $\sim 10$~TeV scale,
such a restriction might seem unwarranted and $q > 3$ may be
considered. However, since the ALP couplings to the gluons and photon
have a dependence on $q$ which appears only through the elements of
the CW transformation matrix, namely $C_{(n>0) N}$, the corresponding
gain is marginal. It can be similarly argued that a change in the
position of the VLQ coupling over the CW lattice ({\em i.e.}, to
assume $j < N$) also does not lead to a substantial gain.

Naively, a second possibility would be to postulate additional VLQs
such that the effective $a_n g g $ coupling may be raised by way of
all the VLQs contributing in the loop diagram. However, the
corresponding change in the axion potential would putatively lead to
the formation of stable domain walls/cosmic strings in the early
Universe and is phenomenologically disfavoured. While one might
attempt to mitigate the issue by invoking some of the solutions
proposed in the literature, e.g. by assuming a soft breaking of the
discrete symmetry or an inflation-induced statistical bias in the
axion potential, the model would, nevertheless, be severely
constrained due to the inherent drawbacks of such solutions
\cite{Beyer:2022ywc}. Consider, instead, the introduction of one (or
more) vector-like leptons (VLLs).  This does not contribute to the
$a_n g g $ couplings (thereby evading the domain wall constraints),
but does contribute to the $a_n \gamma \gamma $ couplings, which now
get enhanced by a factor of
\[
1 + \frac{1}{3 Y_\Psi^2} \sum N_L Q_L^2
\]
where $N_L$ is the number of VLLs of charge $Q_L$. Even the simplest
choice of a single VLL of charge $Q=-1$ enhances the coupling by a
factor of $1.75$, thereby allowing for a similar enhancement in
$f$ without changing the rates. This, in turn, allows for a smaller
Yukawa $\lambda_\Psi$, and postponement of the strong coupling phase
until a cutoff scale $\mu \gtrsim 10^7$ GeV.  As for the VLL
phenomenology, the LHC limits are understandably much weaker.  For it
to decay, it must have at least a small Yukawa coupling with a SM
lepton, and postulating this to be the $\tau$ would not only escape
low-energy constraints, but also allow for intriguing signals at the
LHC.

An entirely different mechanism of raising $f$ without suppressing the
signal is afforded by the choice of the $U(1)_N$ charge $\xi$ for the
$\Psi_L$ and $\Phi_N$. Increasing $\xi$ and $f$ proportionately keeps
$f_{\rm eff}$ (and, therefore, the $a_n g g$ coupling) unchanged.

Whatever such mechanisms may be, it is quite apparent that the
clockwork-axion-ALP scenario would be manifested at the LHC in terms
of interesting signals, perhaps the most interesting being the case of
the possibility of a strong coupling regime.

In addition to diphoton channels, several other prospective final
states could enhance our search efforts at the collider, offering
complementary avenues to test the model. Of particular interest are
processes involving the heavy VLQ or the radial scalars which, while
lying below the current LHC sensitivities, can be probed at the
upcoming and proposed high-intensity and high-energy upgrades. For
instance, as previously mentioned, the exotic decay of the VLQ to ALPs
in association with a SM quark, \emph{e.g.} $T \to t + a_n ( \to
\gamma \gamma)$, presents an interesting channel to study. Processes
mediated by the heavy radial scalars $\phi_n$ such as $pp \to \phi_n
\to a_n ( \to 2 \gamma / 2 j) a_n ( \to 2\gamma / 2j)$ are also worth
exploring, more so because of the recent interest in the search for
boosted diphotons \cite{CMS:2024vjn}. Another noteworthy process is
the axion mediated production of a $Z$ boson in association with a
photon, with the $Z$ subsequently decaying into dijets $(p p \to a_n
\to Z (\to j j) \, \gamma )$. While the cross section for this process
is low for the chosen benchmark points in the parameter space and the
background is substantial, it remains a promising channel for probing
the model at future colliders, particularly for benchmark point III.
  
 This apart, there also exist other intriguing channels, the
 nonstandard decay $Z \to a_n \gamma$ being one interesting
 example. Although the process is hardly relevant for an investigation
 at the LHC, given the strong limits placed on the $Z$ exotic decay
 fractions at LEP, it could be well within the projected sensitivities
 of future lepton colliders, \emph{e.g.} the proposed ILC running in
 the Giga-$Z$ mode \cite{Steinberg:2021iay}. Furthermore, while the
 minimal setup considered in this work does not engender coupling of
 the ALPs to the light quarks or leptons even at the one-loop level,
 one could envision extending or modifying the model to include mixing
 of the VLQ with the light quarks and introduce additional VLLs that
 mix with the SM leptons in order to induce ALP couplings with the
 light SM fermions. In that case, several interesting channels open
 up, most notable of them being $h \to a_n a_n \to b \Bar{b} \, \mu
 \Bar{\mu}$ (with both the ATLAS and CMS experiments having observed
 intriguing signals \cite{ATLAS:2021hbr,CMS:2018nsh} in this channel)
 and $h \to Z \, a_n, \, a_n \to \, \mbox{hadrons}$
 \cite{Schmieden:2021pvm}. This would also have interesting
 ramifications for flavour physics, including tests for lepton flavour
 universality in $B$-decays.

 A complementary test of the model can also be found in the context of
 the direct detection of the light axion. A quick inspection shows
 that the masses and the effective photon couplings of the QCD axion
 in BP I and II(a) are close to the current sensitivities of several
 cavity haloscope experiments such as ADMX, CAPP, HAYSTAC, QUAX,
 etc. Thus, a collider probe of the characteristic multi-ALP
 signatures, which would largely constrain the parameters $m$ and
 $f_{\rm eff}$, alongwith the complementary probe of the light axion
 at the current or next generation of direct detection experiments,
 which is sensitive to the parameters $f_{\rm eff}$, $q$ and $N$,
 present a robust experimental platform to test the model in the near
 future.

\begin{acknowledgments}
The authors thank Satyaki Bhattacharya, Tuhin S. Roy and Shilpi Jain for helpful
discussions. D.C. acknowledges IoE, University of Delhi grant
IoE/2024-25/12/FRP and grant no. CRG/2023/008234 of the ANRF, India.
S.M. acknowledges research Grant
No. CRG/2018/004889 of the SERB, India. T.S. acknowledges the support
from the Dr. D.S.  Kothari Postdoctoral fellowship scheme
no. F.4-2/2006 (BSR)/PH/20-21/0163.
\end{acknowledgments}

\section*{Appendices}
\appendix

\section{The axial anomaly} \label{sec:appanomaly}

In the KSVZ model described in this work, the left and the
right-handed projections of the heavy fermion $\Psi$ are chirally
charged under the global symmetry $U(1)_N$. As is usually the case
in KSVZ models, the chiral (or axial) $U(1)_N$ (analogous to
$U(1)_{PQ}$ in the original setup) symmetry is anomalous which, in turn,
results in the couplings of the pseudoscalar $\pi_N$ to gluons and the
hypercharge boson given in eq.(\ref{eq:lagpivv}). This is most
succinctly realized in the path integral formulation of the KSVZ
theory, with the partition function given by

\beq
Z[0]_{A} = \int \mathcal{D}\Bar{\Psi} \mathcal{D}\Psi \exp{\left[ i\int d^4 x \, \mathcal{L}_{\Psi}\right]}.
\eeq
Here, the subscript $A$ denotes a fixed background with respect to the
gauge fields and $\mathcal{L}_{\Psi}[\Psi,\Bar{\Psi},D_{\mu}\Psi,
D_{\mu}\Bar{\Psi}]$ marks the full Lagrangian for $\Psi$ including the
kinetic terms as well. Post SSB (at the scale $f$), the Yukawa term for
$\Psi$ given in eq.(\ref{eq:lagpsi2}) (ignoring the mixing terms with
the SM fermions for brevity) can be rewritten in the convenient
form
\beq
-\frac{\lambda_{\Psi}}{\sqrt{2}}(\phi + f)e^{i\xi \pi_N \gamma_5/f}\Bar{\Psi}\Psi.
\eeq

The anomalous nature of $U(1)_{N}$ is apparent when we see that while the tree level action is invariant under the chiral transformation 
\beq
\Psi'=U \Psi = e^{-i\xi \beta \gamma_5/(2f)}\Psi \, , \quad \pi_N \to  \pi_N + \beta ,
\eeq
where $\beta$ is a constant transformation parameter, the measure of the path integral is not. The measure would transform as,

\beq
\mathcal{D}\Bar{\Psi} \mathcal{D}\Psi \to \mathcal{J}^{-2} \mathcal{D}\Bar{\Psi} \mathcal{D}\Psi\, ,
\eeq
where $\mathcal{J}$ is the Jacobian corresponding to the chiral transformation. The Jacobian, after appropriate regularisation, is given by \cite{Schwartz_2013, Bilal:2008qx}
\beq
\mathcal{J}=\exp{\left[i\int d^4 x \,   \frac{1}{32 \pi^2}\frac{\xi \, \beta}{f} \,  \mbox{Tr}\{ F^{\mu \nu} \Tilde{F}_{\mu \nu}\} \right]}.
\eeq
where $F^{\mu \nu} \Tilde{F}_{\mu \nu} \equiv F^{A \mu \nu} \Tilde{F}^{B}_{ \mu \nu} T^A T^B$ which stands for the gauge fields for all the local symmetries under which $\Psi$ is charged with $T^{A}$ being the respective generators. For brevity, we define $T^{A}$ such that they include the corresponding gauge couplings. Tr denotes trace over the gauge representation of the field $\Psi$. Therefore, the anomaly term which appears in the full Lagrangian is given by
\beq
\mathcal{A}(x)= -\frac{\xi \, \beta}{16 \pi^2 f} \,  \mbox{Tr}\{ F^{\mu \nu} \Tilde{F}_{\mu \nu}\}.
\eeq

Similarly, if we use the freedom to redefine the field $\Psi$ so as to absorb the pseudoscalar field through the chiral transformation
\beq
\Psi \to e^{-i\xi \pi_N \gamma_5/(2f)}\Psi,
\eeq
we obtain the corresponding anomaly term in the Lagrangian given by

\beq
\mathcal{A}(x)= -\frac{\xi \,\pi_N}{16 \pi^2 f} \mbox{Tr}\{ F^{\mu \nu} \Tilde{F}_{\mu \nu}\}.
\eeq
A straightforward evaluation of the trace in the preceding expression leads to the terms in eq.(\ref{eq:lagpivv}).

\section{VLQ decay widths} \label{sec:appdecay}
In the following we list the decay width expressions for the heavy
fermion $\Psi$ with the labels $q$ and $V$ denoting the SM
quarks $(b, t)$ and the EW gauge bosons $(W,Z)$ respectively.

\begin{widetext}
\beq
\begin{split}
     \Gamma(T \to q V)  = & |g^{V}_L|^2 \;
     \frac{\sqrt{m_q^4 - 2m_q^2 \left(m_T^2 + m_V^2\right) + \left(m_T^2 - m_V^2\right)^2}}{32 \pi \, m_T^3 \, m_V^2}
    \Big[\left(m_T^2 - m_q^2\right)^2 + m_V^2\left(m_T^2 + m_q^2\right) - 2 m_V^4\Big]  \\[1.5ex]
    \approx & \frac{|g^{V}_L|^2 }{32 \pi}\frac{m_T^3}{m_V^2} \quad (\mbox{for} \,\, m_T \gg m_{q, V}) \quad .
\end{split}
\eeq
\end{widetext}
Here, the couplings $g^{V}_{L}$ are given by
\beq \label{eq:vectcoup}
\dis g^W_L = \frac{g}{\sqrt{2}}U_{tT}\ , \qquad \qquad
g^Z_{L} = \frac{1}{2}\frac{g}{c_w}U^{*}_{t t}U_{t T} \ ,
\eeq
whereas the corresponding (off-diagonal)
right-handed couplings vanish identically thanks to an analogue of the GIM
mechanism.

For decays to scalars we ignore the small mixing between the SM Higgs and the heavy scalars $\phi_n$. The corresponding expressions are

\begin{widetext}
\begin{subequations}
\beq
\begin{split}
    \Gamma(T \to t \, h ) & = \frac{\sqrt{m_h^4 - 2m_h^2 \left(m_T^2 + m_t^2\right) + \left(m_T^2 - m_t^2\right)^2}}{64 \pi \, m_T^3}\\
    & \times \left\{\left( |y^{h}_{Tt}|^2 + |y^{h}_{tT}|^2\right)\left[ m_T^2 + m_t^2 - m_{h}^2\right] +4\mbox{Re}\left( y^{h *}_{Tt}y^{h}_{tT}\right)m_T m_t\right\} \\[1.5ex]
    & \approx \frac{|y^h_{Tt}|^2 + |y^h_{tT}|^2}{64 \pi}m_T \quad (\mbox{for} \,\, m_T \gg m_{t,h}) \ ,
\end{split}
\eeq

\beq
\begin{split}
    \Gamma(T \to t \, \phi_N ) & = \frac{\sqrt{m_{\phi_N}^4 - 2m_{\phi_N}^2 \left(m_T^2 + m_t^2\right) + \left(m_T^2 - m_t^2\right)^2}}{64 \pi \, m_T^3}\\
    & \times \left\{\left( |y^{\Phi}_{Tt}|^2 + |y^{\Phi}_{tT}|^2\right)\left[ m_T^2 + m_t^2 - m_{\phi_N}^2\right] +4\mbox{Re}\left( y^{\Phi *}_{Tt}y^{\Phi}_{tT}\right)m_T m_t\right\}\\[1.5ex]
    & \approx \frac{|y^{\Phi}_{Tt}|^2 + |y^{\Phi}_{tT}|^2}{64 \pi}m_T\left( 1 -  \frac{m_{\phi_N}^2}{m_T^2}\right)^2 \quad (\mbox{for} \,\, m_{T,\phi_N} \gg m_{t})
\end{split}
\eeq
and 
\beq
\barr{rcl}
    \Gamma(T \to t \, a_n ) & = & \dis
    \frac{\xi^2 C_{n N}^2\sqrt{m_{n}^4 - 2m_{n}^2 \left(m_T^2 + m_t^2\right) + \left(m_T^2 - m_t^2\right)^2}}{32 \pi f^2 \, m_T^3}\\[1.5ex]
    & & \dis  \left( |U_{TT}|^2 |U_{Tt}|^2\right)\left[ \left( m_T^2 - m_t^2\right)^2 - \left(m_T^2 + m_t^2\right)m_n^2\right]\\[2.5ex]
    & \approx & \dis \frac{\lambda_{\Psi}^2 \, \xi^2 \, C_{n N}^2 |U_{TT}|^2 |U_{Tt}|^2}{32 \pi} \, m_T \qquad (\mbox{for} \,\, m_T \gg m_{t,a_n}) \quad ,
\earr
\eeq
\end{subequations}
\end{widetext}
where the effective Yukawa couplings are given by
\beq
\begin{split}
y^h_{Tt} &\approx \dis \lambda_h U^*_{Tt}V_{tt} \ ,\\
y^h_{tT} &=  \dis \lambda_h U^*_{tt}V_{tT}+y_{\Psi} U^*_{tt}V_{TT} \ ,\\
y^{\Phi}_{tT} &\approx  \dis \lambda_{\Psi} U^*_{tT}V_{TT}\ , \\
y^{\Phi}_{Tt} &=  \dis \lambda_{\Psi} U^*_{TT}V_{Tt}+y'_{\Psi} U^*_{TT}V_{tt}  \ .
\end{split}
\eeq




\newpage

\bibliographystyle{apsrev4-1} 
\bibliography{references.bib} 

\begin{thebibliography}{141}%
\makeatletter
\providecommand \@ifxundefined [1]{%
 \@ifx{#1\undefined}
}%
\providecommand \@ifnum [1]{%
 \ifnum #1\expandafter \@firstoftwo
 \else \expandafter \@secondoftwo
 \fi
}%
\providecommand \@ifx [1]{%
 \ifx #1\expandafter \@firstoftwo
 \else \expandafter \@secondoftwo
 \fi
}%
\providecommand \natexlab [1]{#1}%
\providecommand \enquote  [1]{``#1''}%
\providecommand \bibnamefont  [1]{#1}%
\providecommand \bibfnamefont [1]{#1}%
\providecommand \citenamefont [1]{#1}%
\providecommand \href@noop [0]{\@secondoftwo}%
\providecommand \href [0]{\begingroup \@sanitize@url \@href}%
\providecommand \@href[1]{\@@startlink{#1}\@@href}%
\providecommand \@@href[1]{\endgroup#1\@@endlink}%
\providecommand \@sanitize@url [0]{\catcode `\\12\catcode `\$12\catcode `\&12\catcode `\#12\catcode `\^12\catcode `\_12\catcode `\%12\relax}%
\providecommand \@@startlink[1]{}%
\providecommand \@@endlink[0]{}%
\providecommand \url  [0]{\begingroup\@sanitize@url \@url }%
\providecommand \@url [1]{\endgroup\@href {#1}{\urlprefix }}%
\providecommand \urlprefix  [0]{URL }%
\providecommand \Eprint [0]{\href }%
\providecommand \doibase [0]{http://dx.doi.org/}%
\providecommand \selectlanguage [0]{\@gobble}%
\providecommand \bibinfo  [0]{\@secondoftwo}%
\providecommand \bibfield  [0]{\@secondoftwo}%
\providecommand \translation [1]{[#1]}%
\providecommand \BibitemOpen [0]{}%
\providecommand \bibitemStop [0]{}%
\providecommand \bibitemNoStop [0]{.\EOS\space}%
\providecommand \EOS [0]{\spacefactor3000\relax}%
\providecommand \BibitemShut  [1]{\csname bibitem#1\endcsname}%
\let\auto@bib@innerbib\@empty
\bibitem [{\citenamefont {Peccei}\ and\ \citenamefont {Quinn}(1977)}]{PQPhysRevLett}%
  \BibitemOpen
  \bibfield  {author} {\bibinfo {author} {\bibfnamefont {R.~D.}\ \bibnamefont {Peccei}}\ and\ \bibinfo {author} {\bibfnamefont {H.~R.}\ \bibnamefont {Quinn}},\ }\href {\doibase 10.1103/PhysRevLett.38.1440} {\bibfield  {journal} {\bibinfo  {journal} {Phys. Rev. Lett.}\ }\textbf {\bibinfo {volume} {38}},\ \bibinfo {pages} {1440} (\bibinfo {year} {1977})}\BibitemShut {NoStop}%
\bibitem [{\citenamefont {Weinberg}(1978)}]{WeinbergPhysRevLett}%
  \BibitemOpen
  \bibfield  {author} {\bibinfo {author} {\bibfnamefont {S.}~\bibnamefont {Weinberg}},\ }\href {\doibase 10.1103/PhysRevLett.40.223} {\bibfield  {journal} {\bibinfo  {journal} {Phys. Rev. Lett.}\ }\textbf {\bibinfo {volume} {40}},\ \bibinfo {pages} {223} (\bibinfo {year} {1978})}\BibitemShut {NoStop}%
\bibitem [{\citenamefont {Wilczek}(1978)}]{WilczekPhysRevLett}%
  \BibitemOpen
  \bibfield  {author} {\bibinfo {author} {\bibfnamefont {F.}~\bibnamefont {Wilczek}},\ }\href {\doibase 10.1103/PhysRevLett.40.279} {\bibfield  {journal} {\bibinfo  {journal} {Phys. Rev. Lett.}\ }\textbf {\bibinfo {volume} {40}},\ \bibinfo {pages} {279} (\bibinfo {year} {1978})}\BibitemShut {NoStop}%
\bibitem [{\citenamefont {Georgi}\ and\ \citenamefont {McArthur}(1981)}]{Georgi:1981be}%
  \BibitemOpen
  \bibfield  {author} {\bibinfo {author} {\bibfnamefont {H.}~\bibnamefont {Georgi}}\ and\ \bibinfo {author} {\bibfnamefont {I.~N.}\ \bibnamefont {McArthur}},\ }\href@noop {} {\bibfield  {journal} {\bibinfo  {journal} {{\emph{INSTANTONS AND THE $u$ QUARK MASS}}}\ } (\bibinfo {year} {1981})}\BibitemShut {NoStop}%
\bibitem [{\citenamefont {Kaplan}\ and\ \citenamefont {Manohar}(1986)}]{Kaplan:1986ru}%
  \BibitemOpen
  \bibfield  {author} {\bibinfo {author} {\bibfnamefont {D.~B.}\ \bibnamefont {Kaplan}}\ and\ \bibinfo {author} {\bibfnamefont {A.~V.}\ \bibnamefont {Manohar}},\ }\href {\doibase 10.1103/PhysRevLett.56.2004} {\bibfield  {journal} {\bibinfo  {journal} {Phys. Rev. Lett.}\ }\textbf {\bibinfo {volume} {56}},\ \bibinfo {pages} {2004} (\bibinfo {year} {1986})}\BibitemShut {NoStop}%
\bibitem [{\citenamefont {Choi}\ \emph {et~al.}(1988)\citenamefont {Choi}, \citenamefont {Kim},\ and\ \citenamefont {Sze}}]{Choi:1988sy}%
  \BibitemOpen
  \bibfield  {author} {\bibinfo {author} {\bibfnamefont {K.}~\bibnamefont {Choi}}, \bibinfo {author} {\bibfnamefont {C.~W.}\ \bibnamefont {Kim}}, \ and\ \bibinfo {author} {\bibfnamefont {W.~K.}\ \bibnamefont {Sze}},\ }\href {\doibase 10.1103/PhysRevLett.61.794} {\bibfield  {journal} {\bibinfo  {journal} {Phys. Rev. Lett.}\ }\textbf {\bibinfo {volume} {61}},\ \bibinfo {pages} {794} (\bibinfo {year} {1988})}\BibitemShut {NoStop}%
\bibitem [{\citenamefont {Banks}\ \emph {et~al.}(1994)\citenamefont {Banks}, \citenamefont {Nir},\ and\ \citenamefont {Seiberg}}]{Banks:1994yg}%
  \BibitemOpen
  \bibfield  {author} {\bibinfo {author} {\bibfnamefont {T.}~\bibnamefont {Banks}}, \bibinfo {author} {\bibfnamefont {Y.}~\bibnamefont {Nir}}, \ and\ \bibinfo {author} {\bibfnamefont {N.}~\bibnamefont {Seiberg}},\ }in\ \href@noop {} {\emph {\bibinfo {booktitle} {{2nd IFT Workshop on Yukawa Couplings and the Origins of Mass}}}}\ (\bibinfo {year} {1994})\ pp.\ \bibinfo {pages} {26--41},\ \Eprint {http://arxiv.org/abs/hep-ph/9403203} {arXiv:hep-ph/9403203} \BibitemShut {NoStop}%
\bibitem [{\citenamefont {Dine}\ \emph {et~al.}(2015)\citenamefont {Dine}, \citenamefont {Draper},\ and\ \citenamefont {Festuccia}}]{Dine:2014dga}%
  \BibitemOpen
  \bibfield  {author} {\bibinfo {author} {\bibfnamefont {M.}~\bibnamefont {Dine}}, \bibinfo {author} {\bibfnamefont {P.}~\bibnamefont {Draper}}, \ and\ \bibinfo {author} {\bibfnamefont {G.}~\bibnamefont {Festuccia}},\ }\href {\doibase 10.1103/PhysRevD.92.054004} {\bibfield  {journal} {\bibinfo  {journal} {Phys. Rev. D}\ }\textbf {\bibinfo {volume} {92}},\ \bibinfo {pages} {054004} (\bibinfo {year} {2015})},\ \Eprint {http://arxiv.org/abs/1410.8505} {arXiv:1410.8505 [hep-ph]} \BibitemShut {NoStop}%
\bibitem [{\citenamefont {Nelson}(1984)}]{Nelson:1983zb}%
  \BibitemOpen
  \bibfield  {author} {\bibinfo {author} {\bibfnamefont {A.~E.}\ \bibnamefont {Nelson}},\ }\href {\doibase 10.1016/0370-2693(84)92025-2} {\bibfield  {journal} {\bibinfo  {journal} {Phys. Lett. B}\ }\textbf {\bibinfo {volume} {136}},\ \bibinfo {pages} {387} (\bibinfo {year} {1984})}\BibitemShut {NoStop}%
\bibitem [{\citenamefont {Barr}(1984)}]{Barr:1984qx}%
  \BibitemOpen
  \bibfield  {author} {\bibinfo {author} {\bibfnamefont {S.~M.}\ \bibnamefont {Barr}},\ }\href {\doibase 10.1103/PhysRevLett.53.329} {\bibfield  {journal} {\bibinfo  {journal} {Phys. Rev. Lett.}\ }\textbf {\bibinfo {volume} {53}},\ \bibinfo {pages} {329} (\bibinfo {year} {1984})}\BibitemShut {NoStop}%
\bibitem [{\citenamefont {Alexandrou}\ \emph {et~al.}(2020)\citenamefont {Alexandrou}, \citenamefont {Finkenrath}, \citenamefont {Funcke}, \citenamefont {Jansen}, \citenamefont {Kostrzewa}, \citenamefont {Pittler},\ and\ \citenamefont {Urbach}}]{Alexandrou:2020bkd}%
  \BibitemOpen
  \bibfield  {author} {\bibinfo {author} {\bibfnamefont {C.}~\bibnamefont {Alexandrou}}, \bibinfo {author} {\bibfnamefont {J.}~\bibnamefont {Finkenrath}}, \bibinfo {author} {\bibfnamefont {L.}~\bibnamefont {Funcke}}, \bibinfo {author} {\bibfnamefont {K.}~\bibnamefont {Jansen}}, \bibinfo {author} {\bibfnamefont {B.}~\bibnamefont {Kostrzewa}}, \bibinfo {author} {\bibfnamefont {F.}~\bibnamefont {Pittler}}, \ and\ \bibinfo {author} {\bibfnamefont {C.}~\bibnamefont {Urbach}},\ }\href {\doibase 10.1103/PhysRevLett.125.232001} {\bibfield  {journal} {\bibinfo  {journal} {Phys. Rev. Lett.}\ }\textbf {\bibinfo {volume} {125}},\ \bibinfo {pages} {232001} (\bibinfo {year} {2020})},\ \Eprint {http://arxiv.org/abs/2002.07802} {arXiv:2002.07802 [hep-lat]} \BibitemShut {NoStop}%
\bibitem [{\citenamefont {Dine}\ and\ \citenamefont {Draper}(2015)}]{Dine:2015jga}%
  \BibitemOpen
  \bibfield  {author} {\bibinfo {author} {\bibfnamefont {M.}~\bibnamefont {Dine}}\ and\ \bibinfo {author} {\bibfnamefont {P.}~\bibnamefont {Draper}},\ }\href {\doibase 10.1007/JHEP08(2015)132} {\bibfield  {journal} {\bibinfo  {journal} {JHEP}\ }\textbf {\bibinfo {volume} {08}},\ \bibinfo {pages} {132} (\bibinfo {year} {2015})},\ \Eprint {http://arxiv.org/abs/1506.05433} {arXiv:1506.05433 [hep-ph]} \BibitemShut {NoStop}%
\bibitem [{\citenamefont {Lazarides}\ and\ \citenamefont {Shafi}(1982)}]{Lazarides:1982tw}%
  \BibitemOpen
  \bibfield  {author} {\bibinfo {author} {\bibfnamefont {G.}~\bibnamefont {Lazarides}}\ and\ \bibinfo {author} {\bibfnamefont {Q.}~\bibnamefont {Shafi}},\ }\href {\doibase 10.1016/0370-2693(82)90506-8} {\bibfield  {journal} {\bibinfo  {journal} {Phys. Lett. B}\ }\textbf {\bibinfo {volume} {115}},\ \bibinfo {pages} {21} (\bibinfo {year} {1982})}\BibitemShut {NoStop}%
\bibitem [{\citenamefont {Chatterjee}\ \emph {et~al.}(2020)\citenamefont {Chatterjee}, \citenamefont {Higaki},\ and\ \citenamefont {Nitta}}]{Chatterjee:2019rch}%
  \BibitemOpen
  \bibfield  {author} {\bibinfo {author} {\bibfnamefont {C.}~\bibnamefont {Chatterjee}}, \bibinfo {author} {\bibfnamefont {T.}~\bibnamefont {Higaki}}, \ and\ \bibinfo {author} {\bibfnamefont {M.}~\bibnamefont {Nitta}},\ }\href {\doibase 10.1103/PhysRevD.101.075026} {\bibfield  {journal} {\bibinfo  {journal} {Phys. Rev. D}\ }\textbf {\bibinfo {volume} {101}},\ \bibinfo {pages} {075026} (\bibinfo {year} {2020})},\ \Eprint {http://arxiv.org/abs/1903.11753} {arXiv:1903.11753 [hep-ph]} \BibitemShut {NoStop}%
\bibitem [{\citenamefont {Ibe}\ \emph {et~al.}(2020)\citenamefont {Ibe}, \citenamefont {Kobayashi}, \citenamefont {Suzuki},\ and\ \citenamefont {Yanagida}}]{Ibe:2019yew}%
  \BibitemOpen
  \bibfield  {author} {\bibinfo {author} {\bibfnamefont {M.}~\bibnamefont {Ibe}}, \bibinfo {author} {\bibfnamefont {S.}~\bibnamefont {Kobayashi}}, \bibinfo {author} {\bibfnamefont {M.}~\bibnamefont {Suzuki}}, \ and\ \bibinfo {author} {\bibfnamefont {T.~T.}\ \bibnamefont {Yanagida}},\ }\href {\doibase 10.1103/PhysRevD.101.035029} {\bibfield  {journal} {\bibinfo  {journal} {Phys. Rev. D}\ }\textbf {\bibinfo {volume} {101}},\ \bibinfo {pages} {035029} (\bibinfo {year} {2020})},\ \Eprint {http://arxiv.org/abs/1909.01604} {arXiv:1909.01604 [hep-ph]} \BibitemShut {NoStop}%
\bibitem [{\citenamefont {Zhang}(2024)}]{Zhang:2023gfu}%
  \BibitemOpen
  \bibfield  {author} {\bibinfo {author} {\bibfnamefont {Y.}~\bibnamefont {Zhang}},\ }\href {\doibase 10.1103/PhysRevLett.132.081003} {\bibfield  {journal} {\bibinfo  {journal} {Phys. Rev. Lett.}\ }\textbf {\bibinfo {volume} {132}},\ \bibinfo {pages} {081003} (\bibinfo {year} {2024})},\ \Eprint {http://arxiv.org/abs/2305.15495} {arXiv:2305.15495 [hep-ph]} \BibitemShut {NoStop}%
\bibitem [{\citenamefont {Duffy}\ and\ \citenamefont {van Bibber}(2009)}]{Duffy:2009ig}%
  \BibitemOpen
  \bibfield  {author} {\bibinfo {author} {\bibfnamefont {L.~D.}\ \bibnamefont {Duffy}}\ and\ \bibinfo {author} {\bibfnamefont {K.}~\bibnamefont {van Bibber}},\ }\href {\doibase 10.1088/1367-2630/11/10/105008} {\bibfield  {journal} {\bibinfo  {journal} {New J. Phys.}\ }\textbf {\bibinfo {volume} {11}},\ \bibinfo {pages} {105008} (\bibinfo {year} {2009})},\ \Eprint {http://arxiv.org/abs/0904.3346} {arXiv:0904.3346 [hep-ph]} \BibitemShut {NoStop}%
\bibitem [{\citenamefont {Chadha-Day}\ \emph {et~al.}(2022)\citenamefont {Chadha-Day}, \citenamefont {Ellis},\ and\ \citenamefont {Marsh}}]{Chadha-Day:2021szb}%
  \BibitemOpen
  \bibfield  {author} {\bibinfo {author} {\bibfnamefont {F.}~\bibnamefont {Chadha-Day}}, \bibinfo {author} {\bibfnamefont {J.}~\bibnamefont {Ellis}}, \ and\ \bibinfo {author} {\bibfnamefont {D.~J.~E.}\ \bibnamefont {Marsh}},\ }\href {\doibase 10.1126/sciadv.abj3618} {\bibfield  {journal} {\bibinfo  {journal} {Sci. Adv.}\ }\textbf {\bibinfo {volume} {8}},\ \bibinfo {pages} {abj3618} (\bibinfo {year} {2022})},\ \Eprint {http://arxiv.org/abs/2105.01406} {arXiv:2105.01406 [hep-ph]} \BibitemShut {NoStop}%
\bibitem [{\citenamefont {Navas}\ \emph {et~al.}(2024)\citenamefont {Navas} \emph {et~al.}}]{ParticleDataGroup:2024cfk}%
  \BibitemOpen
  \bibfield  {author} {\bibinfo {author} {\bibfnamefont {S.}~\bibnamefont {Navas}} \emph {et~al.} (\bibinfo {collaboration} {Particle Data Group}),\ }\href {\doibase 10.1103/PhysRevD.110.030001} {\bibfield  {journal} {\bibinfo  {journal} {Phys. Rev. D}\ }\textbf {\bibinfo {volume} {110}},\ \bibinfo {pages} {030001} (\bibinfo {year} {2024})}\BibitemShut {NoStop}%
\bibitem [{\citenamefont {O'Hare}(2020)}]{AxionLimits}%
  \BibitemOpen
  \bibfield  {author} {\bibinfo {author} {\bibfnamefont {C.}~\bibnamefont {O'Hare}},\ }\href {\doibase 10.5281/zenodo.3932430} {\enquote {\bibinfo {title} {cajohare/axionlimits: Axionlimits},}\ }\bibinfo {howpublished} {\url{https://cajohare.github.io/AxionLimits/}} (\bibinfo {year} {2020})\BibitemShut {NoStop}%
\bibitem [{\citenamefont {Bauer}\ \emph {et~al.}(2019)\citenamefont {Bauer}, \citenamefont {Heiles}, \citenamefont {Neubert},\ and\ \citenamefont {Thamm}}]{Bauer:2018uxu}%
  \BibitemOpen
  \bibfield  {author} {\bibinfo {author} {\bibfnamefont {M.}~\bibnamefont {Bauer}}, \bibinfo {author} {\bibfnamefont {M.}~\bibnamefont {Heiles}}, \bibinfo {author} {\bibfnamefont {M.}~\bibnamefont {Neubert}}, \ and\ \bibinfo {author} {\bibfnamefont {A.}~\bibnamefont {Thamm}},\ }\href {\doibase 10.1140/epjc/s10052-019-6587-9} {\bibfield  {journal} {\bibinfo  {journal} {Eur. Phys. J. C}\ }\textbf {\bibinfo {volume} {79}},\ \bibinfo {pages} {74} (\bibinfo {year} {2019})},\ \Eprint {http://arxiv.org/abs/1808.10323} {arXiv:1808.10323 [hep-ph]} \BibitemShut {NoStop}%
\bibitem [{\citenamefont {d'Enterria}(2021)}]{dEnterria:2021ljz}%
  \BibitemOpen
  \bibfield  {author} {\bibinfo {author} {\bibfnamefont {D.}~\bibnamefont {d'Enterria}},\ }in\ \href@noop {} {\emph {\bibinfo {booktitle} {{Workshop on Feebly Interacting Particles}}}}\ (\bibinfo {year} {2021})\ \Eprint {http://arxiv.org/abs/2102.08971} {arXiv:2102.08971 [hep-ex]} \BibitemShut {NoStop}%
\bibitem [{\citenamefont {Fl\'orez}\ \emph {et~al.}(2021)\citenamefont {Fl\'orez}, \citenamefont {Gurrola}, \citenamefont {Johns}, \citenamefont {Sheldon}, \citenamefont {Sheridan}, \citenamefont {Sinha},\ and\ \citenamefont {Soubasis}}]{Florez:2021zoo}%
  \BibitemOpen
  \bibfield  {author} {\bibinfo {author} {\bibfnamefont {A.}~\bibnamefont {Fl\'orez}}, \bibinfo {author} {\bibfnamefont {A.}~\bibnamefont {Gurrola}}, \bibinfo {author} {\bibfnamefont {W.}~\bibnamefont {Johns}}, \bibinfo {author} {\bibfnamefont {P.}~\bibnamefont {Sheldon}}, \bibinfo {author} {\bibfnamefont {E.}~\bibnamefont {Sheridan}}, \bibinfo {author} {\bibfnamefont {K.}~\bibnamefont {Sinha}}, \ and\ \bibinfo {author} {\bibfnamefont {B.}~\bibnamefont {Soubasis}},\ }\href {\doibase 10.1103/PhysRevD.103.095001} {\bibfield  {journal} {\bibinfo  {journal} {Phys. Rev. D}\ }\textbf {\bibinfo {volume} {103}},\ \bibinfo {pages} {095001} (\bibinfo {year} {2021})},\ \Eprint {http://arxiv.org/abs/2101.11119} {arXiv:2101.11119 [hep-ph]} \BibitemShut {NoStop}%
\bibitem [{\citenamefont {Rubakov}(1997)}]{Rubakov:1997vp}%
  \BibitemOpen
  \bibfield  {author} {\bibinfo {author} {\bibfnamefont {V.~A.}\ \bibnamefont {Rubakov}},\ }\href {\doibase 10.1134/1.567390} {\bibfield  {journal} {\bibinfo  {journal} {JETP Lett.}\ }\textbf {\bibinfo {volume} {65}},\ \bibinfo {pages} {621} (\bibinfo {year} {1997})},\ \Eprint {http://arxiv.org/abs/hep-ph/9703409} {arXiv:hep-ph/9703409} \BibitemShut {NoStop}%
\bibitem [{\citenamefont {Berezhiani}\ \emph {et~al.}(2001)\citenamefont {Berezhiani}, \citenamefont {Gianfagna},\ and\ \citenamefont {Giannotti}}]{Berezhiani:2000gh}%
  \BibitemOpen
  \bibfield  {author} {\bibinfo {author} {\bibfnamefont {Z.}~\bibnamefont {Berezhiani}}, \bibinfo {author} {\bibfnamefont {L.}~\bibnamefont {Gianfagna}}, \ and\ \bibinfo {author} {\bibfnamefont {M.}~\bibnamefont {Giannotti}},\ }\href {\doibase 10.1016/S0370-2693(00)01392-7} {\bibfield  {journal} {\bibinfo  {journal} {Phys. Lett. B}\ }\textbf {\bibinfo {volume} {500}},\ \bibinfo {pages} {286} (\bibinfo {year} {2001})},\ \Eprint {http://arxiv.org/abs/hep-ph/0009290} {arXiv:hep-ph/0009290} \BibitemShut {NoStop}%
\bibitem [{\citenamefont {Fukuda}\ \emph {et~al.}(2015)\citenamefont {Fukuda}, \citenamefont {Harigaya}, \citenamefont {Ibe},\ and\ \citenamefont {Yanagida}}]{Fukuda:2015ana}%
  \BibitemOpen
  \bibfield  {author} {\bibinfo {author} {\bibfnamefont {H.}~\bibnamefont {Fukuda}}, \bibinfo {author} {\bibfnamefont {K.}~\bibnamefont {Harigaya}}, \bibinfo {author} {\bibfnamefont {M.}~\bibnamefont {Ibe}}, \ and\ \bibinfo {author} {\bibfnamefont {T.~T.}\ \bibnamefont {Yanagida}},\ }\href {\doibase 10.1103/PhysRevD.92.015021} {\bibfield  {journal} {\bibinfo  {journal} {Phys. Rev. D}\ }\textbf {\bibinfo {volume} {92}},\ \bibinfo {pages} {015021} (\bibinfo {year} {2015})},\ \Eprint {http://arxiv.org/abs/1504.06084} {arXiv:1504.06084 [hep-ph]} \BibitemShut {NoStop}%
\bibitem [{\citenamefont {Hook}\ \emph {et~al.}(2020)\citenamefont {Hook}, \citenamefont {Kumar}, \citenamefont {Liu},\ and\ \citenamefont {Sundrum}}]{Hook:2019qoh}%
  \BibitemOpen
  \bibfield  {author} {\bibinfo {author} {\bibfnamefont {A.}~\bibnamefont {Hook}}, \bibinfo {author} {\bibfnamefont {S.}~\bibnamefont {Kumar}}, \bibinfo {author} {\bibfnamefont {Z.}~\bibnamefont {Liu}}, \ and\ \bibinfo {author} {\bibfnamefont {R.}~\bibnamefont {Sundrum}},\ }\href {\doibase 10.1103/PhysRevLett.124.221801} {\bibfield  {journal} {\bibinfo  {journal} {Phys. Rev. Lett.}\ }\textbf {\bibinfo {volume} {124}},\ \bibinfo {pages} {221801} (\bibinfo {year} {2020})},\ \Eprint {http://arxiv.org/abs/1911.12364} {arXiv:1911.12364 [hep-ph]} \BibitemShut {NoStop}%
\bibitem [{\citenamefont {Agrawal}\ and\ \citenamefont {Howe}(2018)}]{Agrawal:2017ksf}%
  \BibitemOpen
  \bibfield  {author} {\bibinfo {author} {\bibfnamefont {P.}~\bibnamefont {Agrawal}}\ and\ \bibinfo {author} {\bibfnamefont {K.}~\bibnamefont {Howe}},\ }\href {\doibase 10.1007/JHEP12(2018)029} {\bibfield  {journal} {\bibinfo  {journal} {JHEP}\ }\textbf {\bibinfo {volume} {12}},\ \bibinfo {pages} {029} (\bibinfo {year} {2018})},\ \Eprint {http://arxiv.org/abs/1710.04213} {arXiv:1710.04213 [hep-ph]} \BibitemShut {NoStop}%
\bibitem [{\citenamefont {Gherghetta}\ \emph {et~al.}(2020)\citenamefont {Gherghetta}, \citenamefont {Khoze}, \citenamefont {Pomarol},\ and\ \citenamefont {Shirman}}]{Gherghetta:2020keg}%
  \BibitemOpen
  \bibfield  {author} {\bibinfo {author} {\bibfnamefont {T.}~\bibnamefont {Gherghetta}}, \bibinfo {author} {\bibfnamefont {V.~V.}\ \bibnamefont {Khoze}}, \bibinfo {author} {\bibfnamefont {A.}~\bibnamefont {Pomarol}}, \ and\ \bibinfo {author} {\bibfnamefont {Y.}~\bibnamefont {Shirman}},\ }\href {\doibase 10.1007/JHEP03(2020)063} {\bibfield  {journal} {\bibinfo  {journal} {JHEP}\ }\textbf {\bibinfo {volume} {03}},\ \bibinfo {pages} {063} (\bibinfo {year} {2020})},\ \Eprint {http://arxiv.org/abs/2001.05610} {arXiv:2001.05610 [hep-ph]} \BibitemShut {NoStop}%
\bibitem [{\citenamefont {Freese}\ \emph {et~al.}(1990)\citenamefont {Freese}, \citenamefont {Frieman},\ and\ \citenamefont {Olinto}}]{Freese:1990rb}%
  \BibitemOpen
  \bibfield  {author} {\bibinfo {author} {\bibfnamefont {K.}~\bibnamefont {Freese}}, \bibinfo {author} {\bibfnamefont {J.~A.}\ \bibnamefont {Frieman}}, \ and\ \bibinfo {author} {\bibfnamefont {A.~V.}\ \bibnamefont {Olinto}},\ }\href {\doibase 10.1103/PhysRevLett.65.3233} {\bibfield  {journal} {\bibinfo  {journal} {Phys. Rev. Lett.}\ }\textbf {\bibinfo {volume} {65}},\ \bibinfo {pages} {3233} (\bibinfo {year} {1990})}\BibitemShut {NoStop}%
\bibitem [{\citenamefont {Adams}\ \emph {et~al.}(1993)\citenamefont {Adams}, \citenamefont {Bond}, \citenamefont {Freese}, \citenamefont {Frieman},\ and\ \citenamefont {Olinto}}]{Adams:1992bn}%
  \BibitemOpen
  \bibfield  {author} {\bibinfo {author} {\bibfnamefont {F.~C.}\ \bibnamefont {Adams}}, \bibinfo {author} {\bibfnamefont {J.~R.}\ \bibnamefont {Bond}}, \bibinfo {author} {\bibfnamefont {K.}~\bibnamefont {Freese}}, \bibinfo {author} {\bibfnamefont {J.~A.}\ \bibnamefont {Frieman}}, \ and\ \bibinfo {author} {\bibfnamefont {A.~V.}\ \bibnamefont {Olinto}},\ }\href {\doibase 10.1103/PhysRevD.47.426} {\bibfield  {journal} {\bibinfo  {journal} {Phys. Rev. D}\ }\textbf {\bibinfo {volume} {47}},\ \bibinfo {pages} {426} (\bibinfo {year} {1993})},\ \Eprint {http://arxiv.org/abs/hep-ph/9207245} {arXiv:hep-ph/9207245} \BibitemShut {NoStop}%
\bibitem [{\citenamefont {Aad}\ \emph {et~al.}(2023{\natexlab{a}})\citenamefont {Aad} \emph {et~al.}}]{ATLAS:2022abz}%
  \BibitemOpen
  \bibfield  {author} {\bibinfo {author} {\bibfnamefont {G.}~\bibnamefont {Aad}} \emph {et~al.} (\bibinfo {collaboration} {ATLAS}),\ }\href {\doibase 10.1007/JHEP07(2023)155} {\bibfield  {journal} {\bibinfo  {journal} {JHEP}\ }\textbf {\bibinfo {volume} {07}},\ \bibinfo {pages} {155} (\bibinfo {year} {2023}{\natexlab{a}})},\ \Eprint {http://arxiv.org/abs/2211.04172} {arXiv:2211.04172 [hep-ex]} \BibitemShut {NoStop}%
\bibitem [{CMS(2023)}]{CMS-PAS-HIG-20-002}%
  \BibitemOpen
  \href {https://cds.cern.ch/record/2852907} {\emph {\bibinfo {title} {{Search for a standard model-like Higgs boson in the mass range between 70 and 110$~\mathrm{GeV}$ in the diphoton final state in proton-proton collisions at $\sqrt{s}=13~\mathrm{TeV}$}}}},\ \bibinfo {type} {Tech. Rep.}\ \bibinfo {number} {CMS-PAS-HIG-20-002}\ (\bibinfo  {institution} {CERN},\ \bibinfo {address} {Geneva},\ \bibinfo {year} {2023})\BibitemShut {NoStop}%
\bibitem [{\citenamefont {Aad}\ \emph {et~al.}(2022)\citenamefont {Aad} \emph {et~al.}}]{ATLAS:2021hbr}%
  \BibitemOpen
  \bibfield  {author} {\bibinfo {author} {\bibfnamefont {G.}~\bibnamefont {Aad}} \emph {et~al.} (\bibinfo {collaboration} {ATLAS}),\ }\href {\doibase 10.1103/PhysRevD.105.012006} {\bibfield  {journal} {\bibinfo  {journal} {Phys. Rev. D}\ }\textbf {\bibinfo {volume} {105}},\ \bibinfo {pages} {012006} (\bibinfo {year} {2022})},\ \Eprint {http://arxiv.org/abs/2110.00313} {arXiv:2110.00313 [hep-ex]} \BibitemShut {NoStop}%
\bibitem [{\citenamefont {Sirunyan}\ \emph {et~al.}(2019)\citenamefont {Sirunyan} \emph {et~al.}}]{CMS:2018nsh}%
  \BibitemOpen
  \bibfield  {author} {\bibinfo {author} {\bibfnamefont {A.~M.}\ \bibnamefont {Sirunyan}} \emph {et~al.} (\bibinfo {collaboration} {CMS}),\ }\href {\doibase 10.1016/j.physletb.2019.06.021} {\bibfield  {journal} {\bibinfo  {journal} {Phys. Lett. B}\ }\textbf {\bibinfo {volume} {795}},\ \bibinfo {pages} {398} (\bibinfo {year} {2019})},\ \Eprint {http://arxiv.org/abs/1812.06359} {arXiv:1812.06359 [hep-ex]} \BibitemShut {NoStop}%
\bibitem [{\citenamefont {Aad}\ \emph {et~al.}(2024{\natexlab{a}})\citenamefont {Aad} \emph {et~al.}}]{ATLAS:2024itc}%
  \BibitemOpen
  \bibfield  {author} {\bibinfo {author} {\bibfnamefont {G.}~\bibnamefont {Aad}} \emph {et~al.} (\bibinfo {collaboration} {ATLAS}),\ }\href@noop {} {\  (\bibinfo {year} {2024}{\natexlab{a}})},\ \Eprint {http://arxiv.org/abs/2405.04914} {arXiv:2405.04914 [hep-ex]} \BibitemShut {NoStop}%
\bibitem [{\citenamefont {Schmieden}(2021)}]{Schmieden:2021pvm}%
  \BibitemOpen
  \bibfield  {author} {\bibinfo {author} {\bibfnamefont {K.}~\bibnamefont {Schmieden}} (\bibinfo {collaboration} {ATLAS, CMS, TOTEM}),\ }\href {\doibase 10.22323/1.397.0005} {\bibfield  {journal} {\bibinfo  {journal} {PoS}\ }\textbf {\bibinfo {volume} {LHCP2021}},\ \bibinfo {pages} {005} (\bibinfo {year} {2021})}\BibitemShut {NoStop}%
\bibitem [{\citenamefont {Kim}\ \emph {et~al.}(2005)\citenamefont {Kim}, \citenamefont {Nilles},\ and\ \citenamefont {Peloso}}]{Kim:2004rp}%
  \BibitemOpen
  \bibfield  {author} {\bibinfo {author} {\bibfnamefont {J.~E.}\ \bibnamefont {Kim}}, \bibinfo {author} {\bibfnamefont {H.~P.}\ \bibnamefont {Nilles}}, \ and\ \bibinfo {author} {\bibfnamefont {M.}~\bibnamefont {Peloso}},\ }\href {\doibase 10.1088/1475-7516/2005/01/005} {\bibfield  {journal} {\bibinfo  {journal} {JCAP}\ }\textbf {\bibinfo {volume} {01}},\ \bibinfo {pages} {005} (\bibinfo {year} {2005})},\ \Eprint {http://arxiv.org/abs/hep-ph/0409138} {arXiv:hep-ph/0409138} \BibitemShut {NoStop}%
\bibitem [{\citenamefont {Choi}\ \emph {et~al.}(2014)\citenamefont {Choi}, \citenamefont {Kim},\ and\ \citenamefont {Yun}}]{Choi:2014rja}%
  \BibitemOpen
  \bibfield  {author} {\bibinfo {author} {\bibfnamefont {K.}~\bibnamefont {Choi}}, \bibinfo {author} {\bibfnamefont {H.}~\bibnamefont {Kim}}, \ and\ \bibinfo {author} {\bibfnamefont {S.}~\bibnamefont {Yun}},\ }\href {\doibase 10.1103/PhysRevD.90.023545} {\bibfield  {journal} {\bibinfo  {journal} {Phys. Rev. D}\ }\textbf {\bibinfo {volume} {90}},\ \bibinfo {pages} {023545} (\bibinfo {year} {2014})},\ \Eprint {http://arxiv.org/abs/1404.6209} {arXiv:1404.6209 [hep-th]} \BibitemShut {NoStop}%
\bibitem [{\citenamefont {Higaki}\ \emph {et~al.}(2016{\natexlab{a}})\citenamefont {Higaki}, \citenamefont {Jeong}, \citenamefont {Kitajima},\ and\ \citenamefont {Takahashi}}]{Higaki:2015jag}%
  \BibitemOpen
  \bibfield  {author} {\bibinfo {author} {\bibfnamefont {T.}~\bibnamefont {Higaki}}, \bibinfo {author} {\bibfnamefont {K.~S.}\ \bibnamefont {Jeong}}, \bibinfo {author} {\bibfnamefont {N.}~\bibnamefont {Kitajima}}, \ and\ \bibinfo {author} {\bibfnamefont {F.}~\bibnamefont {Takahashi}},\ }\href {\doibase 10.1016/j.physletb.2016.01.055} {\bibfield  {journal} {\bibinfo  {journal} {Phys. Lett. B}\ }\textbf {\bibinfo {volume} {755}},\ \bibinfo {pages} {13} (\bibinfo {year} {2016}{\natexlab{a}})},\ \Eprint {http://arxiv.org/abs/1512.05295} {arXiv:1512.05295 [hep-ph]} \BibitemShut {NoStop}%
\bibitem [{\citenamefont {Farina}\ \emph {et~al.}(2017)\citenamefont {Farina}, \citenamefont {Pappadopulo}, \citenamefont {Rompineve},\ and\ \citenamefont {Tesi}}]{Farina:2016tgd}%
  \BibitemOpen
  \bibfield  {author} {\bibinfo {author} {\bibfnamefont {M.}~\bibnamefont {Farina}}, \bibinfo {author} {\bibfnamefont {D.}~\bibnamefont {Pappadopulo}}, \bibinfo {author} {\bibfnamefont {F.}~\bibnamefont {Rompineve}}, \ and\ \bibinfo {author} {\bibfnamefont {A.}~\bibnamefont {Tesi}},\ }\href {\doibase 10.1007/JHEP01(2017)095} {\bibfield  {journal} {\bibinfo  {journal} {JHEP}\ }\textbf {\bibinfo {volume} {01}},\ \bibinfo {pages} {095} (\bibinfo {year} {2017})},\ \Eprint {http://arxiv.org/abs/1611.09855} {arXiv:1611.09855 [hep-ph]} \BibitemShut {NoStop}%
\bibitem [{\citenamefont {Choi}\ and\ \citenamefont {Im}(2016)}]{Choi:2015fiu}%
  \BibitemOpen
  \bibfield  {author} {\bibinfo {author} {\bibfnamefont {K.}~\bibnamefont {Choi}}\ and\ \bibinfo {author} {\bibfnamefont {S.~H.}\ \bibnamefont {Im}},\ }\href {\doibase 10.1007/JHEP01(2016)149} {\bibfield  {journal} {\bibinfo  {journal} {JHEP}\ }\textbf {\bibinfo {volume} {01}},\ \bibinfo {pages} {149} (\bibinfo {year} {2016})},\ \Eprint {http://arxiv.org/abs/1511.00132} {arXiv:1511.00132 [hep-ph]} \BibitemShut {NoStop}%
\bibitem [{\citenamefont {Kaplan}\ and\ \citenamefont {Rattazzi}(2016)}]{Kaplan:2015fuy}%
  \BibitemOpen
  \bibfield  {author} {\bibinfo {author} {\bibfnamefont {D.~E.}\ \bibnamefont {Kaplan}}\ and\ \bibinfo {author} {\bibfnamefont {R.}~\bibnamefont {Rattazzi}},\ }\href {\doibase 10.1103/PhysRevD.93.085007} {\bibfield  {journal} {\bibinfo  {journal} {Phys. Rev. D}\ }\textbf {\bibinfo {volume} {93}},\ \bibinfo {pages} {085007} (\bibinfo {year} {2016})},\ \Eprint {http://arxiv.org/abs/1511.01827} {arXiv:1511.01827 [hep-ph]} \BibitemShut {NoStop}%
\bibitem [{\citenamefont {Giudice}\ and\ \citenamefont {McCullough}(2017)}]{Giudice:2016yja}%
  \BibitemOpen
  \bibfield  {author} {\bibinfo {author} {\bibfnamefont {G.~F.}\ \bibnamefont {Giudice}}\ and\ \bibinfo {author} {\bibfnamefont {M.}~\bibnamefont {McCullough}},\ }\href {\doibase 10.1007/JHEP02(2017)036} {\bibfield  {journal} {\bibinfo  {journal} {JHEP}\ }\textbf {\bibinfo {volume} {02}},\ \bibinfo {pages} {036} (\bibinfo {year} {2017})},\ \Eprint {http://arxiv.org/abs/1610.07962} {arXiv:1610.07962 [hep-ph]} \BibitemShut {NoStop}%
\bibitem [{\citenamefont {Kim}(1979)}]{Kim:1979if}%
  \BibitemOpen
  \bibfield  {author} {\bibinfo {author} {\bibfnamefont {J.~E.}\ \bibnamefont {Kim}},\ }\href {\doibase 10.1103/PhysRevLett.43.103} {\bibfield  {journal} {\bibinfo  {journal} {Phys. Rev. Lett.}\ }\textbf {\bibinfo {volume} {43}},\ \bibinfo {pages} {103} (\bibinfo {year} {1979})}\BibitemShut {NoStop}%
\bibitem [{\citenamefont {Shifman}\ \emph {et~al.}(1980)\citenamefont {Shifman}, \citenamefont {Vainshtein},\ and\ \citenamefont {Zakharov}}]{Shifman:1979if}%
  \BibitemOpen
  \bibfield  {author} {\bibinfo {author} {\bibfnamefont {M.~A.}\ \bibnamefont {Shifman}}, \bibinfo {author} {\bibfnamefont {A.~I.}\ \bibnamefont {Vainshtein}}, \ and\ \bibinfo {author} {\bibfnamefont {V.~I.}\ \bibnamefont {Zakharov}},\ }\href {\doibase 10.1016/0550-3213(80)90209-6} {\bibfield  {journal} {\bibinfo  {journal} {Nucl. Phys. B}\ }\textbf {\bibinfo {volume} {166}},\ \bibinfo {pages} {493} (\bibinfo {year} {1980})}\BibitemShut {NoStop}%
\bibitem [{\citenamefont {Zhitnitsky}(1980)}]{Zhitnitsky:1980tq}%
  \BibitemOpen
  \bibfield  {author} {\bibinfo {author} {\bibfnamefont {A.~R.}\ \bibnamefont {Zhitnitsky}},\ }\href@noop {} {\bibfield  {journal} {\bibinfo  {journal} {Sov. J. Nucl. Phys.}\ }\textbf {\bibinfo {volume} {31}},\ \bibinfo {pages} {260} (\bibinfo {year} {1980})}\BibitemShut {NoStop}%
\bibitem [{\citenamefont {Dine}\ \emph {et~al.}(1981)\citenamefont {Dine}, \citenamefont {Fischler},\ and\ \citenamefont {Srednicki}}]{Dine:1981rt}%
  \BibitemOpen
  \bibfield  {author} {\bibinfo {author} {\bibfnamefont {M.}~\bibnamefont {Dine}}, \bibinfo {author} {\bibfnamefont {W.}~\bibnamefont {Fischler}}, \ and\ \bibinfo {author} {\bibfnamefont {M.}~\bibnamefont {Srednicki}},\ }\href {\doibase 10.1016/0370-2693(81)90590-6} {\bibfield  {journal} {\bibinfo  {journal} {Phys. Lett. B}\ }\textbf {\bibinfo {volume} {104}},\ \bibinfo {pages} {199} (\bibinfo {year} {1981})}\BibitemShut {NoStop}%
\bibitem [{\citenamefont {Giudice}\ \emph {et~al.}(2018)\citenamefont {Giudice}, \citenamefont {Kats}, \citenamefont {McCullough}, \citenamefont {Torre},\ and\ \citenamefont {Urbano}}]{Giudice:2017fmj}%
  \BibitemOpen
  \bibfield  {author} {\bibinfo {author} {\bibfnamefont {G.~F.}\ \bibnamefont {Giudice}}, \bibinfo {author} {\bibfnamefont {Y.}~\bibnamefont {Kats}}, \bibinfo {author} {\bibfnamefont {M.}~\bibnamefont {McCullough}}, \bibinfo {author} {\bibfnamefont {R.}~\bibnamefont {Torre}}, \ and\ \bibinfo {author} {\bibfnamefont {A.}~\bibnamefont {Urbano}},\ }\href {\doibase 10.1007/JHEP06(2018)009} {\bibfield  {journal} {\bibinfo  {journal} {JHEP}\ }\textbf {\bibinfo {volume} {06}},\ \bibinfo {pages} {009} (\bibinfo {year} {2018})},\ \Eprint {http://arxiv.org/abs/1711.08437} {arXiv:1711.08437 [hep-ph]} \BibitemShut {NoStop}%
\bibitem [{\citenamefont {Beauchesne}\ and\ \citenamefont {Kats}(2020)}]{Beauchesne:2019tpx}%
  \BibitemOpen
  \bibfield  {author} {\bibinfo {author} {\bibfnamefont {H.}~\bibnamefont {Beauchesne}}\ and\ \bibinfo {author} {\bibfnamefont {Y.}~\bibnamefont {Kats}},\ }\href {\doibase 10.1140/epjc/s10052-020-7746-8} {\bibfield  {journal} {\bibinfo  {journal} {Eur. Phys. J. C}\ }\textbf {\bibinfo {volume} {80}},\ \bibinfo {pages} {192} (\bibinfo {year} {2020})},\ \Eprint {http://arxiv.org/abs/1907.03676} {arXiv:1907.03676 [hep-ph]} \BibitemShut {NoStop}%
\bibitem [{\citenamefont {Aad}\ \emph {et~al.}(2023{\natexlab{b}})\citenamefont {Aad} \emph {et~al.}}]{ATLAS:2023hbp}%
  \BibitemOpen
  \bibfield  {author} {\bibinfo {author} {\bibfnamefont {G.}~\bibnamefont {Aad}} \emph {et~al.} (\bibinfo {collaboration} {ATLAS}),\ }\href {\doibase 10.1007/JHEP10(2023)079} {\bibfield  {journal} {\bibinfo  {journal} {JHEP}\ }\textbf {\bibinfo {volume} {10}},\ \bibinfo {pages} {079} (\bibinfo {year} {2023}{\natexlab{b}})},\ \Eprint {http://arxiv.org/abs/2305.10894} {arXiv:2305.10894 [hep-ex]} \BibitemShut {NoStop}%
\bibitem [{\citenamefont {Aad}\ \emph {et~al.}(2023{\natexlab{c}})\citenamefont {Aad} \emph {et~al.}}]{ATLAS:2023pja}%
  \BibitemOpen
  \bibfield  {author} {\bibinfo {author} {\bibfnamefont {G.}~\bibnamefont {Aad}} \emph {et~al.} (\bibinfo {collaboration} {ATLAS}),\ }\href {\doibase 10.1007/JHEP08(2023)153} {\bibfield  {journal} {\bibinfo  {journal} {JHEP}\ }\textbf {\bibinfo {volume} {08}},\ \bibinfo {pages} {153} (\bibinfo {year} {2023}{\natexlab{c}})},\ \Eprint {http://arxiv.org/abs/2305.03401} {arXiv:2305.03401 [hep-ex]} \BibitemShut {NoStop}%
\bibitem [{\citenamefont {Kamionkowski}\ and\ \citenamefont {March-Russell}(1992)}]{Kamionkowski:1992mf}%
  \BibitemOpen
  \bibfield  {author} {\bibinfo {author} {\bibfnamefont {M.}~\bibnamefont {Kamionkowski}}\ and\ \bibinfo {author} {\bibfnamefont {J.}~\bibnamefont {March-Russell}},\ }\href {\doibase 10.1016/0370-2693(92)90492-M} {\bibfield  {journal} {\bibinfo  {journal} {Phys. Lett. B}\ }\textbf {\bibinfo {volume} {282}},\ \bibinfo {pages} {137} (\bibinfo {year} {1992})},\ \Eprint {http://arxiv.org/abs/hep-th/9202003} {arXiv:hep-th/9202003} \BibitemShut {NoStop}%
\bibitem [{\citenamefont {Randall}(1992)}]{Randall:1992ut}%
  \BibitemOpen
  \bibfield  {author} {\bibinfo {author} {\bibfnamefont {L.}~\bibnamefont {Randall}},\ }\href {\doibase 10.1016/0370-2693(92)91928-3} {\bibfield  {journal} {\bibinfo  {journal} {Phys. Lett. B}\ }\textbf {\bibinfo {volume} {284}},\ \bibinfo {pages} {77} (\bibinfo {year} {1992})}\BibitemShut {NoStop}%
\bibitem [{\citenamefont {Coy}\ \emph {et~al.}(2017)\citenamefont {Coy}, \citenamefont {Frigerio},\ and\ \citenamefont {Ibe}}]{Coy:2017yex}%
  \BibitemOpen
  \bibfield  {author} {\bibinfo {author} {\bibfnamefont {R.}~\bibnamefont {Coy}}, \bibinfo {author} {\bibfnamefont {M.}~\bibnamefont {Frigerio}}, \ and\ \bibinfo {author} {\bibfnamefont {M.}~\bibnamefont {Ibe}},\ }\href {\doibase 10.1007/JHEP10(2017)002} {\bibfield  {journal} {\bibinfo  {journal} {JHEP}\ }\textbf {\bibinfo {volume} {10}},\ \bibinfo {pages} {002} (\bibinfo {year} {2017})},\ \Eprint {http://arxiv.org/abs/1706.04529} {arXiv:1706.04529 [hep-ph]} \BibitemShut {NoStop}%
\bibitem [{\citenamefont {Ben-Dayan}(2019)}]{Ben-Dayan:2017rvr}%
  \BibitemOpen
  \bibfield  {author} {\bibinfo {author} {\bibfnamefont {I.}~\bibnamefont {Ben-Dayan}},\ }\href {\doibase 10.1103/PhysRevD.99.096006} {\bibfield  {journal} {\bibinfo  {journal} {Phys. Rev. D}\ }\textbf {\bibinfo {volume} {99}},\ \bibinfo {pages} {096006} (\bibinfo {year} {2019})},\ \Eprint {http://arxiv.org/abs/1706.05308} {arXiv:1706.05308 [hep-ph]} \BibitemShut {NoStop}%
\bibitem [{\citenamefont {Alonso-\'Alvarez}\ \emph {et~al.}(2019)\citenamefont {Alonso-\'Alvarez}, \citenamefont {Gavela},\ and\ \citenamefont {Quilez}}]{Alonso-Alvarez:2018irt}%
  \BibitemOpen
  \bibfield  {author} {\bibinfo {author} {\bibfnamefont {G.}~\bibnamefont {Alonso-\'Alvarez}}, \bibinfo {author} {\bibfnamefont {M.~B.}\ \bibnamefont {Gavela}}, \ and\ \bibinfo {author} {\bibfnamefont {P.}~\bibnamefont {Quilez}},\ }\href {\doibase 10.1140/epjc/s10052-019-6732-5} {\bibfield  {journal} {\bibinfo  {journal} {Eur. Phys. J. C}\ }\textbf {\bibinfo {volume} {79}},\ \bibinfo {pages} {223} (\bibinfo {year} {2019})},\ \Eprint {http://arxiv.org/abs/1811.05466} {arXiv:1811.05466 [hep-ph]} \BibitemShut {NoStop}%
\bibitem [{\citenamefont {Banerjee}\ \emph {et~al.}(2018)\citenamefont {Banerjee}, \citenamefont {Ghosh},\ and\ \citenamefont {Ray}}]{Banerjee:2018grm}%
  \BibitemOpen
  \bibfield  {author} {\bibinfo {author} {\bibfnamefont {A.}~\bibnamefont {Banerjee}}, \bibinfo {author} {\bibfnamefont {S.}~\bibnamefont {Ghosh}}, \ and\ \bibinfo {author} {\bibfnamefont {T.~S.}\ \bibnamefont {Ray}},\ }\href {\doibase 10.1007/JHEP11(2018)075} {\bibfield  {journal} {\bibinfo  {journal} {JHEP}\ }\textbf {\bibinfo {volume} {11}},\ \bibinfo {pages} {075} (\bibinfo {year} {2018})},\ \Eprint {http://arxiv.org/abs/1808.04010} {arXiv:1808.04010 [hep-ph]} \BibitemShut {NoStop}%
\bibitem [{\citenamefont {{Di Vecchia}}\ and\ \citenamefont {Veneziano}(1980)}]{DIVECCHIA1980253}%
  \BibitemOpen
  \bibfield  {author} {\bibinfo {author} {\bibfnamefont {P.}~\bibnamefont {{Di Vecchia}}}\ and\ \bibinfo {author} {\bibfnamefont {G.}~\bibnamefont {Veneziano}},\ }\href {\doibase https://doi.org/10.1016/0550-3213(80)90370-3} {\bibfield  {journal} {\bibinfo  {journal} {Nuclear Physics B}\ }\textbf {\bibinfo {volume} {171}},\ \bibinfo {pages} {253} (\bibinfo {year} {1980})}\BibitemShut {NoStop}%
\bibitem [{\citenamefont {Grilli~di Cortona}\ \emph {et~al.}(2016)\citenamefont {Grilli~di Cortona}, \citenamefont {Hardy}, \citenamefont {Pardo~Vega},\ and\ \citenamefont {Villadoro}}]{GrillidiCortona:2015jxo}%
  \BibitemOpen
  \bibfield  {author} {\bibinfo {author} {\bibfnamefont {G.}~\bibnamefont {Grilli~di Cortona}}, \bibinfo {author} {\bibfnamefont {E.}~\bibnamefont {Hardy}}, \bibinfo {author} {\bibfnamefont {J.}~\bibnamefont {Pardo~Vega}}, \ and\ \bibinfo {author} {\bibfnamefont {G.}~\bibnamefont {Villadoro}},\ }\href {\doibase 10.1007/JHEP01(2016)034} {\bibfield  {journal} {\bibinfo  {journal} {JHEP}\ }\textbf {\bibinfo {volume} {01}},\ \bibinfo {pages} {034} (\bibinfo {year} {2016})},\ \Eprint {http://arxiv.org/abs/1511.02867} {arXiv:1511.02867 [hep-ph]} \BibitemShut {NoStop}%
\bibitem [{\citenamefont {Workman}\ \emph {et~al.}(2022)\citenamefont {Workman} \emph {et~al.}}]{ParticleDataGroup:2022pth}%
  \BibitemOpen
  \bibfield  {author} {\bibinfo {author} {\bibfnamefont {R.~L.}\ \bibnamefont {Workman}} \emph {et~al.} (\bibinfo {collaboration} {Particle Data Group}),\ }\href {\doibase 10.1093/ptep/ptac097} {\bibfield  {journal} {\bibinfo  {journal} {PTEP}\ }\textbf {\bibinfo {volume} {2022}},\ \bibinfo {pages} {083C01} (\bibinfo {year} {2022})}\BibitemShut {NoStop}%
\bibitem [{\citenamefont {Schulthess}\ \emph {et~al.}(2022)\citenamefont {Schulthess} \emph {et~al.}}]{Schulthess:2022pbp}%
  \BibitemOpen
  \bibfield  {author} {\bibinfo {author} {\bibfnamefont {I.}~\bibnamefont {Schulthess}} \emph {et~al.},\ }\href {\doibase 10.1103/PhysRevLett.129.191801} {\bibfield  {journal} {\bibinfo  {journal} {Phys. Rev. Lett.}\ }\textbf {\bibinfo {volume} {129}},\ \bibinfo {pages} {191801} (\bibinfo {year} {2022})},\ \Eprint {http://arxiv.org/abs/2204.01454} {arXiv:2204.01454 [hep-ex]} \BibitemShut {NoStop}%
\bibitem [{\citenamefont {Abel}\ \emph {et~al.}(2017)\citenamefont {Abel} \emph {et~al.}}]{Abel:2017rtm}%
  \BibitemOpen
  \bibfield  {author} {\bibinfo {author} {\bibfnamefont {C.}~\bibnamefont {Abel}} \emph {et~al.},\ }\href {\doibase 10.1103/PhysRevX.7.041034} {\bibfield  {journal} {\bibinfo  {journal} {Phys. Rev. X}\ }\textbf {\bibinfo {volume} {7}},\ \bibinfo {pages} {041034} (\bibinfo {year} {2017})},\ \Eprint {http://arxiv.org/abs/1708.06367} {arXiv:1708.06367 [hep-ph]} \BibitemShut {NoStop}%
\bibitem [{\citenamefont {Roussy}\ \emph {et~al.}(2021)\citenamefont {Roussy} \emph {et~al.}}]{Roussy:2020ily}%
  \BibitemOpen
  \bibfield  {author} {\bibinfo {author} {\bibfnamefont {T.~S.}\ \bibnamefont {Roussy}} \emph {et~al.},\ }\href {\doibase 10.1103/PhysRevLett.126.171301} {\bibfield  {journal} {\bibinfo  {journal} {Phys. Rev. Lett.}\ }\textbf {\bibinfo {volume} {126}},\ \bibinfo {pages} {171301} (\bibinfo {year} {2021})},\ \Eprint {http://arxiv.org/abs/2006.15787} {arXiv:2006.15787 [hep-ph]} \BibitemShut {NoStop}%
\bibitem [{\citenamefont {Madge}\ \emph {et~al.}(2024)\citenamefont {Madge}, \citenamefont {Perez},\ and\ \citenamefont {Meir}}]{Madge:2024aot}%
  \BibitemOpen
  \bibfield  {author} {\bibinfo {author} {\bibfnamefont {E.}~\bibnamefont {Madge}}, \bibinfo {author} {\bibfnamefont {G.}~\bibnamefont {Perez}}, \ and\ \bibinfo {author} {\bibfnamefont {Z.}~\bibnamefont {Meir}},\ }\href@noop {} {\  (\bibinfo {year} {2024})},\ \Eprint {http://arxiv.org/abs/2404.00616} {arXiv:2404.00616 [physics.atom-ph]} \BibitemShut {NoStop}%
\bibitem [{\citenamefont {Zhang}\ \emph {et~al.}(2023)\citenamefont {Zhang}, \citenamefont {Banerjee}, \citenamefont {Leyser}, \citenamefont {Perez}, \citenamefont {Schiller}, \citenamefont {Budker},\ and\ \citenamefont {Antypas}}]{Zhang:2022ewz}%
  \BibitemOpen
  \bibfield  {author} {\bibinfo {author} {\bibfnamefont {X.}~\bibnamefont {Zhang}}, \bibinfo {author} {\bibfnamefont {A.}~\bibnamefont {Banerjee}}, \bibinfo {author} {\bibfnamefont {M.}~\bibnamefont {Leyser}}, \bibinfo {author} {\bibfnamefont {G.}~\bibnamefont {Perez}}, \bibinfo {author} {\bibfnamefont {S.}~\bibnamefont {Schiller}}, \bibinfo {author} {\bibfnamefont {D.}~\bibnamefont {Budker}}, \ and\ \bibinfo {author} {\bibfnamefont {D.}~\bibnamefont {Antypas}},\ }\href {\doibase 10.1103/PhysRevLett.130.251002} {\bibfield  {journal} {\bibinfo  {journal} {Phys. Rev. Lett.}\ }\textbf {\bibinfo {volume} {130}},\ \bibinfo {pages} {251002} (\bibinfo {year} {2023})},\ \Eprint {http://arxiv.org/abs/2212.04413} {arXiv:2212.04413 [physics.atom-ph]} \BibitemShut {NoStop}%
\bibitem [{\citenamefont {Fox}\ \emph {et~al.}(2023)\citenamefont {Fox}, \citenamefont {Weiner},\ and\ \citenamefont {Xiao}}]{Fox:2023xgx}%
  \BibitemOpen
  \bibfield  {author} {\bibinfo {author} {\bibfnamefont {P.~J.}\ \bibnamefont {Fox}}, \bibinfo {author} {\bibfnamefont {N.}~\bibnamefont {Weiner}}, \ and\ \bibinfo {author} {\bibfnamefont {H.}~\bibnamefont {Xiao}},\ }\href {\doibase 10.1103/PhysRevD.108.095043} {\bibfield  {journal} {\bibinfo  {journal} {Phys. Rev. D}\ }\textbf {\bibinfo {volume} {108}},\ \bibinfo {pages} {095043} (\bibinfo {year} {2023})},\ \Eprint {http://arxiv.org/abs/2302.00685} {arXiv:2302.00685 [hep-ph]} \BibitemShut {NoStop}%
\bibitem [{\citenamefont {Blum}\ \emph {et~al.}(2014)\citenamefont {Blum}, \citenamefont {D'Agnolo}, \citenamefont {Lisanti},\ and\ \citenamefont {Safdi}}]{Blum:2014vsa}%
  \BibitemOpen
  \bibfield  {author} {\bibinfo {author} {\bibfnamefont {K.}~\bibnamefont {Blum}}, \bibinfo {author} {\bibfnamefont {R.~T.}\ \bibnamefont {D'Agnolo}}, \bibinfo {author} {\bibfnamefont {M.}~\bibnamefont {Lisanti}}, \ and\ \bibinfo {author} {\bibfnamefont {B.~R.}\ \bibnamefont {Safdi}},\ }\href {\doibase 10.1016/j.physletb.2014.07.059} {\bibfield  {journal} {\bibinfo  {journal} {Phys. Lett. B}\ }\textbf {\bibinfo {volume} {737}},\ \bibinfo {pages} {30} (\bibinfo {year} {2014})},\ \Eprint {http://arxiv.org/abs/1401.6460} {arXiv:1401.6460 [hep-ph]} \BibitemShut {NoStop}%
\bibitem [{\citenamefont {Mehta}\ \emph {et~al.}(2020)\citenamefont {Mehta}, \citenamefont {Demirtas}, \citenamefont {Long}, \citenamefont {Marsh}, \citenamefont {Mcallister},\ and\ \citenamefont {Stott}}]{Mehta:2020kwu}%
  \BibitemOpen
  \bibfield  {author} {\bibinfo {author} {\bibfnamefont {V.~M.}\ \bibnamefont {Mehta}}, \bibinfo {author} {\bibfnamefont {M.}~\bibnamefont {Demirtas}}, \bibinfo {author} {\bibfnamefont {C.}~\bibnamefont {Long}}, \bibinfo {author} {\bibfnamefont {D.~J.~E.}\ \bibnamefont {Marsh}}, \bibinfo {author} {\bibfnamefont {L.}~\bibnamefont {Mcallister}}, \ and\ \bibinfo {author} {\bibfnamefont {M.~J.}\ \bibnamefont {Stott}},\ }\href@noop {} {\  (\bibinfo {year} {2020})},\ \Eprint {http://arxiv.org/abs/2011.08693} {arXiv:2011.08693 [hep-th]} \BibitemShut {NoStop}%
\bibitem [{\citenamefont {Baryakhtar}\ \emph {et~al.}(2021)\citenamefont {Baryakhtar}, \citenamefont {Galanis}, \citenamefont {Lasenby},\ and\ \citenamefont {Simon}}]{Baryakhtar:2020gao}%
  \BibitemOpen
  \bibfield  {author} {\bibinfo {author} {\bibfnamefont {M.}~\bibnamefont {Baryakhtar}}, \bibinfo {author} {\bibfnamefont {M.}~\bibnamefont {Galanis}}, \bibinfo {author} {\bibfnamefont {R.}~\bibnamefont {Lasenby}}, \ and\ \bibinfo {author} {\bibfnamefont {O.}~\bibnamefont {Simon}},\ }\href {\doibase 10.1103/PhysRevD.103.095019} {\bibfield  {journal} {\bibinfo  {journal} {Phys. Rev. D}\ }\textbf {\bibinfo {volume} {103}},\ \bibinfo {pages} {095019} (\bibinfo {year} {2021})},\ \Eprint {http://arxiv.org/abs/2011.11646} {arXiv:2011.11646 [hep-ph]} \BibitemShut {NoStop}%
\bibitem [{\citenamefont {\"Unal}\ \emph {et~al.}(2021)\citenamefont {\"Unal}, \citenamefont {Pacucci},\ and\ \citenamefont {Loeb}}]{Unal:2020jiy}%
  \BibitemOpen
  \bibfield  {author} {\bibinfo {author} {\bibfnamefont {C.}~\bibnamefont {\"Unal}}, \bibinfo {author} {\bibfnamefont {F.}~\bibnamefont {Pacucci}}, \ and\ \bibinfo {author} {\bibfnamefont {A.}~\bibnamefont {Loeb}},\ }\href {\doibase 10.1088/1475-7516/2021/05/007} {\bibfield  {journal} {\bibinfo  {journal} {JCAP}\ }\textbf {\bibinfo {volume} {05}},\ \bibinfo {pages} {007} (\bibinfo {year} {2021})},\ \Eprint {http://arxiv.org/abs/2012.12790} {arXiv:2012.12790 [hep-ph]} \BibitemShut {NoStop}%
\bibitem [{\citenamefont {Hook}\ and\ \citenamefont {Huang}(2018)}]{Hook:2017psm}%
  \BibitemOpen
  \bibfield  {author} {\bibinfo {author} {\bibfnamefont {A.}~\bibnamefont {Hook}}\ and\ \bibinfo {author} {\bibfnamefont {J.}~\bibnamefont {Huang}},\ }\href {\doibase 10.1007/JHEP06(2018)036} {\bibfield  {journal} {\bibinfo  {journal} {JHEP}\ }\textbf {\bibinfo {volume} {06}},\ \bibinfo {pages} {036} (\bibinfo {year} {2018})},\ \Eprint {http://arxiv.org/abs/1708.08464} {arXiv:1708.08464 [hep-ph]} \BibitemShut {NoStop}%
\bibitem [{\citenamefont {Zhang}\ \emph {et~al.}(2021)\citenamefont {Zhang}, \citenamefont {Lyu}, \citenamefont {Huang}, \citenamefont {Johnson}, \citenamefont {Sagunski}, \citenamefont {Sakellariadou},\ and\ \citenamefont {Yang}}]{Zhang:2021mks}%
  \BibitemOpen
  \bibfield  {author} {\bibinfo {author} {\bibfnamefont {J.}~\bibnamefont {Zhang}}, \bibinfo {author} {\bibfnamefont {Z.}~\bibnamefont {Lyu}}, \bibinfo {author} {\bibfnamefont {J.}~\bibnamefont {Huang}}, \bibinfo {author} {\bibfnamefont {M.~C.}\ \bibnamefont {Johnson}}, \bibinfo {author} {\bibfnamefont {L.}~\bibnamefont {Sagunski}}, \bibinfo {author} {\bibfnamefont {M.}~\bibnamefont {Sakellariadou}}, \ and\ \bibinfo {author} {\bibfnamefont {H.}~\bibnamefont {Yang}},\ }\href {\doibase 10.1103/PhysRevLett.127.161101} {\bibfield  {journal} {\bibinfo  {journal} {Phys. Rev. Lett.}\ }\textbf {\bibinfo {volume} {127}},\ \bibinfo {pages} {161101} (\bibinfo {year} {2021})},\ \Eprint {http://arxiv.org/abs/2105.13963} {arXiv:2105.13963 [hep-ph]} \BibitemShut {NoStop}%
\bibitem [{\citenamefont {Caloni}\ \emph {et~al.}(2022)\citenamefont {Caloni}, \citenamefont {Gerbino}, \citenamefont {Lattanzi},\ and\ \citenamefont {Visinelli}}]{Caloni:2022uya}%
  \BibitemOpen
  \bibfield  {author} {\bibinfo {author} {\bibfnamefont {L.}~\bibnamefont {Caloni}}, \bibinfo {author} {\bibfnamefont {M.}~\bibnamefont {Gerbino}}, \bibinfo {author} {\bibfnamefont {M.}~\bibnamefont {Lattanzi}}, \ and\ \bibinfo {author} {\bibfnamefont {L.}~\bibnamefont {Visinelli}},\ }\href {\doibase 10.1088/1475-7516/2022/09/021} {\bibfield  {journal} {\bibinfo  {journal} {JCAP}\ }\textbf {\bibinfo {volume} {09}},\ \bibinfo {pages} {021} (\bibinfo {year} {2022})},\ \Eprint {http://arxiv.org/abs/2205.01637} {arXiv:2205.01637 [astro-ph.CO]} \BibitemShut {NoStop}%
\bibitem [{\citenamefont {Lucente}\ \emph {et~al.}(2022)\citenamefont {Lucente}, \citenamefont {Mastrototaro}, \citenamefont {Carenza}, \citenamefont {Di~Luzio}, \citenamefont {Giannotti},\ and\ \citenamefont {Mirizzi}}]{Lucente:2022vuo}%
  \BibitemOpen
  \bibfield  {author} {\bibinfo {author} {\bibfnamefont {G.}~\bibnamefont {Lucente}}, \bibinfo {author} {\bibfnamefont {L.}~\bibnamefont {Mastrototaro}}, \bibinfo {author} {\bibfnamefont {P.}~\bibnamefont {Carenza}}, \bibinfo {author} {\bibfnamefont {L.}~\bibnamefont {Di~Luzio}}, \bibinfo {author} {\bibfnamefont {M.}~\bibnamefont {Giannotti}}, \ and\ \bibinfo {author} {\bibfnamefont {A.}~\bibnamefont {Mirizzi}},\ }\href {\doibase 10.1103/PhysRevD.105.123020} {\bibfield  {journal} {\bibinfo  {journal} {Phys. Rev. D}\ }\textbf {\bibinfo {volume} {105}},\ \bibinfo {pages} {123020} (\bibinfo {year} {2022})},\ \Eprint {http://arxiv.org/abs/2203.15812} {arXiv:2203.15812 [hep-ph]} \BibitemShut {NoStop}%
\bibitem [{\citenamefont {Springmann}\ \emph {et~al.}(2024)\citenamefont {Springmann}, \citenamefont {Stadlbauer}, \citenamefont {Stelzl},\ and\ \citenamefont {Weiler}}]{Springmann:2024ret}%
  \BibitemOpen
  \bibfield  {author} {\bibinfo {author} {\bibfnamefont {K.}~\bibnamefont {Springmann}}, \bibinfo {author} {\bibfnamefont {M.}~\bibnamefont {Stadlbauer}}, \bibinfo {author} {\bibfnamefont {S.}~\bibnamefont {Stelzl}}, \ and\ \bibinfo {author} {\bibfnamefont {A.}~\bibnamefont {Weiler}},\ }\href@noop {} {\  (\bibinfo {year} {2024})},\ \Eprint {http://arxiv.org/abs/2410.19902} {arXiv:2410.19902 [hep-ph]} \BibitemShut {NoStop}%
\bibitem [{\citenamefont {Balkin}\ \emph {et~al.}(2024)\citenamefont {Balkin}, \citenamefont {Serra}, \citenamefont {Springmann}, \citenamefont {Stelzl},\ and\ \citenamefont {Weiler}}]{Balkin:2022qer}%
  \BibitemOpen
  \bibfield  {author} {\bibinfo {author} {\bibfnamefont {R.}~\bibnamefont {Balkin}}, \bibinfo {author} {\bibfnamefont {J.}~\bibnamefont {Serra}}, \bibinfo {author} {\bibfnamefont {K.}~\bibnamefont {Springmann}}, \bibinfo {author} {\bibfnamefont {S.}~\bibnamefont {Stelzl}}, \ and\ \bibinfo {author} {\bibfnamefont {A.}~\bibnamefont {Weiler}},\ }\href {\doibase 10.1103/PhysRevD.109.095032} {\bibfield  {journal} {\bibinfo  {journal} {Phys. Rev. D}\ }\textbf {\bibinfo {volume} {109}},\ \bibinfo {pages} {095032} (\bibinfo {year} {2024})},\ \Eprint {http://arxiv.org/abs/2211.02661} {arXiv:2211.02661 [hep-ph]} \BibitemShut {NoStop}%
\bibitem [{\citenamefont {G\'omez-Ba\~n\'on}\ \emph {et~al.}(2024)\citenamefont {G\'omez-Ba\~n\'on}, \citenamefont {Bartnick}, \citenamefont {Springmann},\ and\ \citenamefont {Pons}}]{Gomez-Banon:2024oux}%
  \BibitemOpen
  \bibfield  {author} {\bibinfo {author} {\bibfnamefont {A.}~\bibnamefont {G\'omez-Ba\~n\'on}}, \bibinfo {author} {\bibfnamefont {K.}~\bibnamefont {Bartnick}}, \bibinfo {author} {\bibfnamefont {K.}~\bibnamefont {Springmann}}, \ and\ \bibinfo {author} {\bibfnamefont {J.~A.}\ \bibnamefont {Pons}},\ }\href@noop {} {\  (\bibinfo {year} {2024})},\ \Eprint {http://arxiv.org/abs/2408.07740} {arXiv:2408.07740 [hep-ph]} \BibitemShut {NoStop}%
\bibitem [{\citenamefont {Kumamoto}\ \emph {et~al.}(2024)\citenamefont {Kumamoto}, \citenamefont {Huang}, \citenamefont {Drischler}, \citenamefont {Baryakhtar},\ and\ \citenamefont {Reddy}}]{Kumamoto:2024wjd}%
  \BibitemOpen
  \bibfield  {author} {\bibinfo {author} {\bibfnamefont {M.}~\bibnamefont {Kumamoto}}, \bibinfo {author} {\bibfnamefont {J.}~\bibnamefont {Huang}}, \bibinfo {author} {\bibfnamefont {C.}~\bibnamefont {Drischler}}, \bibinfo {author} {\bibfnamefont {M.}~\bibnamefont {Baryakhtar}}, \ and\ \bibinfo {author} {\bibfnamefont {S.}~\bibnamefont {Reddy}},\ }\href@noop {} {\  (\bibinfo {year} {2024})},\ \Eprint {http://arxiv.org/abs/2410.21590} {arXiv:2410.21590 [hep-ph]} \BibitemShut {NoStop}%
\bibitem [{\citenamefont {Long}(2018)}]{Long:2018nsl}%
  \BibitemOpen
  \bibfield  {author} {\bibinfo {author} {\bibfnamefont {A.~J.}\ \bibnamefont {Long}},\ }\href {\doibase 10.1007/JHEP07(2018)066} {\bibfield  {journal} {\bibinfo  {journal} {JHEP}\ }\textbf {\bibinfo {volume} {07}},\ \bibinfo {pages} {066} (\bibinfo {year} {2018})},\ \Eprint {http://arxiv.org/abs/1803.07086} {arXiv:1803.07086 [hep-ph]} \BibitemShut {NoStop}%
\bibitem [{\citenamefont {Higaki}\ \emph {et~al.}(2016{\natexlab{b}})\citenamefont {Higaki}, \citenamefont {Jeong}, \citenamefont {Kitajima}, \citenamefont {Sekiguchi},\ and\ \citenamefont {Takahashi}}]{Higaki:2016jjh}%
  \BibitemOpen
  \bibfield  {author} {\bibinfo {author} {\bibfnamefont {T.}~\bibnamefont {Higaki}}, \bibinfo {author} {\bibfnamefont {K.~S.}\ \bibnamefont {Jeong}}, \bibinfo {author} {\bibfnamefont {N.}~\bibnamefont {Kitajima}}, \bibinfo {author} {\bibfnamefont {T.}~\bibnamefont {Sekiguchi}}, \ and\ \bibinfo {author} {\bibfnamefont {F.}~\bibnamefont {Takahashi}},\ }\href {\doibase 10.1007/JHEP08(2016)044} {\bibfield  {journal} {\bibinfo  {journal} {JHEP}\ }\textbf {\bibinfo {volume} {08}},\ \bibinfo {pages} {044} (\bibinfo {year} {2016}{\natexlab{b}})},\ \Eprint {http://arxiv.org/abs/1606.05552} {arXiv:1606.05552 [hep-ph]} \BibitemShut {NoStop}%
\bibitem [{\citenamefont {Lee}\ \emph {et~al.}(2024)\citenamefont {Lee}, \citenamefont {Murai}, \citenamefont {Takahashi},\ and\ \citenamefont {Yin}}]{Lee:2024toz}%
  \BibitemOpen
  \bibfield  {author} {\bibinfo {author} {\bibfnamefont {J.}~\bibnamefont {Lee}}, \bibinfo {author} {\bibfnamefont {K.}~\bibnamefont {Murai}}, \bibinfo {author} {\bibfnamefont {F.}~\bibnamefont {Takahashi}}, \ and\ \bibinfo {author} {\bibfnamefont {W.}~\bibnamefont {Yin}},\ }\href@noop {} {\  (\bibinfo {year} {2024})},\ \Eprint {http://arxiv.org/abs/2409.09749} {arXiv:2409.09749 [hep-ph]} \BibitemShut {NoStop}%
\bibitem [{\citenamefont {Barr}\ \emph {et~al.}(1987)\citenamefont {Barr}, \citenamefont {Choi},\ and\ \citenamefont {Kim}}]{Barr:1986hs}%
  \BibitemOpen
  \bibfield  {author} {\bibinfo {author} {\bibfnamefont {S.~M.}\ \bibnamefont {Barr}}, \bibinfo {author} {\bibfnamefont {K.}~\bibnamefont {Choi}}, \ and\ \bibinfo {author} {\bibfnamefont {J.~E.}\ \bibnamefont {Kim}},\ }\href {\doibase 10.1016/0550-3213(87)90288-4} {\bibfield  {journal} {\bibinfo  {journal} {Nucl. Phys. B}\ }\textbf {\bibinfo {volume} {283}},\ \bibinfo {pages} {591} (\bibinfo {year} {1987})}\BibitemShut {NoStop}%
\bibitem [{\citenamefont {Kim}(1987)}]{Kim:1986ax}%
  \BibitemOpen
  \bibfield  {author} {\bibinfo {author} {\bibfnamefont {J.~E.}\ \bibnamefont {Kim}},\ }\href {\doibase 10.1016/0370-1573(87)90017-2} {\bibfield  {journal} {\bibinfo  {journal} {Phys. Rept.}\ }\textbf {\bibinfo {volume} {150}},\ \bibinfo {pages} {1} (\bibinfo {year} {1987})}\BibitemShut {NoStop}%
\bibitem [{\citenamefont {Chang}\ \emph {et~al.}(1999)\citenamefont {Chang}, \citenamefont {Hagmann},\ and\ \citenamefont {Sikivie}}]{Chang:1998tb}%
  \BibitemOpen
  \bibfield  {author} {\bibinfo {author} {\bibfnamefont {S.}~\bibnamefont {Chang}}, \bibinfo {author} {\bibfnamefont {C.}~\bibnamefont {Hagmann}}, \ and\ \bibinfo {author} {\bibfnamefont {P.}~\bibnamefont {Sikivie}},\ }\href {\doibase 10.1103/PhysRevD.59.023505} {\bibfield  {journal} {\bibinfo  {journal} {Phys. Rev. D}\ }\textbf {\bibinfo {volume} {59}},\ \bibinfo {pages} {023505} (\bibinfo {year} {1999})},\ \Eprint {http://arxiv.org/abs/hep-ph/9807374} {arXiv:hep-ph/9807374} \BibitemShut {NoStop}%
\bibitem [{\citenamefont {Sikivie}(1982)}]{Sikivie:1982qv}%
  \BibitemOpen
  \bibfield  {author} {\bibinfo {author} {\bibfnamefont {P.}~\bibnamefont {Sikivie}},\ }\href {\doibase 10.1103/PhysRevLett.48.1156} {\bibfield  {journal} {\bibinfo  {journal} {Phys. Rev. Lett.}\ }\textbf {\bibinfo {volume} {48}},\ \bibinfo {pages} {1156} (\bibinfo {year} {1982})}\BibitemShut {NoStop}%
\bibitem [{\citenamefont {Fox}\ \emph {et~al.}(2004)\citenamefont {Fox}, \citenamefont {Pierce},\ and\ \citenamefont {Thomas}}]{Fox:2004kb}%
  \BibitemOpen
  \bibfield  {author} {\bibinfo {author} {\bibfnamefont {P.}~\bibnamefont {Fox}}, \bibinfo {author} {\bibfnamefont {A.}~\bibnamefont {Pierce}}, \ and\ \bibinfo {author} {\bibfnamefont {S.~D.}\ \bibnamefont {Thomas}},\ }\href@noop {} {\  (\bibinfo {year} {2004})},\ \Eprint {http://arxiv.org/abs/hep-th/0409059} {arXiv:hep-th/0409059} \BibitemShut {NoStop}%
\bibitem [{\citenamefont {Semertzidis}\ and\ \citenamefont {Youn}(2022)}]{Semertzidis:2021rxs}%
  \BibitemOpen
  \bibfield  {author} {\bibinfo {author} {\bibfnamefont {Y.~K.}\ \bibnamefont {Semertzidis}}\ and\ \bibinfo {author} {\bibfnamefont {S.}~\bibnamefont {Youn}},\ }\href {\doibase 10.1126/sciadv.abm9928} {\bibfield  {journal} {\bibinfo  {journal} {Sci. Adv.}\ }\textbf {\bibinfo {volume} {8}},\ \bibinfo {pages} {abm9928} (\bibinfo {year} {2022})},\ \Eprint {http://arxiv.org/abs/2104.14831} {arXiv:2104.14831 [hep-ph]} \BibitemShut {NoStop}%
\bibitem [{\citenamefont {Mariotti}\ \emph {et~al.}(2018)\citenamefont {Mariotti}, \citenamefont {Redigolo}, \citenamefont {Sala},\ and\ \citenamefont {Tobioka}}]{Mariotti:2017vtv}%
  \BibitemOpen
  \bibfield  {author} {\bibinfo {author} {\bibfnamefont {A.}~\bibnamefont {Mariotti}}, \bibinfo {author} {\bibfnamefont {D.}~\bibnamefont {Redigolo}}, \bibinfo {author} {\bibfnamefont {F.}~\bibnamefont {Sala}}, \ and\ \bibinfo {author} {\bibfnamefont {K.}~\bibnamefont {Tobioka}},\ }\href {\doibase 10.1016/j.physletb.2018.06.039} {\bibfield  {journal} {\bibinfo  {journal} {Phys. Lett. B}\ }\textbf {\bibinfo {volume} {783}},\ \bibinfo {pages} {13} (\bibinfo {year} {2018})},\ \Eprint {http://arxiv.org/abs/1710.01743} {arXiv:1710.01743 [hep-ph]} \BibitemShut {NoStop}%
\bibitem [{\citenamefont {Ball}\ \emph {et~al.}(2013)\citenamefont {Ball}, \citenamefont {Bonvini}, \citenamefont {Forte}, \citenamefont {Marzani},\ and\ \citenamefont {Ridolfi}}]{Ball:2013bra}%
  \BibitemOpen
  \bibfield  {author} {\bibinfo {author} {\bibfnamefont {R.~D.}\ \bibnamefont {Ball}}, \bibinfo {author} {\bibfnamefont {M.}~\bibnamefont {Bonvini}}, \bibinfo {author} {\bibfnamefont {S.}~\bibnamefont {Forte}}, \bibinfo {author} {\bibfnamefont {S.}~\bibnamefont {Marzani}}, \ and\ \bibinfo {author} {\bibfnamefont {G.}~\bibnamefont {Ridolfi}},\ }\href {\doibase 10.1016/j.nuclphysb.2013.06.012} {\bibfield  {journal} {\bibinfo  {journal} {Nucl. Phys. B}\ }\textbf {\bibinfo {volume} {874}},\ \bibinfo {pages} {746} (\bibinfo {year} {2013})},\ \Eprint {http://arxiv.org/abs/1303.3590} {arXiv:1303.3590 [hep-ph]} \BibitemShut {NoStop}%
\bibitem [{\citenamefont {Bonvini}\ \emph {et~al.}(2014)\citenamefont {Bonvini}, \citenamefont {Ball}, \citenamefont {Forte}, \citenamefont {Marzani},\ and\ \citenamefont {Ridolfi}}]{Bonvini:2014jma}%
  \BibitemOpen
  \bibfield  {author} {\bibinfo {author} {\bibfnamefont {M.}~\bibnamefont {Bonvini}}, \bibinfo {author} {\bibfnamefont {R.~D.}\ \bibnamefont {Ball}}, \bibinfo {author} {\bibfnamefont {S.}~\bibnamefont {Forte}}, \bibinfo {author} {\bibfnamefont {S.}~\bibnamefont {Marzani}}, \ and\ \bibinfo {author} {\bibfnamefont {G.}~\bibnamefont {Ridolfi}},\ }\href {\doibase 10.1088/0954-3899/41/9/095002} {\bibfield  {journal} {\bibinfo  {journal} {J. Phys. G}\ }\textbf {\bibinfo {volume} {41}},\ \bibinfo {pages} {095002} (\bibinfo {year} {2014})},\ \Eprint {http://arxiv.org/abs/1404.3204} {arXiv:1404.3204 [hep-ph]} \BibitemShut {NoStop}%
\bibitem [{\citenamefont {Bonvini}\ \emph {et~al.}(2016)\citenamefont {Bonvini}, \citenamefont {Marzani}, \citenamefont {Muselli},\ and\ \citenamefont {Rottoli}}]{Bonvini:2016frm}%
  \BibitemOpen
  \bibfield  {author} {\bibinfo {author} {\bibfnamefont {M.}~\bibnamefont {Bonvini}}, \bibinfo {author} {\bibfnamefont {S.}~\bibnamefont {Marzani}}, \bibinfo {author} {\bibfnamefont {C.}~\bibnamefont {Muselli}}, \ and\ \bibinfo {author} {\bibfnamefont {L.}~\bibnamefont {Rottoli}},\ }\href {\doibase 10.1007/JHEP08(2016)105} {\bibfield  {journal} {\bibinfo  {journal} {JHEP}\ }\textbf {\bibinfo {volume} {08}},\ \bibinfo {pages} {105} (\bibinfo {year} {2016})},\ \Eprint {http://arxiv.org/abs/1603.08000} {arXiv:1603.08000 [hep-ph]} \BibitemShut {NoStop}%
\bibitem [{\citenamefont {Ahmed}\ \emph {et~al.}(2016)\citenamefont {Ahmed}, \citenamefont {Bonvini}, \citenamefont {Kumar}, \citenamefont {Mathews}, \citenamefont {Rana}, \citenamefont {Ravindran},\ and\ \citenamefont {Rottoli}}]{Ahmed:2016otz}%
  \BibitemOpen
  \bibfield  {author} {\bibinfo {author} {\bibfnamefont {T.}~\bibnamefont {Ahmed}}, \bibinfo {author} {\bibfnamefont {M.}~\bibnamefont {Bonvini}}, \bibinfo {author} {\bibfnamefont {M.~C.}\ \bibnamefont {Kumar}}, \bibinfo {author} {\bibfnamefont {P.}~\bibnamefont {Mathews}}, \bibinfo {author} {\bibfnamefont {N.}~\bibnamefont {Rana}}, \bibinfo {author} {\bibfnamefont {V.}~\bibnamefont {Ravindran}}, \ and\ \bibinfo {author} {\bibfnamefont {L.}~\bibnamefont {Rottoli}},\ }\href {\doibase 10.1140/epjc/s10052-016-4510-1} {\bibfield  {journal} {\bibinfo  {journal} {Eur. Phys. J. C}\ }\textbf {\bibinfo {volume} {76}},\ \bibinfo {pages} {663} (\bibinfo {year} {2016})},\ \Eprint {http://arxiv.org/abs/1606.00837} {arXiv:1606.00837 [hep-ph]} \BibitemShut {NoStop}%
\bibitem [{\citenamefont {Djouadi}(2008)}]{Djouadi:2005gi}%
  \BibitemOpen
  \bibfield  {author} {\bibinfo {author} {\bibfnamefont {A.}~\bibnamefont {Djouadi}},\ }\href {\doibase 10.1016/j.physrep.2007.10.004} {\bibfield  {journal} {\bibinfo  {journal} {Phys. Rept.}\ }\textbf {\bibinfo {volume} {457}},\ \bibinfo {pages} {1} (\bibinfo {year} {2008})},\ \Eprint {http://arxiv.org/abs/hep-ph/0503172} {arXiv:hep-ph/0503172} \BibitemShut {NoStop}%
\bibitem [{\citenamefont {Aad}\ \emph {et~al.}(2021)\citenamefont {Aad} \emph {et~al.}}]{ATLAS:2021uiz}%
  \BibitemOpen
  \bibfield  {author} {\bibinfo {author} {\bibfnamefont {G.}~\bibnamefont {Aad}} \emph {et~al.} (\bibinfo {collaboration} {ATLAS}),\ }\href {\doibase 10.1016/j.physletb.2021.136651} {\bibfield  {journal} {\bibinfo  {journal} {Phys. Lett. B}\ }\textbf {\bibinfo {volume} {822}},\ \bibinfo {pages} {136651} (\bibinfo {year} {2021})},\ \Eprint {http://arxiv.org/abs/2102.13405} {arXiv:2102.13405 [hep-ex]} \BibitemShut {NoStop}%
\bibitem [{\citenamefont {Aad}\ \emph {et~al.}(2024{\natexlab{b}})\citenamefont {Aad} \emph {et~al.}}]{ATLAS:2024bjr}%
  \BibitemOpen
  \bibfield  {author} {\bibinfo {author} {\bibfnamefont {G.}~\bibnamefont {Aad}} \emph {et~al.} (\bibinfo {collaboration} {ATLAS}),\ }\href@noop {} {\  (\bibinfo {year} {2024}{\natexlab{b}})},\ \Eprint {http://arxiv.org/abs/2407.07546} {arXiv:2407.07546 [hep-ex]} \BibitemShut {NoStop}%
\bibitem [{CMS(2024)}]{CMS-PAS-EXO-22-024}%
  \BibitemOpen
  \href {https://cds.cern.ch/record/2893028} {\emph {\bibinfo {title} {{Search for new physics in high-mass diphoton events from proton-proton collisions at $\sqrt{s}=13\,\mathrm{TeV}$}}}},\ \bibinfo {type} {Tech. Rep.}\ \bibinfo {number} {CMS-PAS-EXO-22-024}\ (\bibinfo  {institution} {CERN},\ \bibinfo {address} {Geneva},\ \bibinfo {year} {2024})\BibitemShut {NoStop}%
\bibitem [{\citenamefont {Cid~Vidal}\ \emph {et~al.}(2019)\citenamefont {Cid~Vidal}, \citenamefont {Mariotti}, \citenamefont {Redigolo}, \citenamefont {Sala},\ and\ \citenamefont {Tobioka}}]{CidVidal:2018blh}%
  \BibitemOpen
  \bibfield  {author} {\bibinfo {author} {\bibfnamefont {X.}~\bibnamefont {Cid~Vidal}}, \bibinfo {author} {\bibfnamefont {A.}~\bibnamefont {Mariotti}}, \bibinfo {author} {\bibfnamefont {D.}~\bibnamefont {Redigolo}}, \bibinfo {author} {\bibfnamefont {F.}~\bibnamefont {Sala}}, \ and\ \bibinfo {author} {\bibfnamefont {K.}~\bibnamefont {Tobioka}},\ }\href {\doibase 10.1007/JHEP01(2019)113} {\bibfield  {journal} {\bibinfo  {journal} {JHEP}\ }\textbf {\bibinfo {volume} {01}},\ \bibinfo {pages} {113} (\bibinfo {year} {2019})},\ \bibinfo {note} {[Erratum: JHEP 06, 141 (2020)]},\ \Eprint {http://arxiv.org/abs/1810.09452} {arXiv:1810.09452 [hep-ph]} \BibitemShut {NoStop}%
\bibitem [{\citenamefont {Acanfora}\ \emph {et~al.}(2024)\citenamefont {Acanfora}, \citenamefont {Franceschini}, \citenamefont {Mastroddi},\ and\ \citenamefont {Redigolo}}]{Acanfora:2024spi}%
  \BibitemOpen
  \bibfield  {author} {\bibinfo {author} {\bibfnamefont {F.}~\bibnamefont {Acanfora}}, \bibinfo {author} {\bibfnamefont {R.}~\bibnamefont {Franceschini}}, \bibinfo {author} {\bibfnamefont {A.}~\bibnamefont {Mastroddi}}, \ and\ \bibinfo {author} {\bibfnamefont {D.}~\bibnamefont {Redigolo}},\ }\href@noop {} {\  (\bibinfo {year} {2024})},\ \Eprint {http://arxiv.org/abs/2406.14614} {arXiv:2406.14614 [hep-ph]} \BibitemShut {NoStop}%
\bibitem [{\citenamefont {Alonso-\'Alvarez}\ \emph {et~al.}(2023)\citenamefont {Alonso-\'Alvarez}, \citenamefont {Cline},\ and\ \citenamefont {Xiao}}]{Alonso-Alvarez:2023wig}%
  \BibitemOpen
  \bibfield  {author} {\bibinfo {author} {\bibfnamefont {G.}~\bibnamefont {Alonso-\'Alvarez}}, \bibinfo {author} {\bibfnamefont {J.~M.}\ \bibnamefont {Cline}}, \ and\ \bibinfo {author} {\bibfnamefont {T.}~\bibnamefont {Xiao}},\ }\href {\doibase 10.1007/JHEP07(2023)187} {\bibfield  {journal} {\bibinfo  {journal} {JHEP}\ }\textbf {\bibinfo {volume} {07}},\ \bibinfo {pages} {187} (\bibinfo {year} {2023})},\ \Eprint {http://arxiv.org/abs/2305.00018} {arXiv:2305.00018 [hep-ph]} \BibitemShut {NoStop}%
\bibitem [{\citenamefont {Ball}\ \emph {et~al.}(2015)\citenamefont {Ball} \emph {et~al.}}]{NNPDF:2014otw}%
  \BibitemOpen
  \bibfield  {author} {\bibinfo {author} {\bibfnamefont {R.~D.}\ \bibnamefont {Ball}} \emph {et~al.} (\bibinfo {collaboration} {NNPDF}),\ }\href {\doibase 10.1007/JHEP04(2015)040} {\bibfield  {journal} {\bibinfo  {journal} {JHEP}\ }\textbf {\bibinfo {volume} {04}},\ \bibinfo {pages} {040} (\bibinfo {year} {2015})},\ \Eprint {http://arxiv.org/abs/1410.8849} {arXiv:1410.8849 [hep-ph]} \BibitemShut {NoStop}%
\bibitem [{\citenamefont {Alloul}\ \emph {et~al.}(2014)\citenamefont {Alloul}, \citenamefont {Christensen}, \citenamefont {Degrande}, \citenamefont {Duhr},\ and\ \citenamefont {Fuks}}]{Alloul:2013bka}%
  \BibitemOpen
  \bibfield  {author} {\bibinfo {author} {\bibfnamefont {A.}~\bibnamefont {Alloul}}, \bibinfo {author} {\bibfnamefont {N.~D.}\ \bibnamefont {Christensen}}, \bibinfo {author} {\bibfnamefont {C.}~\bibnamefont {Degrande}}, \bibinfo {author} {\bibfnamefont {C.}~\bibnamefont {Duhr}}, \ and\ \bibinfo {author} {\bibfnamefont {B.}~\bibnamefont {Fuks}},\ }\href {\doibase 10.1016/j.cpc.2014.04.012} {\bibfield  {journal} {\bibinfo  {journal} {Comput. Phys. Commun.}\ }\textbf {\bibinfo {volume} {185}},\ \bibinfo {pages} {2250} (\bibinfo {year} {2014})},\ \Eprint {http://arxiv.org/abs/1310.1921} {arXiv:1310.1921 [hep-ph]} \BibitemShut {NoStop}%
\bibitem [{\citenamefont {Alwall}\ \emph {et~al.}(2014)\citenamefont {Alwall}, \citenamefont {Frederix}, \citenamefont {Frixione}, \citenamefont {Hirschi}, \citenamefont {Maltoni}, \citenamefont {Mattelaer}, \citenamefont {Shao}, \citenamefont {Stelzer}, \citenamefont {Torrielli},\ and\ \citenamefont {Zaro}}]{Alwall:2014hca}%
  \BibitemOpen
  \bibfield  {author} {\bibinfo {author} {\bibfnamefont {J.}~\bibnamefont {Alwall}}, \bibinfo {author} {\bibfnamefont {R.}~\bibnamefont {Frederix}}, \bibinfo {author} {\bibfnamefont {S.}~\bibnamefont {Frixione}}, \bibinfo {author} {\bibfnamefont {V.}~\bibnamefont {Hirschi}}, \bibinfo {author} {\bibfnamefont {F.}~\bibnamefont {Maltoni}}, \bibinfo {author} {\bibfnamefont {O.}~\bibnamefont {Mattelaer}}, \bibinfo {author} {\bibfnamefont {H.~S.}\ \bibnamefont {Shao}}, \bibinfo {author} {\bibfnamefont {T.}~\bibnamefont {Stelzer}}, \bibinfo {author} {\bibfnamefont {P.}~\bibnamefont {Torrielli}}, \ and\ \bibinfo {author} {\bibfnamefont {M.}~\bibnamefont {Zaro}},\ }\href {\doibase 10.1007/JHEP07(2014)079} {\bibfield  {journal} {\bibinfo  {journal} {JHEP}\ }\textbf {\bibinfo {volume} {07}},\ \bibinfo {pages} {079} (\bibinfo {year} {2014})},\ \Eprint {http://arxiv.org/abs/1405.0301} {arXiv:1405.0301 [hep-ph]} \BibitemShut {NoStop}%
\bibitem [{\citenamefont {Sj\"ostrand}\ \emph {et~al.}(2015)\citenamefont {Sj\"ostrand}, \citenamefont {Ask}, \citenamefont {Christiansen}, \citenamefont {Corke}, \citenamefont {Desai}, \citenamefont {Ilten}, \citenamefont {Mrenna}, \citenamefont {Prestel}, \citenamefont {Rasmussen},\ and\ \citenamefont {Skands}}]{Sjostrand:2014zea}%
  \BibitemOpen
  \bibfield  {author} {\bibinfo {author} {\bibfnamefont {T.}~\bibnamefont {Sj\"ostrand}}, \bibinfo {author} {\bibfnamefont {S.}~\bibnamefont {Ask}}, \bibinfo {author} {\bibfnamefont {J.~R.}\ \bibnamefont {Christiansen}}, \bibinfo {author} {\bibfnamefont {R.}~\bibnamefont {Corke}}, \bibinfo {author} {\bibfnamefont {N.}~\bibnamefont {Desai}}, \bibinfo {author} {\bibfnamefont {P.}~\bibnamefont {Ilten}}, \bibinfo {author} {\bibfnamefont {S.}~\bibnamefont {Mrenna}}, \bibinfo {author} {\bibfnamefont {S.}~\bibnamefont {Prestel}}, \bibinfo {author} {\bibfnamefont {C.~O.}\ \bibnamefont {Rasmussen}}, \ and\ \bibinfo {author} {\bibfnamefont {P.~Z.}\ \bibnamefont {Skands}},\ }\href {\doibase 10.1016/j.cpc.2015.01.024} {\bibfield  {journal} {\bibinfo  {journal} {Comput. Phys. Commun.}\ }\textbf {\bibinfo {volume} {191}},\ \bibinfo {pages} {159} (\bibinfo {year} {2015})},\ \Eprint {http://arxiv.org/abs/1410.3012} {arXiv:1410.3012 [hep-ph]} \BibitemShut {NoStop}%
\bibitem [{\citenamefont {Bierlich}\ \emph {et~al.}(2022)\citenamefont {Bierlich} \emph {et~al.}}]{Bierlich:2022pfr}%
  \BibitemOpen
  \bibfield  {author} {\bibinfo {author} {\bibfnamefont {C.}~\bibnamefont {Bierlich}} \emph {et~al.},\ }\href {\doibase 10.21468/SciPostPhysCodeb.8} {\bibfield  {journal} {\bibinfo  {journal} {SciPost Phys. Codeb.}\ }\textbf {\bibinfo {volume} {2022}},\ \bibinfo {pages} {8} (\bibinfo {year} {2022})},\ \Eprint {http://arxiv.org/abs/2203.11601} {arXiv:2203.11601 [hep-ph]} \BibitemShut {NoStop}%
\bibitem [{\citenamefont {de~Favereau}\ \emph {et~al.}(2014)\citenamefont {de~Favereau}, \citenamefont {Delaere}, \citenamefont {Demin}, \citenamefont {Giammanco}, \citenamefont {Lema\^\i{}tre}, \citenamefont {Mertens},\ and\ \citenamefont {Selvaggi}}]{deFavereau:2013fsa}%
  \BibitemOpen
  \bibfield  {author} {\bibinfo {author} {\bibfnamefont {J.}~\bibnamefont {de~Favereau}}, \bibinfo {author} {\bibfnamefont {C.}~\bibnamefont {Delaere}}, \bibinfo {author} {\bibfnamefont {P.}~\bibnamefont {Demin}}, \bibinfo {author} {\bibfnamefont {A.}~\bibnamefont {Giammanco}}, \bibinfo {author} {\bibfnamefont {V.}~\bibnamefont {Lema\^\i{}tre}}, \bibinfo {author} {\bibfnamefont {A.}~\bibnamefont {Mertens}}, \ and\ \bibinfo {author} {\bibfnamefont {M.}~\bibnamefont {Selvaggi}} (\bibinfo {collaboration} {DELPHES 3}),\ }\href {\doibase 10.1007/JHEP02(2014)057} {\bibfield  {journal} {\bibinfo  {journal} {JHEP}\ }\textbf {\bibinfo {volume} {02}},\ \bibinfo {pages} {057} (\bibinfo {year} {2014})},\ \Eprint {http://arxiv.org/abs/1307.6346} {arXiv:1307.6346 [hep-ex]} \BibitemShut {NoStop}%
\bibitem [{\citenamefont {Artoisenet}\ \emph {et~al.}(2013)\citenamefont {Artoisenet}, \citenamefont {Frederix}, \citenamefont {Mattelaer},\ and\ \citenamefont {Rietkerk}}]{Artoisenet:2012st}%
  \BibitemOpen
  \bibfield  {author} {\bibinfo {author} {\bibfnamefont {P.}~\bibnamefont {Artoisenet}}, \bibinfo {author} {\bibfnamefont {R.}~\bibnamefont {Frederix}}, \bibinfo {author} {\bibfnamefont {O.}~\bibnamefont {Mattelaer}}, \ and\ \bibinfo {author} {\bibfnamefont {R.}~\bibnamefont {Rietkerk}},\ }\href {\doibase 10.1007/JHEP03(2013)015} {\bibfield  {journal} {\bibinfo  {journal} {JHEP}\ }\textbf {\bibinfo {volume} {03}},\ \bibinfo {pages} {015} (\bibinfo {year} {2013})},\ \Eprint {http://arxiv.org/abs/1212.3460} {arXiv:1212.3460 [hep-ph]} \BibitemShut {NoStop}%
\bibitem [{\citenamefont {Cacciari}\ \emph {et~al.}(2012)\citenamefont {Cacciari}, \citenamefont {Salam},\ and\ \citenamefont {Soyez}}]{Cacciari:2011ma}%
  \BibitemOpen
  \bibfield  {author} {\bibinfo {author} {\bibfnamefont {M.}~\bibnamefont {Cacciari}}, \bibinfo {author} {\bibfnamefont {G.~P.}\ \bibnamefont {Salam}}, \ and\ \bibinfo {author} {\bibfnamefont {G.}~\bibnamefont {Soyez}},\ }\href {\doibase 10.1140/epjc/s10052-012-1896-2} {\bibfield  {journal} {\bibinfo  {journal} {Eur. Phys. J. C}\ }\textbf {\bibinfo {volume} {72}},\ \bibinfo {pages} {1896} (\bibinfo {year} {2012})},\ \Eprint {http://arxiv.org/abs/1111.6097} {arXiv:1111.6097 [hep-ph]} \BibitemShut {NoStop}%
\bibitem [{\citenamefont {Cacciari}\ \emph {et~al.}(2008)\citenamefont {Cacciari}, \citenamefont {Salam},\ and\ \citenamefont {Soyez}}]{Cacciari:2008gp}%
  \BibitemOpen
  \bibfield  {author} {\bibinfo {author} {\bibfnamefont {M.}~\bibnamefont {Cacciari}}, \bibinfo {author} {\bibfnamefont {G.~P.}\ \bibnamefont {Salam}}, \ and\ \bibinfo {author} {\bibfnamefont {G.}~\bibnamefont {Soyez}},\ }\href {\doibase 10.1088/1126-6708/2008/04/063} {\bibfield  {journal} {\bibinfo  {journal} {JHEP}\ }\textbf {\bibinfo {volume} {04}},\ \bibinfo {pages} {063} (\bibinfo {year} {2008})},\ \Eprint {http://arxiv.org/abs/0802.1189} {arXiv:0802.1189 [hep-ph]} \BibitemShut {NoStop}%
\bibitem [{\citenamefont {Gershtein}\ \emph {et~al.}(2021)\citenamefont {Gershtein}, \citenamefont {Knapen},\ and\ \citenamefont {Redigolo}}]{Gershtein:2020mwi}%
  \BibitemOpen
  \bibfield  {author} {\bibinfo {author} {\bibfnamefont {Y.}~\bibnamefont {Gershtein}}, \bibinfo {author} {\bibfnamefont {S.}~\bibnamefont {Knapen}}, \ and\ \bibinfo {author} {\bibfnamefont {D.}~\bibnamefont {Redigolo}},\ }\href {\doibase 10.1016/j.physletb.2021.136758} {\bibfield  {journal} {\bibinfo  {journal} {Phys. Lett. B}\ }\textbf {\bibinfo {volume} {823}},\ \bibinfo {pages} {136758} (\bibinfo {year} {2021})},\ \Eprint {http://arxiv.org/abs/2012.07864} {arXiv:2012.07864 [hep-ph]} \BibitemShut {NoStop}%
\bibitem [{\citenamefont {Conte}\ \emph {et~al.}(2013)\citenamefont {Conte}, \citenamefont {Fuks},\ and\ \citenamefont {Serret}}]{Conte:2012fm}%
  \BibitemOpen
  \bibfield  {author} {\bibinfo {author} {\bibfnamefont {E.}~\bibnamefont {Conte}}, \bibinfo {author} {\bibfnamefont {B.}~\bibnamefont {Fuks}}, \ and\ \bibinfo {author} {\bibfnamefont {G.}~\bibnamefont {Serret}},\ }\href {\doibase 10.1016/j.cpc.2012.09.009} {\bibfield  {journal} {\bibinfo  {journal} {Comput. Phys. Commun.}\ }\textbf {\bibinfo {volume} {184}},\ \bibinfo {pages} {222} (\bibinfo {year} {2013})},\ \Eprint {http://arxiv.org/abs/1206.1599} {arXiv:1206.1599 [hep-ph]} \BibitemShut {NoStop}%
\bibitem [{\citenamefont {{M. Aharrouche, et al, ATLAS Electromagnetic Barrel Calorimeter Collaboration}}(2006)}]{AHARROUCHE2006601}%
  \BibitemOpen
  \bibfield  {author} {\bibinfo {author} {\bibnamefont {{M. Aharrouche, et al, ATLAS Electromagnetic Barrel Calorimeter Collaboration}}},\ }\href {\doibase https://doi.org/10.1016/j.nima.2006.07.053} {\bibfield  {journal} {\bibinfo  {journal} {Nuclear Instruments and Methods in Physics Research Section A: Accelerators, Spectrometers, Detectors and Associated Equipment}\ }\textbf {\bibinfo {volume} {568}},\ \bibinfo {pages} {601} (\bibinfo {year} {2006})}\BibitemShut {NoStop}%
\bibitem [{\citenamefont {Salvio}\ \emph {et~al.}(2016)\citenamefont {Salvio}, \citenamefont {Staub}, \citenamefont {Strumia},\ and\ \citenamefont {Urbano}}]{Salvio:2016hnf}%
  \BibitemOpen
  \bibfield  {author} {\bibinfo {author} {\bibfnamefont {A.}~\bibnamefont {Salvio}}, \bibinfo {author} {\bibfnamefont {F.}~\bibnamefont {Staub}}, \bibinfo {author} {\bibfnamefont {A.}~\bibnamefont {Strumia}}, \ and\ \bibinfo {author} {\bibfnamefont {A.}~\bibnamefont {Urbano}},\ }\href {\doibase 10.1007/JHEP03(2016)214} {\bibfield  {journal} {\bibinfo  {journal} {JHEP}\ }\textbf {\bibinfo {volume} {03}},\ \bibinfo {pages} {214} (\bibinfo {year} {2016})},\ \Eprint {http://arxiv.org/abs/1602.01460} {arXiv:1602.01460 [hep-ph]} \BibitemShut {NoStop}%
\bibitem [{\citenamefont {Franceschini}\ \emph {et~al.}(2016)\citenamefont {Franceschini}, \citenamefont {Giudice}, \citenamefont {Kamenik}, \citenamefont {McCullough}, \citenamefont {Pomarol}, \citenamefont {Rattazzi}, \citenamefont {Redi}, \citenamefont {Riva}, \citenamefont {Strumia},\ and\ \citenamefont {Torre}}]{Franceschini:2015kwy}%
  \BibitemOpen
  \bibfield  {author} {\bibinfo {author} {\bibfnamefont {R.}~\bibnamefont {Franceschini}}, \bibinfo {author} {\bibfnamefont {G.~F.}\ \bibnamefont {Giudice}}, \bibinfo {author} {\bibfnamefont {J.~F.}\ \bibnamefont {Kamenik}}, \bibinfo {author} {\bibfnamefont {M.}~\bibnamefont {McCullough}}, \bibinfo {author} {\bibfnamefont {A.}~\bibnamefont {Pomarol}}, \bibinfo {author} {\bibfnamefont {R.}~\bibnamefont {Rattazzi}}, \bibinfo {author} {\bibfnamefont {M.}~\bibnamefont {Redi}}, \bibinfo {author} {\bibfnamefont {F.}~\bibnamefont {Riva}}, \bibinfo {author} {\bibfnamefont {A.}~\bibnamefont {Strumia}}, \ and\ \bibinfo {author} {\bibfnamefont {R.}~\bibnamefont {Torre}},\ }\href {\doibase 10.1007/JHEP03(2016)144} {\bibfield  {journal} {\bibinfo  {journal} {JHEP}\ }\textbf {\bibinfo {volume} {03}},\ \bibinfo {pages} {144} (\bibinfo {year} {2016})},\ \Eprint {http://arxiv.org/abs/1512.04933} {arXiv:1512.04933 [hep-ph]} \BibitemShut {NoStop}%
\bibitem [{\citenamefont {Alexander}\ \emph {et~al.}(2023)\citenamefont {Alexander}, \citenamefont {Gilmer}, \citenamefont {Manton},\ and\ \citenamefont {McDonough}}]{Alexander:2023wgk}%
  \BibitemOpen
  \bibfield  {author} {\bibinfo {author} {\bibfnamefont {S.}~\bibnamefont {Alexander}}, \bibinfo {author} {\bibfnamefont {H.}~\bibnamefont {Gilmer}}, \bibinfo {author} {\bibfnamefont {T.}~\bibnamefont {Manton}}, \ and\ \bibinfo {author} {\bibfnamefont {E.}~\bibnamefont {McDonough}},\ }\href {\doibase 10.1103/PhysRevD.108.123014} {\bibfield  {journal} {\bibinfo  {journal} {Phys. Rev. D}\ }\textbf {\bibinfo {volume} {108}},\ \bibinfo {pages} {123014} (\bibinfo {year} {2023})},\ \Eprint {http://arxiv.org/abs/2304.11176} {arXiv:2304.11176 [hep-ph]} \BibitemShut {NoStop}%
\bibitem [{\citenamefont {Alexander}\ \emph {et~al.}(2024)\citenamefont {Alexander}, \citenamefont {Manton},\ and\ \citenamefont {McDonough}}]{Alexander:2024nvi}%
  \BibitemOpen
  \bibfield  {author} {\bibinfo {author} {\bibfnamefont {S.}~\bibnamefont {Alexander}}, \bibinfo {author} {\bibfnamefont {T.}~\bibnamefont {Manton}}, \ and\ \bibinfo {author} {\bibfnamefont {E.}~\bibnamefont {McDonough}},\ }\href {\doibase 10.1103/PhysRevD.109.116019} {\bibfield  {journal} {\bibinfo  {journal} {Phys. Rev. D}\ }\textbf {\bibinfo {volume} {109}},\ \bibinfo {pages} {116019} (\bibinfo {year} {2024})},\ \Eprint {http://arxiv.org/abs/2404.11642} {arXiv:2404.11642 [hep-ph]} \BibitemShut {NoStop}%
\bibitem [{\citenamefont {Boughezal}\ \emph {et~al.}(2015)\citenamefont {Boughezal}, \citenamefont {Caola}, \citenamefont {Melnikov}, \citenamefont {Petriello},\ and\ \citenamefont {Schulze}}]{Boughezal:2015dra}%
  \BibitemOpen
  \bibfield  {author} {\bibinfo {author} {\bibfnamefont {R.}~\bibnamefont {Boughezal}}, \bibinfo {author} {\bibfnamefont {F.}~\bibnamefont {Caola}}, \bibinfo {author} {\bibfnamefont {K.}~\bibnamefont {Melnikov}}, \bibinfo {author} {\bibfnamefont {F.}~\bibnamefont {Petriello}}, \ and\ \bibinfo {author} {\bibfnamefont {M.}~\bibnamefont {Schulze}},\ }\href {\doibase 10.1103/PhysRevLett.115.082003} {\bibfield  {journal} {\bibinfo  {journal} {Phys. Rev. Lett.}\ }\textbf {\bibinfo {volume} {115}},\ \bibinfo {pages} {082003} (\bibinfo {year} {2015})},\ \Eprint {http://arxiv.org/abs/1504.07922} {arXiv:1504.07922 [hep-ph]} \BibitemShut {NoStop}%
\bibitem [{\citenamefont {Atre}\ \emph {et~al.}(2011)\citenamefont {Atre}, \citenamefont {Azuelos}, \citenamefont {Carena}, \citenamefont {Han}, \citenamefont {Ozcan}, \citenamefont {Santiago},\ and\ \citenamefont {Unel}}]{Atre:2011ae}%
  \BibitemOpen
  \bibfield  {author} {\bibinfo {author} {\bibfnamefont {A.}~\bibnamefont {Atre}}, \bibinfo {author} {\bibfnamefont {G.}~\bibnamefont {Azuelos}}, \bibinfo {author} {\bibfnamefont {M.}~\bibnamefont {Carena}}, \bibinfo {author} {\bibfnamefont {T.}~\bibnamefont {Han}}, \bibinfo {author} {\bibfnamefont {E.}~\bibnamefont {Ozcan}}, \bibinfo {author} {\bibfnamefont {J.}~\bibnamefont {Santiago}}, \ and\ \bibinfo {author} {\bibfnamefont {G.}~\bibnamefont {Unel}},\ }\href {\doibase 10.1007/JHEP08(2011)080} {\bibfield  {journal} {\bibinfo  {journal} {JHEP}\ }\textbf {\bibinfo {volume} {08}},\ \bibinfo {pages} {080} (\bibinfo {year} {2011})},\ \Eprint {http://arxiv.org/abs/1102.1987} {arXiv:1102.1987 [hep-ph]} \BibitemShut {NoStop}%
\bibitem [{\citenamefont {Buttazzo}\ \emph {et~al.}(2013)\citenamefont {Buttazzo}, \citenamefont {Degrassi}, \citenamefont {Giardino}, \citenamefont {Giudice}, \citenamefont {Sala}, \citenamefont {Salvio},\ and\ \citenamefont {Strumia}}]{Buttazzo:2013uya}%
  \BibitemOpen
  \bibfield  {author} {\bibinfo {author} {\bibfnamefont {D.}~\bibnamefont {Buttazzo}}, \bibinfo {author} {\bibfnamefont {G.}~\bibnamefont {Degrassi}}, \bibinfo {author} {\bibfnamefont {P.~P.}\ \bibnamefont {Giardino}}, \bibinfo {author} {\bibfnamefont {G.~F.}\ \bibnamefont {Giudice}}, \bibinfo {author} {\bibfnamefont {F.}~\bibnamefont {Sala}}, \bibinfo {author} {\bibfnamefont {A.}~\bibnamefont {Salvio}}, \ and\ \bibinfo {author} {\bibfnamefont {A.}~\bibnamefont {Strumia}},\ }\href {\doibase 10.1007/JHEP12(2013)089} {\bibfield  {journal} {\bibinfo  {journal} {JHEP}\ }\textbf {\bibinfo {volume} {12}},\ \bibinfo {pages} {089} (\bibinfo {year} {2013})},\ \Eprint {http://arxiv.org/abs/1307.3536} {arXiv:1307.3536 [hep-ph]} \BibitemShut {NoStop}%
\bibitem [{\citenamefont {De~Luca}\ \emph {et~al.}(2018)\citenamefont {De~Luca}, \citenamefont {Mitridate}, \citenamefont {Redi}, \citenamefont {Smirnov},\ and\ \citenamefont {Strumia}}]{DeLuca:2018mzn}%
  \BibitemOpen
  \bibfield  {author} {\bibinfo {author} {\bibfnamefont {V.}~\bibnamefont {De~Luca}}, \bibinfo {author} {\bibfnamefont {A.}~\bibnamefont {Mitridate}}, \bibinfo {author} {\bibfnamefont {M.}~\bibnamefont {Redi}}, \bibinfo {author} {\bibfnamefont {J.}~\bibnamefont {Smirnov}}, \ and\ \bibinfo {author} {\bibfnamefont {A.}~\bibnamefont {Strumia}},\ }\href {\doibase 10.1103/PhysRevD.97.115024} {\bibfield  {journal} {\bibinfo  {journal} {Phys. Rev. D}\ }\textbf {\bibinfo {volume} {97}},\ \bibinfo {pages} {115024} (\bibinfo {year} {2018})},\ \Eprint {http://arxiv.org/abs/1801.01135} {arXiv:1801.01135 [hep-ph]} \BibitemShut {NoStop}%
\bibitem [{\citenamefont {Baer}\ \emph {et~al.}(1999)\citenamefont {Baer}, \citenamefont {Cheung},\ and\ \citenamefont {Gunion}}]{Baer:1998pg}%
  \BibitemOpen
  \bibfield  {author} {\bibinfo {author} {\bibfnamefont {H.}~\bibnamefont {Baer}}, \bibinfo {author} {\bibfnamefont {K.-m.}\ \bibnamefont {Cheung}}, \ and\ \bibinfo {author} {\bibfnamefont {J.~F.}\ \bibnamefont {Gunion}},\ }\href {\doibase 10.1103/PhysRevD.59.075002} {\bibfield  {journal} {\bibinfo  {journal} {Phys. Rev. D}\ }\textbf {\bibinfo {volume} {59}},\ \bibinfo {pages} {075002} (\bibinfo {year} {1999})},\ \Eprint {http://arxiv.org/abs/hep-ph/9806361} {arXiv:hep-ph/9806361} \BibitemShut {NoStop}%
\bibitem [{\citenamefont {Arkani-Hamed}\ and\ \citenamefont {Dimopoulos}(2005)}]{Arkani-Hamed:2004ymt}%
  \BibitemOpen
  \bibfield  {author} {\bibinfo {author} {\bibfnamefont {N.}~\bibnamefont {Arkani-Hamed}}\ and\ \bibinfo {author} {\bibfnamefont {S.}~\bibnamefont {Dimopoulos}},\ }\href {\doibase 10.1088/1126-6708/2005/06/073} {\bibfield  {journal} {\bibinfo  {journal} {JHEP}\ }\textbf {\bibinfo {volume} {06}},\ \bibinfo {pages} {073} (\bibinfo {year} {2005})},\ \Eprint {http://arxiv.org/abs/hep-th/0405159} {arXiv:hep-th/0405159} \BibitemShut {NoStop}%
\bibitem [{\citenamefont {Giudice}\ and\ \citenamefont {Romanino}(2004)}]{Giudice:2004tc}%
  \BibitemOpen
  \bibfield  {author} {\bibinfo {author} {\bibfnamefont {G.~F.}\ \bibnamefont {Giudice}}\ and\ \bibinfo {author} {\bibfnamefont {A.}~\bibnamefont {Romanino}},\ }\href {\doibase 10.1016/j.nuclphysb.2004.08.001} {\bibfield  {journal} {\bibinfo  {journal} {Nucl. Phys. B}\ }\textbf {\bibinfo {volume} {699}},\ \bibinfo {pages} {65} (\bibinfo {year} {2004})},\ \bibinfo {note} {[Erratum: Nucl.Phys.B 706, 487--487 (2005)]},\ \Eprint {http://arxiv.org/abs/hep-ph/0406088} {arXiv:hep-ph/0406088} \BibitemShut {NoStop}%
\bibitem [{\citenamefont {Kilian}\ \emph {et~al.}(2005)\citenamefont {Kilian}, \citenamefont {Plehn}, \citenamefont {Richardson},\ and\ \citenamefont {Schmidt}}]{Kilian:2004uj}%
  \BibitemOpen
  \bibfield  {author} {\bibinfo {author} {\bibfnamefont {W.}~\bibnamefont {Kilian}}, \bibinfo {author} {\bibfnamefont {T.}~\bibnamefont {Plehn}}, \bibinfo {author} {\bibfnamefont {P.}~\bibnamefont {Richardson}}, \ and\ \bibinfo {author} {\bibfnamefont {E.}~\bibnamefont {Schmidt}},\ }\href {\doibase 10.1140/epjc/s2004-02046-5} {\bibfield  {journal} {\bibinfo  {journal} {Eur. Phys. J. C}\ }\textbf {\bibinfo {volume} {39}},\ \bibinfo {pages} {229} (\bibinfo {year} {2005})},\ \Eprint {http://arxiv.org/abs/hep-ph/0408088} {arXiv:hep-ph/0408088} \BibitemShut {NoStop}%
\bibitem [{\citenamefont {Hewett}\ \emph {et~al.}(2004)\citenamefont {Hewett}, \citenamefont {Lillie}, \citenamefont {Masip},\ and\ \citenamefont {Rizzo}}]{Hewett:2004nw}%
  \BibitemOpen
  \bibfield  {author} {\bibinfo {author} {\bibfnamefont {J.~L.}\ \bibnamefont {Hewett}}, \bibinfo {author} {\bibfnamefont {B.}~\bibnamefont {Lillie}}, \bibinfo {author} {\bibfnamefont {M.}~\bibnamefont {Masip}}, \ and\ \bibinfo {author} {\bibfnamefont {T.~G.}\ \bibnamefont {Rizzo}},\ }\href {\doibase 10.1088/1126-6708/2004/09/070} {\bibfield  {journal} {\bibinfo  {journal} {JHEP}\ }\textbf {\bibinfo {volume} {09}},\ \bibinfo {pages} {070} (\bibinfo {year} {2004})},\ \Eprint {http://arxiv.org/abs/hep-ph/0408248} {arXiv:hep-ph/0408248} \BibitemShut {NoStop}%
\bibitem [{\citenamefont {Arvanitaki}\ \emph {et~al.}(2007)\citenamefont {Arvanitaki}, \citenamefont {Dimopoulos}, \citenamefont {Pierce}, \citenamefont {Rajendran},\ and\ \citenamefont {Wacker}}]{Arvanitaki:2005nq}%
  \BibitemOpen
  \bibfield  {author} {\bibinfo {author} {\bibfnamefont {A.}~\bibnamefont {Arvanitaki}}, \bibinfo {author} {\bibfnamefont {S.}~\bibnamefont {Dimopoulos}}, \bibinfo {author} {\bibfnamefont {A.}~\bibnamefont {Pierce}}, \bibinfo {author} {\bibfnamefont {S.}~\bibnamefont {Rajendran}}, \ and\ \bibinfo {author} {\bibfnamefont {J.~G.}\ \bibnamefont {Wacker}},\ }\href {\doibase 10.1103/PhysRevD.76.055007} {\bibfield  {journal} {\bibinfo  {journal} {Phys. Rev. D}\ }\textbf {\bibinfo {volume} {76}},\ \bibinfo {pages} {055007} (\bibinfo {year} {2007})},\ \Eprint {http://arxiv.org/abs/hep-ph/0506242} {arXiv:hep-ph/0506242} \BibitemShut {NoStop}%
\bibitem [{\citenamefont {Fairbairn}\ \emph {et~al.}(2007)\citenamefont {Fairbairn}, \citenamefont {Kraan}, \citenamefont {Milstead}, \citenamefont {Sjostrand}, \citenamefont {Skands},\ and\ \citenamefont {Sloan}}]{Fairbairn:2006gg}%
  \BibitemOpen
  \bibfield  {author} {\bibinfo {author} {\bibfnamefont {M.}~\bibnamefont {Fairbairn}}, \bibinfo {author} {\bibfnamefont {A.~C.}\ \bibnamefont {Kraan}}, \bibinfo {author} {\bibfnamefont {D.~A.}\ \bibnamefont {Milstead}}, \bibinfo {author} {\bibfnamefont {T.}~\bibnamefont {Sjostrand}}, \bibinfo {author} {\bibfnamefont {P.~Z.}\ \bibnamefont {Skands}}, \ and\ \bibinfo {author} {\bibfnamefont {T.}~\bibnamefont {Sloan}},\ }\href {\doibase 10.1016/j.physrep.2006.10.002} {\bibfield  {journal} {\bibinfo  {journal} {Phys. Rept.}\ }\textbf {\bibinfo {volume} {438}},\ \bibinfo {pages} {1} (\bibinfo {year} {2007})},\ \Eprint {http://arxiv.org/abs/hep-ph/0611040} {arXiv:hep-ph/0611040} \BibitemShut {NoStop}%
\bibitem [{\citenamefont {Diaz-Cruz}\ \emph {et~al.}(2007)\citenamefont {Diaz-Cruz}, \citenamefont {Ellis}, \citenamefont {Olive},\ and\ \citenamefont {Santoso}}]{Diaz-Cruz:2007ewo}%
  \BibitemOpen
  \bibfield  {author} {\bibinfo {author} {\bibfnamefont {J.~L.}\ \bibnamefont {Diaz-Cruz}}, \bibinfo {author} {\bibfnamefont {J.~R.}\ \bibnamefont {Ellis}}, \bibinfo {author} {\bibfnamefont {K.~A.}\ \bibnamefont {Olive}}, \ and\ \bibinfo {author} {\bibfnamefont {Y.}~\bibnamefont {Santoso}},\ }\href {\doibase 10.1088/1126-6708/2007/05/003} {\bibfield  {journal} {\bibinfo  {journal} {JHEP}\ }\textbf {\bibinfo {volume} {05}},\ \bibinfo {pages} {003} (\bibinfo {year} {2007})},\ \Eprint {http://arxiv.org/abs/hep-ph/0701229} {arXiv:hep-ph/0701229} \BibitemShut {NoStop}%
\bibitem [{\citenamefont {Choudhury}\ \emph {et~al.}(2008)\citenamefont {Choudhury}, \citenamefont {Gupta},\ and\ \citenamefont {Mukhopadhyaya}}]{Choudhury:2008gb}%
  \BibitemOpen
  \bibfield  {author} {\bibinfo {author} {\bibfnamefont {D.}~\bibnamefont {Choudhury}}, \bibinfo {author} {\bibfnamefont {S.~K.}\ \bibnamefont {Gupta}}, \ and\ \bibinfo {author} {\bibfnamefont {B.}~\bibnamefont {Mukhopadhyaya}},\ }\href {\doibase 10.1103/PhysRevD.78.015023} {\bibfield  {journal} {\bibinfo  {journal} {Phys. Rev. D}\ }\textbf {\bibinfo {volume} {78}},\ \bibinfo {pages} {015023} (\bibinfo {year} {2008})},\ \Eprint {http://arxiv.org/abs/0804.3560} {arXiv:0804.3560 [hep-ph]} \BibitemShut {NoStop}%
\bibitem [{\citenamefont {Di~Luzio}\ \emph {et~al.}(2015)\citenamefont {Di~Luzio}, \citenamefont {Gr\"ober}, \citenamefont {Kamenik},\ and\ \citenamefont {Nardecchia}}]{DiLuzio:2015oha}%
  \BibitemOpen
  \bibfield  {author} {\bibinfo {author} {\bibfnamefont {L.}~\bibnamefont {Di~Luzio}}, \bibinfo {author} {\bibfnamefont {R.}~\bibnamefont {Gr\"ober}}, \bibinfo {author} {\bibfnamefont {J.~F.}\ \bibnamefont {Kamenik}}, \ and\ \bibinfo {author} {\bibfnamefont {M.}~\bibnamefont {Nardecchia}},\ }\href {\doibase 10.1007/JHEP07(2015)074} {\bibfield  {journal} {\bibinfo  {journal} {JHEP}\ }\textbf {\bibinfo {volume} {07}},\ \bibinfo {pages} {074} (\bibinfo {year} {2015})},\ \Eprint {http://arxiv.org/abs/1504.00359} {arXiv:1504.00359 [hep-ph]} \BibitemShut {NoStop}%
\bibitem [{ATL(2019)}]{ATL-PHYS-PUB-2019-019}%
  \BibitemOpen
  \href {https://cds.cern.ch/record/2676309} {\emph {\bibinfo {title} {{Generation and Simulation of $R$-Hadrons in the ATLAS Experiment}}}},\ \bibinfo {type} {Tech. Rep.}\ (\bibinfo  {institution} {CERN},\ \bibinfo {address} {Geneva},\ \bibinfo {year} {2019})\BibitemShut {NoStop}%
\bibitem [{\citenamefont {Tumasyan}\ \emph {et~al.}(2023)\citenamefont {Tumasyan} \emph {et~al.}}]{CMS:2022fck}%
  \BibitemOpen
  \bibfield  {author} {\bibinfo {author} {\bibfnamefont {A.}~\bibnamefont {Tumasyan}} \emph {et~al.} (\bibinfo {collaboration} {CMS}),\ }\href {\doibase 10.1007/JHEP07(2023)020} {\bibfield  {journal} {\bibinfo  {journal} {JHEP}\ }\textbf {\bibinfo {volume} {07}},\ \bibinfo {pages} {020} (\bibinfo {year} {2023})},\ \Eprint {http://arxiv.org/abs/2209.07327} {arXiv:2209.07327 [hep-ex]} \BibitemShut {NoStop}%
\bibitem [{\citenamefont {Alves}\ \emph {et~al.}(2024)\citenamefont {Alves}, \citenamefont {Branco}, \citenamefont {Cherchiglia}, \citenamefont {Nishi}, \citenamefont {Penedo}, \citenamefont {Pereira}, \citenamefont {Rebelo},\ and\ \citenamefont {Silva-Marcos}}]{Alves:2023ufm}%
  \BibitemOpen
  \bibfield  {author} {\bibinfo {author} {\bibfnamefont {J.~a.~M.}\ \bibnamefont {Alves}}, \bibinfo {author} {\bibfnamefont {G.~C.}\ \bibnamefont {Branco}}, \bibinfo {author} {\bibfnamefont {A.~L.}\ \bibnamefont {Cherchiglia}}, \bibinfo {author} {\bibfnamefont {C.~C.}\ \bibnamefont {Nishi}}, \bibinfo {author} {\bibfnamefont {J.~T.}\ \bibnamefont {Penedo}}, \bibinfo {author} {\bibfnamefont {P.~M.~F.}\ \bibnamefont {Pereira}}, \bibinfo {author} {\bibfnamefont {M.~N.}\ \bibnamefont {Rebelo}}, \ and\ \bibinfo {author} {\bibfnamefont {J.~I.}\ \bibnamefont {Silva-Marcos}},\ }\href {\doibase 10.1016/j.physrep.2023.12.004} {\bibfield  {journal} {\bibinfo  {journal} {Phys. Rept.}\ }\textbf {\bibinfo {volume} {1057}},\ \bibinfo {pages} {1} (\bibinfo {year} {2024})},\ \Eprint {http://arxiv.org/abs/2304.10561} {arXiv:2304.10561 [hep-ph]} \BibitemShut {NoStop}%
\bibitem [{\citenamefont {Ghosh}\ \emph {et~al.}(2022)\citenamefont {Ghosh}, \citenamefont {Konar},\ and\ \citenamefont {Roshan}}]{Ghosh:2022rta}%
  \BibitemOpen
  \bibfield  {author} {\bibinfo {author} {\bibfnamefont {A.}~\bibnamefont {Ghosh}}, \bibinfo {author} {\bibfnamefont {P.}~\bibnamefont {Konar}}, \ and\ \bibinfo {author} {\bibfnamefont {R.}~\bibnamefont {Roshan}},\ }\href {\doibase 10.1007/JHEP12(2022)167} {\bibfield  {journal} {\bibinfo  {journal} {JHEP}\ }\textbf {\bibinfo {volume} {12}},\ \bibinfo {pages} {167} (\bibinfo {year} {2022})},\ \Eprint {http://arxiv.org/abs/2207.00487} {arXiv:2207.00487 [hep-ph]} \BibitemShut {NoStop}%
\bibitem [{\citenamefont {Ghosh}\ and\ \citenamefont {Konar}(2023)}]{Ghosh:2023xhs}%
  \BibitemOpen
  \bibfield  {author} {\bibinfo {author} {\bibfnamefont {A.}~\bibnamefont {Ghosh}}\ and\ \bibinfo {author} {\bibfnamefont {P.}~\bibnamefont {Konar}},\ }\href@noop {} {\  (\bibinfo {year} {2023})},\ \Eprint {http://arxiv.org/abs/2305.08662} {arXiv:2305.08662 [hep-ph]} \BibitemShut {NoStop}%
\bibitem [{\citenamefont {Ghosh}\ and\ \citenamefont {Konar}(2024)}]{Ghosh:2024boo}%
  \BibitemOpen
  \bibfield  {author} {\bibinfo {author} {\bibfnamefont {A.}~\bibnamefont {Ghosh}}\ and\ \bibinfo {author} {\bibfnamefont {P.}~\bibnamefont {Konar}},\ }\href@noop {} {\  (\bibinfo {year} {2024})},\ \Eprint {http://arxiv.org/abs/2407.01415} {arXiv:2407.01415 [hep-ph]} \BibitemShut {NoStop}%
\bibitem [{\citenamefont {Beyer}\ and\ \citenamefont {Sarkar}(2023)}]{Beyer:2022ywc}%
  \BibitemOpen
  \bibfield  {author} {\bibinfo {author} {\bibfnamefont {K.~A.}\ \bibnamefont {Beyer}}\ and\ \bibinfo {author} {\bibfnamefont {S.}~\bibnamefont {Sarkar}},\ }\href {\doibase 10.21468/SciPostPhys.15.1.003} {\bibfield  {journal} {\bibinfo  {journal} {SciPost Phys.}\ }\textbf {\bibinfo {volume} {15}},\ \bibinfo {pages} {003} (\bibinfo {year} {2023})},\ \Eprint {http://arxiv.org/abs/2211.14635} {arXiv:2211.14635 [hep-ph]} \BibitemShut {NoStop}%
\bibitem [{\citenamefont {Hayrapetyan}\ \emph {et~al.}(2024)\citenamefont {Hayrapetyan} \emph {et~al.}}]{CMS:2024vjn}%
  \BibitemOpen
  \bibfield  {author} {\bibinfo {author} {\bibfnamefont {A.}~\bibnamefont {Hayrapetyan}} \emph {et~al.} (\bibinfo {collaboration} {CMS}),\ }\href@noop {} {\  (\bibinfo {year} {2024})},\ \Eprint {http://arxiv.org/abs/2405.00834} {arXiv:2405.00834 [hep-ex]} \BibitemShut {NoStop}%
\bibitem [{\citenamefont {Steinberg}\ and\ \citenamefont {Wells}(2021)}]{Steinberg:2021iay}%
  \BibitemOpen
  \bibfield  {author} {\bibinfo {author} {\bibfnamefont {N.}~\bibnamefont {Steinberg}}\ and\ \bibinfo {author} {\bibfnamefont {J.~D.}\ \bibnamefont {Wells}},\ }\href {\doibase 10.1007/JHEP08(2021)120} {\bibfield  {journal} {\bibinfo  {journal} {JHEP}\ }\textbf {\bibinfo {volume} {08}},\ \bibinfo {pages} {120} (\bibinfo {year} {2021})},\ \Eprint {http://arxiv.org/abs/2101.00520} {arXiv:2101.00520 [hep-ph]} \BibitemShut {NoStop}%
\bibitem [{\citenamefont {Schwartz}(2013)}]{Schwartz_2013}%
  \BibitemOpen
  \bibfield  {author} {\bibinfo {author} {\bibfnamefont {M.~D.}\ \bibnamefont {Schwartz}},\ }\href@noop {} {\emph {\bibinfo {title} {Quantum Field Theory and the Standard Model}}}\ (\bibinfo  {publisher} {Cambridge University Press},\ \bibinfo {year} {2013})\BibitemShut {NoStop}%
\bibitem [{\citenamefont {Bilal}(2008)}]{Bilal:2008qx}%
  \BibitemOpen
  \bibfield  {author} {\bibinfo {author} {\bibfnamefont {A.}~\bibnamefont {Bilal}},\ }\href@noop {} {\  (\bibinfo {year} {2008})},\ \Eprint {http://arxiv.org/abs/0802.0634} {arXiv:0802.0634 [hep-th]} \BibitemShut {NoStop}%
\end{thebibliography}%

\end{document}